
 %
\input harvmac.tex
 %
\catcode`@=11
\def\rlx{\relax\leavevmode}                  
 %
 %
 %
\font\tenmib=cmmib10
\font\sevenmib=cmmib10 at 7pt 
\font\fivemib=cmmib10 at 5pt  
\font\tenbsy=cmbsy10
\font\sevenbsy=cmbsy10 at 7pt 
\font\fivebsy=cmbsy10 at 5pt  
\def\BMfont{\textfont0\tenbf \scriptfont0\sevenbf
                              \scriptscriptfont0\fivebf
            \textfont1\tenmib \scriptfont1\sevenmib
                               \scriptscriptfont1\fivemib
            \textfont2\tenbsy \scriptfont2\sevenbsy
                               \scriptscriptfont2\fivebsy}
\def\BM#1{\rlx\ifmmode\mathchoice
                      {\hbox{$\BMfont#1$}}
                      {\hbox{$\BMfont#1$}}
                      {\hbox{$\scriptstyle\BMfont#1$}}
                      {\hbox{$\scriptscriptstyle\BMfont#1$}}
                 \else{$\BMfont#1$}\fi}
 %
 %
 %
 %
\def\inbar{\vrule height1.5ex width.4pt depth0pt}
\def\sinbar{\vrule height1ex width.35pt depth0pt}
\def\ssinbar{\vrule height.7ex width.3pt depth0pt}
\font\cmss=cmss10
\font\cmsss=cmss10 at 7pt
\def\ZZ{\rlx\leavevmode
             \ifmmode\mathchoice
                    {\hbox{\cmss Z\kern-.4em Z}}
                    {\hbox{\cmss Z\kern-.4em Z}}
                    {\lower.9pt\hbox{\cmsss Z\kern-.36em Z}}
                    {\lower1.2pt\hbox{\cmsss Z\kern-.36em Z}}
               \else{\cmss Z\kern-.4em Z}\fi}
\def\Ik{\rlx{\rm I\kern-.18em k}}  
\def\IC{\rlx\leavevmode
             \ifmmode\mathchoice
                    {\hbox{\kern.33em\inbar\kern-.3em{\rm C}}}
                    {\hbox{\kern.33em\inbar\kern-.3em{\rm C}}}
                    {\hbox{\kern.28em\sinbar\kern-.25em{\sevenrm C}}}
                    {\hbox{\kern.25em\ssinbar\kern-.22em{\fiverm C}}}
             \else{\hbox{\kern.3em\inbar\kern-.3em{\rm C}}}\fi}
\def\IP{\rlx{\rm I\kern-.18em P}}
\def\IR{\rlx{\rm I\kern-.18em R}}
\def\Ione{\rlx{\rm 1\kern-2.7pt l}}
 %
 %
\def\boxit#1#2{\hbox{\vrule\vbox{
  \hrule\vskip#1\hbox{\hskip#1\vbox{#2}\hskip#1}%
        \vskip#1\hrule}\vrule}}

\def\intem#1{\par\leavevmode%
              \llap{\hbox to\parindent{\hss{#1}\hfill~}}\ignorespaces}
 %


 %
\newskip\humongous \humongous=0pt plus 1000pt minus 1000pt   
\def\caja{\mathsurround=0pt}
\newif\ifdtup
 %
\def\eqalign#1{\,\vcenter{\openup2\jot \caja
     \ialign{\strut \hfil$\displaystyle{##}$&$
      \displaystyle{{}##}$\hfil\crcr#1\crcr}}\,}
 %

 %
\def\panorama{\global\dtuptrue \openup2\jot \caja
     \everycr{\noalign{\ifdtup \global\dtupfalse
      \vskip-\lineskiplimit \vskip\normallineskiplimit
      \else \penalty\interdisplaylinepenalty \fi}}}
 %
\def\eqalignno#1{\panorama \tabskip=\humongous
     \halign to\displaywidth{\hfil$\displaystyle{##}$
      \tabskip=0pt&$\displaystyle{{}##}$\hfil
       \tabskip=\humongous&\llap{$##$}\tabskip=0pt\crcr#1\crcr}}
 %

 %
\def\twoeqsalignno#1{\panorama \tabskip=\humongous
     \halign to\displaywidth{\hfil$\displaystyle{##}$
      \tabskip=0pt&$\displaystyle{{}##}$\hfil
       \tabskip=0pt&\hfil$\displaystyle{##}$
        \tabskip=0pt&$\displaystyle{{}##}$\hfil
         \tabskip=\humongous&\llap{$##$}\tabskip=0pt\crcr#1\crcr}}
 %

 %
 %
 %
 %
   \let\SS=\S       
\def\,{\hskip1.5pt}           
 %
\let\a=\alpha
\let\b=\beta
\let\c=\chi
\let\d=\delta       \let\vd=\partial             \let\D=\Delta
\let\e=\epsilon     \let\ve=\varepsilon
\let\f=\phi         \let\vf=\varphi              
\let\g=\gamma                                    \let\G=\Gamma
\let\h=\eta
\let\i=\iota
\let\j=\psi                                      
\let\k=\kappa
\let\l=\lambda                                   \let\L=\Lambda
\let\m=\mu
\let\n=\nu
\let\p=\pi          \let\vp=\varpi               \let\P=\Pi
       \let\vq=\vartheta            
\let\r=\rho         
\let\s=\sigma                   \let\S=\Sigma
\let\t=\tau
                                    \let\W=\Omega

 %
 %
\def\Box{\sqcap\llap{$\sqcup$}}
\def\lapp{\lower.4ex\hbox{\rlap{$\sim$}} \raise.4ex\hbox{$<$}}
\def\gapp{\lower.4ex\hbox{\rlap{$\sim$}} \raise.4ex\hbox{$>$}}
\def\con{\ifmmode\raise.1ex\hbox{\bf*}
          \else\raise.1ex\hbox{\bf*}\fi}
\def\bo{{\raise.15ex\hbox{\large$\Box\kern-.39em$}}}

\let\To=\Rightarrow

\def\dual{\relax\leavevmode\lower.9ex\hbox{\titlerms*}}
\def\define{\buildrel\rm def\over =}
\let\id=\equiv
\let\8=\otimes
 %
 %
 %
 %
\let\ba=\overline
\let\2=\underline

\let\Tw=\widetilde
 %
\def\dt#1{{\buildrel{\smash{\lower1pt\hbox{.}}}\over{#1}}}
\def\pd#1#2{{\partial#1\over\partial#2}}
\def\ppd#1#2#3{{\partial^2#1\over\partial#2\partial#3}}
\def\pppd#1#2#3#4{{\partial^3#1\over\partial#2\partial#3\partial#4}}
\font\eightrm=cmr8
\def\6(#1){\relax\leavevmode\hbox{\eightrm(}#1\hbox{\eightrm)}}
\def\0#1{\relax\ifmmode\mathaccent"7017{#1}     
                \else\accent23#1\relax\fi}      
\def\7#1#2{{\mathop{\null#2}\limits^{#1}}}      
\def\5#1#2{{\mathop{\null#2}\limits_{#1}}}      
 %

\def\ket#1{\left| #1\right\rangle}
\def\V#1{\langle#1\rangle}

 %

 %

 %

 %
\newbox\t@b@x
\def\rightarrowfill{$\m@th \mathord- \mkern-6mu
     \cleaders\hbox{$\mkern-2mu \mathord- \mkern-2mu$}\hfill
      \mkern-6mu \mathord\rightarrow$}
\def\tooo#1{\setbox\t@b@x=\hbox{$\scriptstyle#1$}%
             \mathrel{\mathop{\hbox to\wd\t@b@x{\rightarrowfill}}%
              \limits^{#1}}\,}
\def\leftarrowfill{$\m@th \mathord\leftarrow \mkern-6mu
     \cleaders\hbox{$\mkern-2mu \mathord- \mkern-2mu$}\hfill
      \mkern-6mu \mathord-$}
\def\froo#1{\setbox\t@b@x=\hbox{$\scriptstyle#1$}%
             \mathrel{\mathop{\hbox to\wd\t@b@x{\leftarrowfill}}%
              \limits^{#1}}\,}
 %
\def\frac#1#2{{#1\over#2}}
\def\frc#1#2{\relax\ifmmode{\textstyle{#1\over#2}} 
                    \else$#1\over#2$\fi}           
\def\inv#1{\frc{1}{#1}}                            
 %
\def\Claim#1#2#3{\bigskip\begingroup%
                  \xdef #1{\secsym\the\meqno}%
                   \writedef{#1\leftbracket#1}%
                    \global\advance\meqno by1\wrlabeL#1%
                     \noindent{\bf#2}\,#1{}\,:~\sl#3\vskip1mm\endgroup}

\def\QED{\rlx\hfill$\Box$\kern-7pt\raise3pt\hbox{$\surd$}\bigskip}
 %
 %

 %
\def\muthstrut{\vphantom1}
\def\mutrix#1{\null\,\vcenter{\normalbaselines\m@th
        \ialign{\hfil$##$\hfil&&~\hfil$##$\hfill\crcr
            \muthstrut\crcr\noalign{\kern-\baselineskip}
            #1\crcr\muthstrut\crcr\noalign{\kern-\baselineskip}}}\,}

 %
\def\YT#1#2{\vcenter{\hbox{\vbox{\baselineskip0pt\parskip=\medskipamount%
             \def\Box{$\sqcap\llap{$\sqcup$}$\kern-1.2pt}%
              \def\Z{\hfil\vskip-5.8pt}\lineskiplimit0pt\lineskip0pt%
               \setbox0=\hbox{#1}\hsize\wd0\parindent=0pt#2}\,}}}
\def\EU{\rlx\ifmmode \c_{{}_E} \else$\c_{{}_E}$\fi}
\def\TM{\rlx\ifmmode {\cal T_M} \else$\cal T_M$\fi}
\def\TW{\rlx\ifmmode {\cal T_W} \else$\cal T_W$\fi}
\def\CM{\rlx\ifmmode {\cal T\rlap{\bf*}\!\!_M}
             \else$\cal T\rlap{\bf*}\!\!_M$\fi}
\def\hm#1#2{\rlx\ifmmode H^{#1}({\cal M},{#2})
                 \else$H^{#1}({\cal M},{#2})$\fi}
\def\CP#1{\rlx\ifmmode\IP^{#1}\else\IP$^{#1}$\fi}
\def\cP#1{\rlx\ifmmode\IC{\rm P}^{#1}\else$\IC{\rm P}^{#1}$\fi}

\def\sll#1{\rlx\rlap{\,\raise1pt\hbox{/}}{#1}}
\def\Sll#1{\rlx\rlap{\,\kern.6pt\raise1pt\hbox{/}}{#1}\kern-.6pt}

\let\SSS=\scriptstyle
\let\ttt=\textstyle
\let\ddd=\displaystyle
 %
 %
\def\ie{\hbox{\it i.e.}}        

\def\CY{Calabi-\kern-.2em Yau}
\def\LGO{Landau-Ginzburg orbifold}
\def\3{\ifmmode\ldots\else$\ldots$\fi}
\def\Z{\hfil\break\rlx\hbox{}\quad}
\def\3{\ifmmode\ldots\else$\ldots$\fi}
\def\?{d\kern-.3em\raise.64ex\hbox{-}}           
\def\9{\raise.43ex\hbox{-}\kern-.37em D}         
\def\ping{\nobreak\par\centerline{---$\circ$---}\goodbreak\bigskip}
 %
 %

 %
\def\Pre#1{{\it #1\ University report}}

\def\NP#1{{\it Nucl.\,Phys.\,}{\bf#1\,}}
\def\PL#1{{\it Phys.\,Lett.\,}{\bf#1\,}}

\def\MPL#1{{\it Mod.\,Phys.\,Lett.\,}{\bf#1\,}}

\def\CMP#1{{\it Commun.\,Math.\,Phys.\,}{\bf#1\,}}
\def\CQG#1{{\it Class.\,Quant.\,Grav.\,}{\bf#1\,}}
\def\IJMP#1{{\it Int.\,J.\,Mod.\,Phys.\,}{\bf#1\,}}
 %
 %
 %
\baselineskip=13.0861pt plus2pt minus1pt
\parskip=\medskipamount
\let\ft=\foot
\noblackbox
\def\SaveTimber{\abovedisplayskip=1.5ex plus.3ex minus.5ex
                \belowdisplayskip=1.5ex plus.3ex minus.5ex
                \abovedisplayshortskip=.2ex plus.2ex minus.4ex
                \belowdisplayshortskip=1.5ex plus.2ex minus.4ex
                \baselineskip=12pt plus1pt minus.5pt
 \parskip=\smallskipamount
 \def\ft##1{\unskip\,\begingroup\footskip9pt plus1pt minus1pt\setbox%
             \strutbox=\hbox{\vrule height6pt depth4.5pt width0pt}%
              \global\advance\ftno by1\footnote{$^{\the\ftno)}$}{##1}%
               \endgroup}
 \def\listrefs{\footatend\vfill\immediate\closeout\rfile%
                \writestoppt\baselineskip=10pt%
                 \centerline{{\bf References}}%
                  \bigskip{\frenchspacing\parindent=20pt\escapechar=` %
                   \rightskip=0pt plus4em\spaceskip=.3333em%
                    \input refs.tmp\vfill\eject}\nonfrenchspacing}}
 %
\def\Afour{\ifx\answ\bigans
            \hsize=16.5truecm\vsize=24.7truecm
             \else
              \hsize=24.7truecm\vsize=16.5truecm
               \fi}
\catcode`@=12

 %
\def\rd{{\rm d}}
\def\db{\relax\leavevmode\hbox{$\partial$\kern-.4em
               \vrule height1.86ex depth-1.8ex width4pt}\kern.8pt}
\def\\{\hfill\break}
\def\nin{\not\in}
\def\rv{\skew{-2}\vec{r}}
\def\mv{\skew{-2}\vec{m}}
\def\kv{\skew{-2}\vec{k}}
\def\1{\vrule height2ex depth.7ex width0pt}
\def\PGL#1{\ifmmode\hbox{PGL}(#1)\else{PGL$(#1)$}\fi}
\def\stack#1#2#3{\matrix{#1\cr\noalign{\vglue#2pt}#3\cr}}
\def\Stack#1#2#3#4#5{\matrix{#1\cr\noalign{\vglue#2pt}#3\cr
                                  \noalign{\vglue#4pt}#5\cr}}
\def\Cnfg#1#2{\matrix{#1}\mkern-6mu\left[\matrix{#2}\right]}
\def\cnfg#1#2{\hbox{\relax\ninepoint\def\normalbaselines{%
             \baselineskip9pt\lineskip.5pt\lineskiplimit.5pt}
              $\mutrix{#1}\mkern-7mu\left[\mutrix{#2}\right]$}}
\def\phrasing#1#2{\nobreak\vglue-5mm%
                   \leftline{\ninepoint\rm(#1,{\it---#2})}%
                    \nobreak\vglue2mm\noindent\ignorespaces}
\def\phrasing#1#2{\ignorespaces}
\def\oiint{\mathop{\bigcirc\mkern-25mu\int\mkern-14mu\int}}
\def\Res{{\rm Res}}
\def\mtx(#1,#2,#3){\left(\matrix{#1\cr\noalign{\vglue-1.3mm}
                               #2\cr\noalign{\vglue-1.3mm}#3}\right)}
 %
 %
\Title{\vbox{\baselineskip12pt\hbox{HUPAPP-94/103}
                               \hbox{IASSNS-HEP-94/97}
                                \hbox{hep-th/9411131}}}
      {\vbox{\centerline{On a Residue Representation of}
              \vskip3mm
             \centerline{Deformation, Koszul and Chiral Rings}}}
\centerline{\titlerms Per Berglund}
\vskip0mm
 \centerline{\it School of Natural Sciences}              \vskip0mm
 \centerline{\it Institute for Advanced Study}            \vskip0mm
 \centerline{\it Olden Lane, Princeton, NJ 08540}   \vskip0mm
 \centerline{\rm berglund@guinness.ias.edu}               \vskip0mm
\vskip .2in
\centerline{\titlerms and}
\vskip .2in
\centerline{\titlerms Tristan H\"ubsch\footnote{$^{\spadesuit}$}
            {On leave from the ``Ru\?er Bo\v skovi\'c'' Institute,
             Zagreb, Croatia.}}          \vskip0mm
 \centerline{\it Department of Physics and Astronomy}     \vskip0mm
 \centerline{\it Howard University, Washington, DC~20059} \vskip0mm
 \centerline{\rm hubsch@cldc.howard.edu}                  \vskip0mm
\vfill

\centerline{ABSTRACT}\vskip2mm
\vbox{\narrower\narrower\baselineskip=12pt\noindent
 A residue-theoretic representation is given for massless matter fields
in (quotients) of (weighted) \CY\ complete intersection models
and the corresponding chiral operators in \LGO{s}.
 The well known polynomial deformations are thus generalized and the
universal but somewhat abstract Koszul computations acquire a concrete
realization and a general but more heuristic reinterpretation.
 A direct correspondence with a BRST-type analysis of constrained systems
also emerges naturally.}

\Date{November 1994\hfill}
 %
\vfill\eject
\footline{\hss\tenrm--\,\folio\,--\hss}
 %
 %
\lref\rStrResOrb{P.S.~Aspinwall: Resolution of Orbifold Singularities
       in String Theory. {\it Institute for Advanced Study report}
       IASSNS-HEP-94/9.}

\lref\rMinDist{P.S.~Aspinwall: Minimum Distances in Non-Trivial String
       Target Spaces. {\it Institute for Advanced Study report}
       IASSNS-HEP-94/19.}

\lref\rAGM{P.S.~Aspinwall, B.R.~Greene and D.R.~Morrison:
       \NP{B416}(1994)414.}

\lref\rMonDivMap{P.S.~Aspinwall,  B.R.~Greene and D.R.~Morrison:
       The Monomial-Divisor Mirror Map. {\it Institute for Advanced
       Study report} IASSNS-HEP-93/43.}

\lref\rSmallD{P.S.~Aspinwall, B.R.~Greene and D.R.~Morrison:
       \NP{B420}(1994)184.}

\lref\rSpTTopCh{P.S.~Aspinwall, B.R.~Greene and D.R.~Morrison:
       Spacetime Topology Change: The Physics of \CY\ Moduli Space.
       {\it Institute for Advanced Study report} IASSNS-HEP-93/81.}

\lref\rABG{M.~Atiyah, R.~Bott and L.~G{\aa}rding:
       Acta Math.~{\bf 131}(1973)145.}

\lref\rBasEast{R.J.~Baston and M.G.~Eastwood: {\it The Penrose
       Transform---Its Interaction with Representation Theory}
       (Claredon Press, Oxford,1989).}

\lref\rPeriods{P.~Berglund, P.~Candelas, X.~de la Ossa, A.~Font,
       T.~H\"ubsch, D.~Jan\v{c}i\'c and F.~Quevedo: \NP{B419}(1994)352.}

\lref\rSpokes{P.~Berglund, E.~Derrick, T.~H\"ubsch and
       D.~Jan\v{c}i\'{c}: \NP{B420}(1994)268.}

\lref\rCQG{P.~Berglund, B.~Greene and T.~H\"ubsch: \MPL{A7}(1992)1885.}

\lref\rGCM{P.~Berglund and T.~H\"ubsch: \NP{B393}(1993)377, also in
       {\it Essays on Mirror Manifolds}~p.388--407, ed.\ S.-T.~Yau,
       (International Press, Hong Kong, 1992).}

\lref\rCfC{P.~Berglund and T.~H\"ubsch: \NP{B411}(1994)223.}

\lref\rMatter{P.~Berglund, T.~H\"ubsch and L.~Parkes: The Complete
       Matter Sector in a Three-Generation Compactification.
       \CMP{148}(1992)57.}

\lref\rBerKatz{P.~Berglund and S.~Katz: \NP{B420}(1994)289.}

\lref\rCF{A.~Cadavid and S.~Ferrara: \PL{B267}(1991)193.}

\lref\rPhilip{P.~Candelas: \NP{B298}(1988)458. }

\lref\rCdFKM{P.~Candelas, X.~de~la~Ossa, A.~Font, S.~Katz and
       D.R.~Morrison: \NP{B416}(1994)481.}

\lref\rCdGP{P.~Candelas, X.~de la Ossa, P.S.~Green and L.~Parkes:
       \PL{258B}(1991)118, \NP{B359}(1991)21. }

\lref\rCOK{P.~Candelas, X.~de~la~Ossa and S.~Katz: ``Mirror Symmetry for
       Calabi-Yau Hypersurfaces in Weighted $\IP^4$ and an Extension of
       Landau-Ginzburg Theory'', in preparation.}

\lref\rCFKM{P.~Candelas, A.~Font, S.~Katz and D.R.~Morrison:
\NP{B429}(1994)626.}

\lref\rRolling{P.~Candelas, P.S.~Green and T.~H\"ubsch: \NP{B330}(1990)49.}

\lref\rEndT{M.G.~Eastwood and T.~H\"ubsch: \CMP{132}(1990)383\semi
       P.~Berglund, T.~H\"ubsch and L.~Parkes: \MPL{A5}(1990)1485,
       \CMP{148}(1992)57.}

\lref\rFerStr{S.~Ferrara and A.~Strominger: N=2 Spacetime
       Supersymmetry and \CY\ Moduli Space. in the Proceedings of the
       ``Strings '89'' meeting in College Station, Texas. }

\lref\rPDM{P.~Green and T.~H\"ubsch: \CMP{113}(1987)505\semi
       T.~H\"ubsch: in {\it Superstrings, Unifying Theories and
       Cosmology 1987}, p.164, ed.\ G.~Furlan et al.\ (World
       Scientific, Singapore, 1988).}

\lref\rPDM{P.~Green and T.~H\"ubsch: \CMP{113}(1987)505.}

\lref\rCYCY{P.~Green and T.~H\"ubsch: Spacetime Variable Superstring
       Vacua. \IJMP{A9}(1994)3203.}

\lref\rWBRG{B.R.~Greene: \CMP{130}(1990)335.}

\lref\rGVW{B.R.~Greene, C.~Vafa and N.P.~Warner:
       \NP{B324}(1989)371\semi
       E.~Martinec: \PL{217B}(1989)431, also in {\it Physics and
       Mathematics of Strings}, p.389--433, eds.~L.~Brink et.~al\semi
       S.~Cecotti, L.~Girardello and A.~Pasquinucci:
       \NP{B328}(1989)701, \IJMP{A6}(1991)2427.}

\lref\rGr{P.~Griffiths: {\it Bull.A.M.S.}{\bf1}(1979)595.}

\lref\rGrHa{P.~Griffiths and J.~Harris: {\it Principles of Algebraic
       Geometry} (John Wiley, New York, 1978).}

\lref\rHKTYI{S.~Hosono, A.~Klem, S.~Theisen and S.-T.~Yau: Mirror
       Symmetry, Mirror Map and Applications to Calabi-Yau
       Hypersurfaces. \Pre{Harvard} HUTMP-93-0801.}

\lref\rHKTYII{S.~Hosono, A.~Klem, S.~Theisen and S.-T.~Yau: Mirror
       Symmetry, Mirror Map and Applications to Complete
       Intersection Calabi-Yau Spaces. CERN-TH-7369-94.}

\lref\rChaSM{T.~H\"ubsch: \PL{247B}(1990)317.}

\lref\rSpliTwo{T.~H\"ubsch: \CQG{8}(1991)L31.}

\lref\rBeast{T.~H\"ubsch: {\it \CY\ Manifolds---A Bestiary for
       Physicists}\Z (World Scientific, Singapore, 1992).}

\lref\rSKEW{S.~Kachru and E.~Witten: \NP{B407}(1993)637.}

\lref\rMax{M.~Kreuzer and H.~Skarke: CERN preprint CERN-TH-6373/92.}

\lref\rPicFux{W.~Lerche, D.-J.~Smit and N.~Warner: \NP{B372}(1992)87\semi
       D.R.~Morrison: in {\it Essays on Mirror Manifolds}~p.241--264, ed.\
       S.-T.~Yau, (International Press, Hong Kong, 1992).}

\lref\rRolf{R.~Schimmrigk: \PL{193B}(1987)193.}

\lref\rSpetz{A.~Strominger: \CMP{133}(1990)163.}

\lref\rLGO{C.~Vafa: \MPL{A4}(1989)1169\semi
       K.~Intrilligator and C.~Vafa: \NP{B339}(1990)95.}

\lref\rChiRi{C.~Vafa and N.~Warner: \PL{218B}(1989)51\semi
       W.~Lerche, C.~Vafa and N.~Warner: \NP{B324}(1989)427.}

\lref\rPhases{E.~Witten: \NP{B403}(1993)159.}

 %
 %

\newsec{Introduction, Results and Summary}\seclab\SIRS\noindent
 \phrasing{reveille and recapitulation}{rash}
Two rather general techniques for studying the complete massless
spectra of superstring compactifications\ft{By complete we mean including
$E_6$ gauge singlets in addition to the moduli fields associated to
deformations of the K\"ahler and complex structure.} have been developed
over the past  several years. One of these~\rEndT\ relies on the `large
radius' description in terms of the geometry of \CY\ spaces, wherein low
energy particles correspond to elements of $H^1({\cal T})$,
$H^1({\cal T}\con)$ and $H^1({\rm End}{\cal T})$.
 For \CY\ complete intersections in products of projective spaces, these
cohomology groups may be calculated through vigorous application of Koszul
spectral sequences and the Bott-Borel-Weil theorem~\refs{\rMatter,\rBeast}.
In fact, complete intersections in products of any generalized flag spaces
can be analyzed in this manner\ft{In principle, the technique also applies to
(quotients) of complete intersections in products of any weighted flag
varieties and, in particular, weighted projective spaces. While we are not
aware of any readily available weighted generalization of the Bott-Borel-Weil
theorem, we will show that the method which we propose herein naturally
applies to these latter models as well.}; further details of these models and
the details of this technique may be found in Ref.~\rBeast.

Complementing these results, the massless spectra of \LGO{s} with
worldsheet $(2,2)$-supersymmetry have been analyzed from the 2-dimensional
supersymmetric quantum field theory point of view~\refs{\rLGO,\rSKEW}.
Interestingly, these two complementing approaches---the former focusing on
special features of target space geometry while the latter centering on
exceptional characteristics of the underlying world-sheet field
theory---actually produce rather closely related descriptions of the
massless spectra for related models~\refs{\rGVW,\rCQG, \rCfC}.
 Indeed, both Landau-Ginzburg and \CY\ models may be regarded as different
`phases' or regimes of a more general underlying 2-dimensional quantum
field theory; they occur at two opposite special points in a complexified
radial moduli space~\rPhases, and may be connected by a sort of analytic
continuation. The remarkably detailed similarity~\refs{\rCQG,\rCfC} in the
results of these two approaches then follows from the fact that many of the
relevant observables, in particular properties of the complex structure
moduli space, are independent of this complex radial modulus.
 More precisely, a general \CY\ complete intersection does not have a pure
Landau-Ginzburg model as the radially counterpoint `phase', but rather a
so-called gauged Landau-Ginzburg model.
 Nevertheless, the $E_6$ {\bf27}'s and {\bf27*}'s of the low-energy
effective particle theory correspond on the geometrical side to elements of
$H^1({\cal T})$ and $H^1({\cal T}\con)$ respectively, while in the \LGO{s}
framework the corresponding states generate the $(c,c)$- and the $(a,c)$-ring,
respectively~\rChiRi.
 Besides the complex structure and K\"ahler moduli fields, the low-energy
effective particle theory also abounds in matter {\bf1}'s---chargeless with
respect to the $E_6\times E_8$ Yang-Mills gauge interaction. These states
correspond to elements of the $H^1({\rm End}{\cal T})$ cohomology group and
are not chiral-primary, but nevertheless admit a similar analysis~\rSKEW\
based on the $(0,2)$-subgroup of the $(2,2)$-supersymmetry.

Generalizing the analysis of Ref.~\rCdGP, a general machinery has recently
been developed to calculate via Special
Geometry~\refs{\rRolling,\rFerStr,\rSpetz} both the Yukawa couplings and the
kinetic terms for the `{\bf27}'s---and then also the `{\bf27*}'s by using
mirror symmetry---as functions over the entire moduli space, for many
families of models~\refs{\rCdGP,\rPeriods,\rHKTYI,\rSpokes,\rHKTYII}.
 This reveals many global properties of the Yukawa couplings---and with it
certain global properties of the parameter
spaces~\refs{\rAGM,\rCdFKM,\rSmallD,\rStrResOrb,\rSpTTopCh,\rCFKM}.
 In addition, the analysis of Refs.~\refs{\rPeriods,\rSpokes} is well
adapted to the fact that twisted $(c,c)$- and $(a,c)$-states may
(at least sometimes) be represented by certain radical polynomials;
that is, polynomials which also involve roots of certain special and in a
sense universal polynomials~\rCfC.
 Related to these, the analysis of the so-called Picard-Fuchs equations
also reveals a great deal of information about the periods and so the
Yukawa couplings~\refs{\rPicFux,\rHKTYI,\rHKTYII}.
\ping

In the present article we introduce another representation for the
massless states, based on certain residue integrals. Besides a possible
application in its own right, we find this residue map bridging the
unpleasant chasm between the well-accepted polynomial deformation method of
Ref.~\rPhilip, together with the very closely related construction of
marginal operators in a \LGO{}~\refs{\rLGO,\rChiRi,\rSKEW} on one side,
and the universal\ft{Pedantry: the techniques we mention rely either on the
geometrical or on the 2-dimensional field theory interpretation of a given
model, whence `universal' means `universal within the scope of
interpretation'. Since there exist models for which both interpretations
are not known, the two categories overlap significantly but neither
contains the other.} but rather abstract technique of
Koszul spectral sequences~\refs{\rEndT,\rBeast} on the other.
Although we are not at this point able to propose a direct generalization of
the Koszul calculation for (quotients) of complete intersections in weighted
projective spaces, the residue recipe does extend naturally to these models
also.
 The calculations based on Koszul spectral sequences (where applicable)
and the residue mapping also seem to acquire a natural interpretation in
terms of a BRST analysis of the underlying 2-dimensional
$(2,2)$-supersymmetric constrained $\s$-model~\rChaSM\ and the more general
linear quantum field theory~\rPhases.

Each representative obtained through the Koszul computation will be shown to
have a precisely corresponding residue integral, which turns out to be a
straightforward generalization of a well-known result. Indeed, our starting
point is provided by the well-known
Atiyah-Bott-G{\aa}rding-Candelas residue formula~\refs{\rABG,\rPhilip}:
\eqn\eABGC{ \W~~ \define
          ~~\Res_{\cal M}\Big[{(x\rd^n x)\over P}\Big]~. }
This defines the ``nowhere vanishing holomorphic $(n{-}1)$-form'' $\W$ on a
complex $(n{-}1)$-dimensional \CY\ hypersurface
 ${\cal M}\define\{P{=}0\}\subset\CP{n}$ as a residue at $\cal M$ of the
rational differential form $(x\rd^n x)/P$ where
\eqn\ePVol{ (x\rd^n x)~~ \define
            ~~{1\over (n{+}1)!} \e_{\m_0\cdots \m_n}
                         x^{\m_0}\,\rd x^{\m_1}\cdots\rd x^{\m_n}~. }
Note the appearance of the antisymmetric tensor $\e_{\cdots}$ in the above
expression, which will have a crucial role in understanding the exceptional
residues. Since each such top differential transforms at most with an
overall factor, we will refer to them as `covariant'. $\W$ is also called
the ``holomorphic volume-form'', as $\W{\wedge}\ba{\W}$ is a (perhaps
non-standard) volume form on $\cal M$. Explicitly, by means of a contour
integral,
\eqn\eABGCInt{ \W~~ \define ~~\oint_{\G(P)} {(x\rd^n x)\over P}~~. }
Here $\G(P)$ is a contour encircling ${\cal M}=\{P{=}0\}$. That is, $\G(P)$
may be identified with a small circle centered at some point $x\in{\cal M}$
and  which lies in a complex plane in \CP{n}, locally transversal to $\cal
M$ at $x$. Provided $\cal M$ is smooth, i.e., $P$ is transversal, the
integrand has a simple pole and the contour integral picks
out the simple residue at $x$. For the integral to be well defined
as an element of $H^3({\cal M},\IC)$, it must be of homogeneity
zero, which induces the \CY\ condition
\eqn\eCY{ \deg(P) ~=~ \sum_{\m=0}^n \deg(x^\m)~,
          \qquad{\rm over}~\CP{n}~.}
Note immediately that away from ${\cal M}\in\CP{n}$, where $P{\ne}0$, the
integrand is analytic and the value of the contour integral is zero. In
other words, $\W$ in~\eABGC\ is supported precisely on the hypersurface
$\cal M$, where it is nonzero and invariant under holonomy\ft{The holonomy
group of $\cal M$ is generated by parallel transport around closed loops in
$\cal M$, so holonomy-invariance generalizes single-valuedness and is
essential in compactification on $\cal M$~\rBeast. This holonomy-invariance
also coincides with the invariance with respect to a gauge-transformation
discussed in \SS~2.1.}.

While an invariant under holonomy over the given $\cal M$, the $\W$ of
Eq.~\eABGC\ however does depend on all the complex structure moduli and is
therefore the key object for the analysis in
Ref.~\refs{\rCdGP,\rPeriods,\rSpokes}, where it appears through its periods:
\eqn\ePeriods{ \vp_k~~ \define ~~\oint_{\g^k} \W~~, }
where the cycles $\g^k$ form a basis for $H_{n-1}({\cal M})$.
 Alternatively, one notes that the choice of some particular 3-form to be
the {\it holomorphic} 3-form is equivalent to having chosen a particular
complex structure; varying this choice then is equivalent to varying the
complex structure. Therefore
\eqn\eSpecGeom{ \pd{\W}{t^\a} ~=~ K_\a\W~ + ~\vf_\a~, \qquad{\rm or}\qquad
                \vf_\a ~\define~ \nabla_\a\W~, }
where $\nabla_\a\W \define (\vd_\a-K_\a)\W$, and $K_\a\rd t^\a$ is
identified as the connection 1-form. In particular, for deformations of
the complex structure which may be realized as deformations of the defining
polynomial\ft{This does exhaust all deformations of complex structure for
all homogeneous hypersurfaces, but not so for their quasihomogeneous
(weighted) cousins, nor for the homogeneous complete intersections~\rPDM.},
$P(t) \define P_0+t^\a \d P_\a$ and
\eqn\ePDM{ \pd{\W}{t^\a}\Big|_{t=0} ~=~
 -\oint_{\G(P_0)}{(x\rd^{n+1}x)\over P_0}\Big({\d P_\a\over P_0}\Big)~. }
So, up to terms which merely reproduce a multiple of $\W_0$, the
homogeneity-0 quantities $({\d P_\a\over P_0})$, and so also the
polynomials $\d P_\a$ (modulo the defining polynomials' gradients),
represent the deformations of the complex structure around $P_0$.

In certain special ``flat'' local coordinates $t^\a$, the connection 1-form
\eqn\eKahK{ K_\a\rd t^\a~ = ~{\int_{\cal M}\ba\W{\wedge}\rd_t\W \over
                       \int_{\cal M}\ba\W{\wedge}\W} }
is zero. However, for purposes of calculating the Yukawa couplings, the
connection terms may freely be omitted and partial derivatives
suffice~\refs{\rRolling,\rFerStr,\rSpetz} even if the
$t^\a$ are not the ``flat'' local coordinates:
\eqn\eYukC{ \k_{\a\b\g}~ =
            ~\int_{\cal M}\W\wedge\pppd{\W}{t^\a}{t^\b}{t^\g} }
is the (unnormalized) Yukawa coupling. The normalization derives from the
Weil-Petersson-Zamolodchikov metric, for which
\eqn\eKahP{ K~ = ~-\ln\Big(\,i\int_{\cal M}\ba\W\wedge\W\,\Big) }
is the K\"ahler potential.
 Thus, in principle, $\W$ completely determines the special geometry on
the space of complex structures and so also the complete dynamics of the
corresponding low-energy physics matter fields modulo higher loop
corrections to the K\"ahler potential.
\ping

Our main goal is to generalize the result~\ePDM\ so as to obtain
representatives of ``twisted'' massless states in the \LGO{} and of the
``higher cohomology'' contributions in the Koszul calculations, these two
being in a partial but very detailed correspondence wherever both `phases'
of a model are well understood~\refs{\rCQG,\rCfC}\ft{Since the \CY,
Landau-Ginzburg and the various `hybrid' phases are connected through
variations of the complexified radial moduli, results pertaining to the
complex structure will agree in any two phases. However, not so for the
K\"ahler structure: although the number of massless states remains the
same, the large radius limit does not have the `quantum symmetry' selection
rule which is a feature of the \LGO{} and so the K\"ahler structure Yukawa
couplings are different in the two phases.}. Notably, these residue
representatives turn out to be of the general form
\eqn\eGenForm{ \sum_{\rv}~\W^{\rv}_{(q)}\>f_{\rv}(x) }
where $\rv$ is a suitable multi-index and
$\W^{\rv}_{(q)}$ are the nowhere vanishing holomorphic volume-forms on
certain `intermediate' \CY\ $q$-dimensional spaces and the $f_{\rv}(x)$
are holomorphic on the complementary factor of the embedding space. These
turn out to provide a universal generalization of the Jacobian ring
structure of Ref.~\rPhilip --- the well known
 ``polynomials modulo the defining polynomials' gradients'' ring structure.

 The knowledge of periods~\ePeriods\ of the holomorphic volume-form $\W$
and certain monodromy information~\refs{\rCdGP,\rBerKatz,\rCdFKM,\rCFKM}
suffices to calculate both the Yukawa couplings and the kinetic terms,
and no further generalization is in principle necessary.
 However, a variety of simply technical or perhaps more essential obstacles
may thwart such a program. For example, a complete set of cycles $\g^k$
in~\ePeriods\ may be very difficult to find, and the action of the modular
groups may not be known sufficiently well to generate all the periods.
Certain deformations may not be representable as polynomial
deformations of the defining polynomial (which in fact is typical of
complete intersections in products of projective spaces~\rPDM).

 Finally, recall that both the Koszul machinery~\refs{\rEndT,\rBeast} and
also the \LGO{} analysis of Ref.~\rSKEW, each enables a systematic and
complete calculation, covering not only $H^1({\cal T})$ but also
$H^1({\cal T}\con)$ and $H^1({\rm End}{\cal T})$, and within the same
framework.
 By establishing a 1--1 correspondence between the universally valid Koszul
calculation for \CY\ models and the residue integral representations
provided here, we prove that the residue calculations also enjoy the
corresponding generality and completeness.
 A detailed correspondence with the universally valid \LGO{}
analysis of Ref.~\rSKEW---for models where both the \CY\ and the \LGO{}
`phases' are known---seems inviting, but will require a study on its own.
Herein, we content ourselves with some cursory remarks in this regard and
focus more on comparison with the $(2,2)$-supersymmetric \LGO{} analysis of
Refs.~\rLGO, as facilitated by known results~\refs{\rCQG,\rCfC}.
 Throughout the article, we also indicate another detailed 1--1
correspondence: that with the BRST analysis of constraint systems and the
associated ghost,- ghost-for-ghost,- {\it etc}.\ degrees of freedom.
Ultimately, this correspondence should provide a fully developed quantum
field theory generalization of our present results and we hope to return to
that in the future.

On the mathematical side, we show that the abstract and generally quite
difficult to realize cohomology elements obtained in such Koszul spectral
sequences have explicit class representatives in terms of residues such
as~\eABGC. Also, the various mappings---the so-called differentials in the
spectral sequences---will also be realized rather easily and are amenable for
calculation.

The paper is organized as follows. In \SS~2 we motivate and present the
basic paradigm for constructing the residue representatives, illustrated by
several examples in \SS~3. In \SS~4, we present a rather natural
generalization of the residue representation to weighted hypersurfaces and
quotients thereof. The main properties of the resulting ring structure are
discussed in \SS~5, while \SS~6 presents an alternative derivation of the
residue representatives and the `radical deformations'.

\newsec{Reaping Residues}\seclab\SReap\noindent
 \phrasing{a regal rococo recitative}{recte et retro}
It seems most natural to introduce the residue representation in the
context of complete intersections of $K$ hypersurfaces in a product of
$N$ projective spaces:
\eqn\eGCnfg{
 {\cal M} ~~\in~~ \Cnfg{\CP{n_1}_1\cr\vdots\cr\CP{n_N}_N}
                       {d_{11}&\ldots&d_{1K}\cr
                        \vdots&\ddots&\vdots\cr
                        d_{N1}&\ldots&d_{NK}\cr}~, }
defined as the simultaneous zero-set of the system of $K$
homogeneous polynomials
\eqn\eGDefEq{ f^j(x_i) ~=~ 0,\qquad
  d_{ij} \define \deg_{x_i}(f^j)~,\quad j=1,\ldots,K~, }
where $d_{ij} \define \deg_{x_i}(f^j)$ is the degree of the $j^{th}$
defining polynomial with respect to $x_i$, the array of homogeneous
coordinates on $\CP{n_i}_i$.
 The matrix on the r.h.s.\ of~\eGCnfg\ specifies the degrees of homogeneity
$d_{ij}$ and suffices to specify the Chern classes of
${\cal M}$. The vanishing of the first Chern class (for $\cal M$ to be a
\CY\ manifold) is ensured by requiring that
\eqn\eCYcond{ \sum_{j=1}^K d_{ij}~=~n_i+1,\quad i=1,\ldots,N~. }
For the generic model, $\dim{\cal M} = \sum_{i=1}^N n_i-K$; we obtain a
three-dimensional \CY\ complete intersection for $K=\sum_{i=1}^N n_i-3$.
It is merely for reasons of preserving a modicum of sanity with the already
unwieldy notation heavily beset with indices that we refrain from allowing
generalized (unitary) flag spaces\ft{Generalized flag spaces are quotients
$G/H$, where $H$ is a maximal regular subgroup of a finite-dimensional
Lie group $G$~\rBasEast; all of these can be utilized.}
 $\big\{U(n_1{+}\ldots{+}n_F)/\prod_{f=1}^FU(n_f)\big\}$
from appearing as factors in the embedding space~\rBeast. As the projective
space is the simplest flag space,
 $\CP{n}={U(n{+}1)\over U(1){\times}U(n)}$,
 the adventurous reader should have no problems other than notational in
extending our results to this even more general class. Note that at least
some of these models
 (involving Grassmannians $G_{n,k}={U(n+k)\over U(n){\times}U(k)}$)
turn out to be equivalent to certain gauged \LGO{s}~\rPhases.
 For most of the time, we furthermore restrict to ordinary (isotropic)
projective spaces and will discuss the (anisotropic) weighted projective
spaces \SS~4; accordingly, the defining polynomials $f^j(x_i)$ are for the
time being all homogeneous rather than quasihomogeneous.

Note that, apart from the degrees $d_{ij}$, the matrix in the r.h.s.\
of~\eGCnfg\ does not specify the defining polynomials $f^j(x_i)$; the
coefficients in $f^j(x_i)$ are therefore free and serve to parametrize the
deformation family of models represented by the configuration
matrix~\eGCnfg. A generic member of this family, $\cal M$, is smooth and we
write $b_{2,1}$ and $b_{1,1}$ for the number of its independent $(2,1)$- and
$(1,1)$-forms, respectively ($\dim{\cal M}{=}3$).

We start by considering several equivalent expressions for the
residue~\eABGC, adapted here from~\rPhilip,
\eqna\eOm
 $$ \W~\define
 ~\Res_{\cal M}\Big[{ \prod_{i=1}^N(x_i\rd^{n_i} x_i)
                           \over f^1\,f^2\,{\cdots}\,f^K }\Big]~.
\eqno\eOm{a}
 $$
The residue may be calculated by means of a suitable $K$-fold contour
integration:
 $$ \W~=
 ~{1\over(2\p i)^K}\oint_{\G(f^1)}{\cdots}\oint_{\G(f^K)}
    {\prod_{i=1}^N(x_i\rd^{n_i}x_i)\over
                               f^1\,f^2\,{\cdots}\,f^K}~,
\eqno\eOm{b}
 $$
where $\G(f^j)$ is a small loop encircling the complex hypersurface defined
by $f^j$. As usual, the residue integrals are in fact independent of any
specific choice of these poly-contours, in part owing to Eq.~\eCYcond. In
practice, the residues are obtained by taking the limit $\ve_j\to0$ and are
independent of the radii $\ve_j$.

 A somewhat tedious but completely straightforward calculation produces the
result of such integrations. For example, work in the coordinate patch
where $x_i^{\m^i_0}\ne0$, for $i=1,\ldots,N$. Noting that $\sum_in_i=K+3$,
one performs a change of variables
\eqn\eChng{ (x^{\m^1_1}, {\cdots}, x^{\m^1_{n_1}}, {\cdots},
             x^{\m^N_1}, {\cdots}, x^{\m^N_{n_N}})~ \longrightarrow
 ~(x^\n, x^\r, x^\s, f^1, {\cdots}, f^K)~, }
whence~\eOm{b} becomes
 $$ \W~=
 ~{1\over(2\p i)^K}\oint_{\G(f^1)}{\cdots}\oint_{\G(f^K)}
   \Big(\prod_{i=1}^N x_i^{\m^i_0}\Big)
   {\rd x^\n\,\rd x^\r\,\rd x^\s
     \over J^{~\n\r\s}_{(\m^1_0,\ldots,\m^N_0)}}
    \prod_{j=1}^K{\rd f^j\over f^j}~.
\eqno\eOm{c}
 $$
The $K$-fold residue integral is now easily completed to produce
 $$ \W~=
 ~\bigg[\Big(\prod_{i=1}^N x_i^{\m^i_0}\Big)
  {\rd x^\n\,\rd x^\r\,\rd x^\s\over
    J^{~\n\r\s}_{(\m^1_0,\ldots,\m^N_0)}}\bigg]_{f^j=0}~,\qquad
    \cases{&$x_i^{\m_0}\ne0$,\quad $i=1,\ldots,N$~,\cr
           \noalign{\vglue1mm}
           &no sum on $~\n,\r,\s$.\cr}
\eqno\eOm{d}
 $$
Of course, $J^{~\n\r\s}_{(\m^1_0,\ldots,\m^N_0)}$ is the Jacobian
of the inverse of the change of coordinates~\eChng.
That is,
\eqn\eJac{ J^{~\n\r\s}_{(\m^1_0,\ldots,\m^N_0)} \define
 ~\det\Bigg[{\vd(x^\n, x^\r, x^\s, f^1, {\cdots}, f^K)
              \over\vd(x^{\m^1_1}, {\cdots}, x^{\m^N_{n_N}})}
        \Bigg]_{x_i^{\m_0}\ne0}~. }
\bigskip

Most notably, this residue sports several quite remarkable properties.

\item{\bf1.} Despite appearances, $\W$ is independent of any particular
choice of $x^\n,x^\r,x^\s$. This easily follows from the observation that,
under a change of this choice of coordinates, both numerator and denominator
in~\eOm{d} transform as tensor densities and with the same multiplicative
factor which then cancels out.

\item{\bf2.} Division by the Jacobian in~\eOm{b} and~\eOm{c} produces no
singularity on $\cal M$. This follows from the (assumed) smoothness of
$\cal M$, whence the $K$ polynomials $f^j$ form a non-degenerate
set and there always exist precisely three independent coordinates
$x^\n,x^\r,x^\s$ (as $\dim{\cal M}{=}3$) to complete this set; the Jacobian
is then non-zero. That is, wherever $J^{~\n\r\s}_{(\m^1_0,\ldots,\m^N_0)}$
may vanish, the numerator of~\eOm{c} must vanish also, and in such a way
that the ratio remains non-zero and may be calculated upon a change to a
more suitable set of coordinates.
 This may be easier to see, from Eq.~\eOm{b}, as follows. The
hypersurfaces $\{f^j{=}0\}$ must intersect transversely for $\cal M$ to be
smooth. That means that at least at the common zero-set
${\cal M}=\bigcap_{j=1}^K\{f^j{=}0\}$, each of the hypersurfaces
$\{f^j{=}0\}$ must be transversal, \ie, vanish linearly. Division by each
$f^j$ therefore creates a simple pole and the corresponding integration
over $\G(f^j)$ then picks out the (finite) residue.

\item{\bf3.} Since the residues are non-zero only where the defining
polynomials $f^j$ vanish, $\W$ vanishes {\it identically} everywhere on
${\cal X}\define\CP{n_1}_1\times{\cdots}\times\CP{n_N}_N$---except on the
complete intersection, $\cal M$. That is, $\W$ is supported precisely and
exclusively on $\cal M$---it looks somewhat like a finite Dirac
$\d$-function, being non-zero (but then finite) only on the subspace
${\cal M}\subset{\cal X}$.

\item{\bf4.} By construction, $\W$ is homogeneous and of degree 0 with
respect to each $\CP{n_i}_i$. Therefore, $\W$ is the $SU(3)$ holonomy
invariant $\e_{\m\n\r}\rd z^\m\rd z^\n\rd z^\r$, \ie\ $\W$ is a
${\cal O}(0)$-valued 3-form on $\cal M$.
 All of this is equally valid if the isotropic $\CP{n_i}_i$ are replaced by
general flag spaces and/or their weighted cousins; upon replacing the top
differentials $(x_i\rd^{n_i} x_i)$ accordingly, the construction applies
{\it verbatim}.

These properties will be crucial in the subsequent analysis: massless
states will be assigned representatives~\eGenForm, in the form of an
expansion in terms of `nowhere zero holomorphic $q$-forms' (over certain
intermediate $q$-dimensional subspaces) and with holomorphic coefficients.
To that end, however, we need a telegraphic summary and a reinterpretation
of the Koszul spectral sequence computation.

\subsec{Rhyme and reason for residues}\subseclab\sRhyme\noindent
 \phrasing{a ragtime rhapsody}{rustico}
The basic fact underlying the Koszul spectral sequences is related to the
BRST symmetry induced by enforcing the constraints which define $\cal M$ as
a submanifold of the embedding space $\cal X$ where the defining polynomials
$f^j(x_i)$ simultaneously vanish. Typically, the embedding space is chosen
to be a product ${\cal X}\define\prod_{i=1}^N\CP{n_i}_i$ of projective
spaces. The BRST symmetry derives from the fact that all functions (and
hence physical observables in particular) on $\cal M$ then become
equivalence classes of functions over $\cal X$, modulo suitable multiples
of the $f^j(x_i)$. Thus, on $\cal M$, a polynomial $\f(x)$ of degree
$\vec{d}_\f$ is well defined only up to the equivalence\ft{Notation: By
the {\it degree} $\vec{d}_\f$ of a polynomial $\f(x_i)$ we mean the array
of degrees $(d_1,\ldots,d_N)$, where $d_i$ is the degree of homogeneity of
$\f(x_i)$ with respect to the homogeneous coordinates $x_i$ of $\CP{n_i}$.
Similarly, by $\vec{d}_j$ we mean the analogous array of degrees
$(d_{1j},\ldots,d_{Nj})$ of $f^j(x_i)$. We forego the pedantry of calling
this a multi-degree and emphasizing this hereafter.}
\eqn\eEqRel{ \f(x_i) ~~\cong~ \f(x_i)
                    ~+~ \sum_{j=1}^K\l_j(x_i)f^j(x_i)~; }
since the $f^j(x_i)$ vanish on $\cal M$. Clearly, the $\l_j(x)$ must be
complex-analytic and have degree $(\vec{d_\f}{-}\vec{d_j})$ for the
above sum to make sense.

The corresponding inhomogeneous gauge transformation is generated by
\eqn\eGauge{ \d\f(x_i) ~=~ \sum_{j=1}^K\l_j(x_i)f^j(x_i)~, }
whereupon the defining polynomials $f^j(x_i)$ may be regarded as the
generators, and the $\l_j(x_i)$ as the parameters of the associated
symmetry; in the corresponding BRST analysis, the $\l_j(x_i)$ are assigned
ghost variables (anticommutivity will follow naturally, see below).

We note that the gauge transformation~\eGauge\ may be used to `gauge away'
a function $\f(x_i)$ at any particular point of $\cal X$---{\it except} on
$\cal M$, where $f^j(x_i)=0$ and where the gauge transformation~\eGauge\
becomes vacuous. Notably, the $\l_j(x_i)$'s must be chosen to vary over
$\cal X$ in just the right way for this to happen. For the product
$\l_jf^j(x)$ to be in general non-zero except at $\{f^j{=}0\}$, the only
place where $\l_j$ may possibly diverge is the zero set $\{f^j{=}0\}$
itself, whence the $\l_j$ may be regarded as {\it local} gauge parameters
which are holomorphic on ${\cal X}-\{f^j{=}0\}$.
 Now, if $\cal M$ is smooth then $f^j$ vanish to first order in their local
Taylor series at $\cal M$. Around any point $\0x\in\cal M$, we have
$f(x)\approx(x{-}\0x)f'(\0x)$ with $f'(\0x)\ne0$ and near $\0x$,
$\l_j(x)\sim\hbox{\it const.}/(x{-}\0x)^{1-\e}$, with $\e>0$. ($\e<0$ would
spoil the holomorphicity of the polynomials $\f(x_i)$. Allowing $\e=0$
would cause the defining polunomials $f^j(x_i)$ to be equivalent to
polynomials that do not necessarily vanish at $\cal M$---contradicting the
very definition of $\cal M$.)
 For the $\l_j(x_i)$ to be single-valued, $\e$ must be an integer and the
$\l_j$ must in fact be holomorphic\ft{On a singular $\cal M$, some of the
$f^j$ vanish to higher order. This then admits certain meromorphic $\l_j$'s,
enlarging the space of `ghosts', and then also `ghost-for-ghosts',
{\it etc}., allowing for possible additional operators and corresponding
low-energy fields.} on all of $\cal X$, rather than just on the complement
of $\cal M$. In particular, only functions complex-analytically
proportional to some $f^j(x_i)$ are equivalent to zero modulo this
`globally holomorphic' transformation~\eGauge.
\ping

Focus now on the first two constraint polynomials, $f^1(x_i)$ and
$f^2(x_i)$ and set temporarily $\l_j(x_i) = 0$, $j=3,\ldots,K$.
Note that the special choice of the first two `ghost' variables
(assuming it permitted by the degrees of homogeneity)
\eqn\eGhosts{ \l_1(x_i) = \l_{12}(x_i)\,f^2(x_i)~,\qquad
              \l_2(x_i) = \l_{21}(x_i)\,f^1(x_i)~,}
corresponds to a trivial component of the equivalence relation~\eEqRel
\eqn\eEqEqRel{ {\eqalign{
 \f(x_i) &\cong~ \f(x_i) ~+~
  \big(\l_{12}(x_i)\,f^2(x_i)f^1(x_i) ~+~
       \l_{21}(x_i)\,f^1(x_i)f^2(x_i)\big)~, \cr
         &=~ \f(x_i) ~+~
  \big[\l_{12}(x_i)+\l_{21}(x_i)\big]\,f^1(x_i)f^2(x_i)~,\cr
         &\id~ \f(x_i)~,\qquad
          \hbox{precisely if}~~\l_{21}={-}\l_{12}~.}} }
This implies that the two-parameter equivalence relation~\eEqRel\ gauges
away `too much' and the one-parameter degree of freedom
$\l_{[12]}=\inv2(\l_{12}{-}\l_{21})$
compensates for this.  Therefore, the equivalence relations~\eEqRel\ must
themselves be taken modulo the equivalence between the two `ghost'
variables~\eGhosts, as generated by $\l_{[12]}(x_i)$. Such second order
equivalence relations are familiar from the BRST analysis and $\l_{[12]}$
corresponds to a `ghost-for-ghost' variable. For the full set of $K$ first
order `ghost' variables $\l_j$, there exist ${K\choose2}$ second order,
`ghost-for-ghost' variables $\l_{[jl]}(x_i)$.
 Clearly, this produces an avalanche of such higher order equivalence
relations, and a corresponding hierarchy of `ghost-for-ghost-for-\3'
variables $\l_{[jl\cdots]}(x_i)$. Equally obvious should be the fact that
this hierarchy stops at the $K^{th}$ order, with a single
`ghost-for-ghost-for\3' variable $\l_{[j_1\cdots j_K]}$. The antisymmetry
here reflects (although is not identical to) the graded anticommutivity of
the actual BRST ghost fields.

For all but the few simplest cases, an explicit listing of this hierarchy
of equivalence relations and `ghost' variables is plagued by a
proliferation of indices, variables and confusion. Let us therefore introduce a
diagrammatic representation, where an equivalence generated by
\eqn\eOneEq{ \f(x_i) ~~\cong~~ \f(x_i) ~+~ \l_1(x_i) f^1(x) }
is represented by
\eqn\eOneMap{ \l_1(x_i) ~\tooo{~f^1~}~ \f(x_i)~, }
meaning simply that $f^1$-multiples of $\l_1$ may be added to $\f$ at will.
This notation also reminds of the BRST and gauge-theoretic interpretation,
where the $f^1(x_i)$ may be regarded as the generator and $\l_1$ the gauge
parameter for $\d\f(x_i)=\l_1(x_i) f^1(x)$.
 For two constraint polynomials, we obtain the diagram
\eqn\eDiagTwo{ \l_{[12]}(x_i)
                \stack{\nearrow}{0}{\searrow}
                 \stack{\l_2(x_i)}{7}{\l_1(x_i)}
                  \stack{\searrow}{0}{\nearrow}
                   \f(x_i)~ \To ~\f_{\cal M}(x_i)~, }
where $\nearrow$ stands for multiplication by $f^1$ and $\searrow$ by
$f^2$ and ``$\To\f_{\cal M}$'' says that the resulting equivalence class may
be regarded as the corresponding function on $\cal M$.
 Stacked in the first column to the left of $\f(x_i)$ are the `ghost'
variables, and in the (here, a single entry) second column to the left---the
`ghost-for-ghost' variables. The diagram~\eDiagTwo\ stands for the
equivalence relations
\eqna\eEqEq
 $$
\f(x_i) ~~\cong~~ \f(x_i) ~+~ \l_1(x_i) f^1(x) ~+~ \l_2(x_i) f^2(x)~,
 $$
where the $\l_j$ are themselves taken modulo the equivalence
 $$
\l_j(x_i) ~~\cong~ \l_j(x_i) ~+~ \l_{[jl]}(x_i)\,f^l(x_i)~.
 $$

The full hierarchy of these gauge equivalence relations may then be
represented diagrammatically\ft{Such pictures seem quite worth the
${K\over2}2^K$ equivalence relations employing $2^K{-}1$ ghost variables
which they represent; at $K=8$ for example, already the number of
equivalence relations reaches truly proverbial proportions, a
corresponding word count having surpassed it long ago.} as
\eqn\eKoSeq{\vcenter{\vbox{\offinterlineskip
\halign{&#&~\hfil$#$\hfil&$#$\hfil&$#$\hfil&$#$
&\hfil$#$\hfil&$#$&\hfil$#$\hfil&%
                          $#$&\hfil$#$\hfil~&~$#$\hfil\cr
  &{\cal O}(\vec{d}_\f{-}\sum_j\vec{d}_j)
&\matrix{\nearrow\cr\rightarrow\cr\searrow\cr}
&\matrix{{\cal O}(\vec{d}_\f{-}\sum_{j\ne1}\vec{d}_j)\cr\vdots\cr
         {\cal O}(\vec{d}_\f{-}\sum_{j\ne K}\vec{d}_j)\cr}
&\Stack{\to}{-3}{\searrow}{2}{\searrow\vrule width0pt depth15pt}
       \mkern-23mu
\Stack{\vrule height17pt width0pt\nearrow}{2}{\nearrow}{-1}{\to}
&\!\!\Stack{\cdots}{3}{\vdots}{3}{\cdots}
     &\Stack{\to}{-3}{\searrow}{2}{\searrow\vrule width0pt depth15pt}
       \mkern-23mu
      \Stack{\vrule height17pt width0pt\nearrow}{2}{\nearrow}{-1}{\to}
&\matrix{{\cal O}(\vec{d}_\f{-}\vec{d}_K)\cr\vdots\cr
         {\cal O}(\vec{d}_\f{-}\vec{d}_1)\cr}
&\matrix{\searrow\cr\rightarrow\cr\nearrow\cr}
      &{\cal O}(\vec{d}_\f)&\To~{\cal O_M}(\vec{d}_\f)&\cr}}}~, }
where ${\cal O}(\vec{d}_\f)$ denotes (the sheaf of) functions of degree
$\vec{d}_\f$ over ${\cal X}$ and the arrows represent multiplication with
the $f^j(x_i)$. The subscript on ${\cal O_M}(\vec{d}_\f)$ of course denotes
restriction to the submanifold $\cal M$---which is what we are after. The
process summarized in the sequence~\eKoSeq\ is in fact the underlying one in
the Koszul calculations and we will refer to it throughout. Of course, the
various arrows represent multiplication by the $f^j(x_i)$; as the degrees
typically specify this fairly well, we forego labeling the arrows.
\ping

The first task in analyzing the function ring on a submanifold is the
determination of scalars---the degree-$\vec0$ objects.
 To that end, set $\vec{d}_\f=\vec0$ in the sequence~\eKoSeq. Note that
whereas functions of negative degree clearly cannot by themselves provide
complex-analytic functions, they {\it will\/} be useful in constructing
differential forms. This in fact is not at all novel: the best known but
often overlooked example is the combination of $x\over x^2{+}y^2$ and
$y\over x^2{+}y^2$, multiplied respectively by $\rd y$ and ${-}\rd x$, to
produce the differential of the polar angle $\rd\vf$---perfectly regular on
a submanifold encircling but not including the origin.

A systematic and complete listing of non-zero such forms is possible owing
to the `master' theorem by Bott, Borel and Weil~\refs{\rEndT,\rBeast}. A
straightforward application of this theorem yields the resulting cohomology
(on $\cal X$) displayed in the lower left quadrant of the chart below:
\eqn\eKoch{\vcenter{\vbox{\offinterlineskip
\halign{&#&~\hfil$#$\hfil&$#$&\hfil$#$\hfil&$#$&\hfil$#$\hfil
&$#$&\hfil$#$\hfil&%
                          $#$&\hfil$#$\hfil~&\vrule#&~$#$\hfil\cr
  &(\otimes_j {\cal E}^*_j)
   &\matrix{\nearrow\cr\rightarrow\cr\searrow\cr}
    &\matrix{(\otimes_{j\neq1}{\cal E}^*_j)\cr\vdots\cr
             (\otimes_{j\neq K}{\cal E}^*_j)\cr}
&\Stack{\to}{-3}{\searrow}{2}{\searrow\vrule width0pt depth15pt}
       \mkern-23mu
\Stack{\vrule height17pt width0pt\nearrow}{2}{\nearrow}{-1}{\to}
    &\!\!\Stack{\cdots}{3}{\vdots}{3}{\cdots}
&\Stack{\to}{-3}{\searrow}{2}{\searrow\vrule width0pt depth15pt}
       \mkern-23mu
\Stack{\vrule height17pt width0pt\nearrow}{2}{\nearrow}{-1}{\to}
    &\matrix{{\cal E}^*_K\cr\vdots\cr
             {\cal E}^*_1\cr}
     &\matrix{\searrow\cr\rightarrow\cr\nearrow\cr}
      &{\cal O}_{\cal X}&&\To{\cal O_M}&\cr
  &\omit&\omit&\omit&\omit&\omit&\omit&\omit&\omit&\omit&height6pt&\omit&\cr
 \noalign{\hrule}
  &\omit&\omit&\omit&\omit&\omit&\omit&\omit&\omit&\omit&height6pt&\omit&\cr
  &0&&0&&\ldots&&0&&H^0\approx\IC&&\To H^0({\cal M})&\cr
  &\omit&\omit&\omit&\omit&\omit&\omit&\omit&\omit&\omit&height6pt&\omit&\cr
  &0&&0&&\ldots&&0&&0&&\To H^1({\cal M})&\cr
  &\omit&\omit&\omit&\omit&\omit&\omit&\omit&\omit&\omit&height6pt&\omit&\cr
  &0&&0&&\ldots&&0&&0&&\To H^2({\cal M})&\cr
  &\omit&\omit&\omit&\omit&\omit&\omit&\omit&\omit&\omit&height6pt&\omit&\cr
  &0&&0&&\ldots&&0&&0&&\To H^3({\cal M})&\cr
  &\omit&\omit&\omit&\omit&\omit&\omit&\omit&\omit&\omit&height6pt&\omit&\cr
  &0&&0&&\ldots&&0&&0&&\To H^4({\cal M})\id0&\cr
  &\omit&\omit&\omit&\omit&\omit&\omit&\omit&\omit&\omit&height6pt&\omit&\cr
  &\vdots&&\vdots&&\vdots&&\vdots&&\vdots&&\vdots&\cr
  &\omit&\omit&\omit&\omit&\omit&\omit&\omit&\omit&\omit&height6pt&\omit&\cr
  &H^X\approx\IC&&0&&\ldots&&0&&0&&\To H^X({\cal M})\id0&\cr
  &\omit&\omit&\omit&\omit&\omit&\omit&\omit&\omit&\omit&height6pt&\omit&\cr
}}}}
where ${\cal E}^*_j\simeq{\cal O}(-\vec{d}_j)$ is the normal bundle
associated to $f_j$.
This chart is typical of Koszul calculations and it contains the
cohomology groups on $\cal X$, $H^0,\ldots,H^{X}$ (with $X=\dim {\cal X}$),
valued in the bundle under which they are stacked.
 Through a process called `filtration' (see below), the cohomology groups on
$\cal X$ give rise to the cohomology groups on $\cal M$, stacked in the
lower right quadrant.

The above chart thus presents the result that there are only two types of
degree-$\vec0$ holomorphic forms on the \CY\ manifold~\eGCnfg:
\item{1.} the restriction of degree-$\vec0$ 0-forms on the embedding space
          ${\cal X}$ to the subspace $\cal M$,
\item{2.} the restriction of $X$-forms with coefficients of
          degree $\big({-}\sum_j\vec{d}_j\big)$.\vglue0mm\noindent
Back in the upper left quadrant, this degree-$({-}\sum_j\vec{d}_j)$ in $2.$
function must become a complex-analytic degree-$\vec0$ function upon
multiplication by the $f^j$ as indicated by the arrows in the sequence on
the top. Therefore, this degree-$({-}\sum_j\vec{d}_j)$ function must be
a constant multiple of $[\prod_{i=1}^N f^j]^{{-}1}$.
Note now that the condition for $\cal M$ to be a \CY\ 3-fold is
 $\sum_j\vec{d}_j=(n_1{+}1,\ldots,n_N{+}1)\define(\vec{n}_i+1)$.

{\it Precisely because the \CY\ condition is satisfied}, we can make a
degree-$\vec0$ differential from such a function through multiplying by the
top differential $\prod_{i=1}^N(x_i\rd^{n_i} x_i)$, very much in analogy
with $\rd\vf={y\rd x{-}x\rd y\over x^2+y^2}$.
This degree-$\vec0$ $X$-form appears in the bottom left of the above chart.
Finally, this will yield a holomorphic 3-form on $\cal M$ upon the
$K$-fold contour integration:
 $$
 \W ~\define~
 ~{1\over(2\p i)^K}\oint_{\G(f^1)}{\cdots}\oint_{\G(f^K)}
    {\prod_{i=1}^N(x_i\rd^{n_i}x_i)\over f^1\,f^2\,{\cdots}\,f^K}~,
\eqno\eOm{b}
 $$
as discussed for Eqs.~\eOm{}. Each contour integration may be identified,
in the chart~\eKoch, as taking one step upwards (each one integration
`cancels' one differential) and one step to the right (the residue
evaluation `cancels' the pole which produced the residue). The
concatenation of such diagonal steps leads to the cohomology on $\cal M$
where the contribution eventually ends up and represents what is called
`filtration' in the general theory of spectral sequences. So,
\eqn\eKoCh{\vcenter{\vbox{\offinterlineskip
\halign{&#&~\hfil$#$\hfil&$#$&\hfil$#$\hfil&$#$&\hfil$#$\hfil
&$#$&\hfil$#$\hfil&%
                          $#$&\hfil$#$\hfil~&\vrule#&~$#$\hfil\cr
  &(\otimes_i {\cal E}^*_i)
   &\matrix{\nearrow\cr\rightarrow\cr\searrow\cr}
    &\matrix{(\otimes_{i\neq1}{\cal E}^*_i)\cr\vdots\cr
             (\otimes_{i\neq K}{\cal E}^*_i)\cr}
&\Stack{\to}{-3}{\searrow}{2}{\searrow\vrule width0pt depth15pt}
       \mkern-23mu
\Stack{\vrule height17pt width0pt\nearrow}{2}{\nearrow}{-1}{\to}
    &\!\!\Stack{\cdots}{3}{\vdots}{3}{\cdots}
&\Stack{\to}{-3}{\searrow}{2}{\searrow\vrule width0pt depth15pt}
       \mkern-23mu
\Stack{\vrule height17pt width0pt\nearrow}{2}{\nearrow}{-1}{\to}
    &\matrix{{\cal E}^*_K\cr\vdots\cr
             {\cal E}^*_1\cr}
     &\matrix{\searrow\cr\rightarrow\cr\nearrow\cr}
      &{\cal O}_{\cal X}&&\To{\cal O_M}&\cr
  &\omit&\omit&\omit&\omit&\omit&\omit&\omit&\omit&\omit&height6pt&\omit&\cr
 \noalign{\hrule}
  &\omit&\omit&\omit&\omit&\omit&\omit&\omit&\omit&\omit&height6pt&\omit&\cr
  &0&&0&&\ldots&&0&&\h&&\To\h\in H^0({\cal M})&\cr
  &\omit&\omit&\omit&\omit&\omit&\omit&\omit&\omit&\omit&height6pt&\omit&\cr
  &0&&0&&\ldots&&0&&0&&\To 0=H^1({\cal M})&\cr
  &\omit&\omit&\omit&\omit&\omit&\omit&\omit&\omit&\omit&height6pt&\omit&\cr
  &0&&0&&\ldots&&0&&0&&\To 0=H^2({\cal M})&\cr
  &\omit&\omit&\omit&\omit&\omit&\omit&\omit&\omit&\omit&height6pt&\omit&\cr
  &0&&0&&\ldots&&0&&0&&\To \W\in H^3({\cal M})&\cr
  &\omit&\omit&\omit&\omit&\omit&\omit&\omit&\omit&\omit&height6pt&\omit&\cr
  &0&&0&&\ldots&&0&&0&&\To H^4({\cal M})\id0&\cr
  &\omit&\omit&\omit&\omit&\omit&\omit&\omit&\omit&\omit&height6pt&\omit&\cr
  &\vdots&&\vdots&&\vdots&&\vdots&&\vdots&&\vdots&\cr
  &\omit&\omit&\omit&\omit&\omit&\omit&\omit&\omit&\omit&height6pt&\omit&\cr
  &\W&&0&&\ldots&&0&&0&&\To 0\id H^X({\cal M})&\cr
  &\omit&\omit&\omit&\omit&\omit&\omit&\omit&\omit&\omit&height6pt&\omit&\cr
}}}}
where the contributions are placed as {\it before} the `filtration': the
$\W$ here `filters' along the SW--NE diagonal to the fourth element in the
right-most column of the lower left quadrant, thereby contributing to
$H^3({\cal M})$.

As the subsequent discussions and examples will hopefully clarify, the
identification of the content of charts~\eKoch\ and~\eKoCh\ defines the
`residue map' (`residue operator' or `residue symbol'), which assigns
residue representatives in $H^\star({\cal M},\3)$ to certain rational
forms on the embedding space. Indeed, our analysis pertains to
generalizations\ft{While perhaps straightforward in principle, the
generalizations presented here have not been reported heretofore, to the
best of our knowledge. Owing to an easy corollary of Murphy's Law, the
results most dearly sought are not readily available in the literature.} of
the  so-called Poincar\'e residue symbol~\refs{\rGrHa,\rGr}, of which the
formula~\eABGC\ is a sample application.

Admittedly, we have not derived anything new so far. Instead, we have
related the major ingredients in Koszul calculations with the simple residue
recipe~\eOm{}. Our aim now is to use this insight and obtain all other
cohomology straightforwardly from such residue considerations. As a
byproduct, this will effectively rederive the required results of the
Bott-Borel-Weil theorem and should also better specify the residue map.

\subsec{A residue refinement}\subseclab\sTheRes\noindent
 \phrasing{the recurring riff}{rinforzando}
Motivated by Eqs.~\eOm{}, a natural generalization comes to mind, and which
will turn out to be essential in the future:
\eqna\eRes
 $$
 \Res_Q^S\big[\f\big]~
 \define~{1\over(2\p i)^{|Q|}}
           \oint{\cdots}\oint_{\prod_{j\in Q}\G(f^j)}
           {\prod_{i\in S} (x_i\rd^{n_i}x_i)\over
             \prod_{j\in Q} f^j}~\f(x_1,\ldots,x_N)~,
\eqno\eRes{a}
 $$
where $\f(x_i)$ is chosen so that
 $$ \twoeqsalignno{
 \deg_i(\f)
 &=~\sum_{j\in Q}\deg_i(f^j)-\sum_{\m=0}^{n_i}\deg_i(x_i^\m)~\geq0~,\quad
 &\quad {\rm for~all}~i&\in S~, & \eRes{b} \cr
 \deg_i(\f)
 &\geq ~\sum_{j\in Q}\deg_i(f^j)~\geq0~,\quad
 &\quad {\rm for~all}~i&\nin S~, & \eRes{c} \cr}
 $$
and where $Q$ labels a subset of the defining polynomials which occur in
the denominator of the integrand in Eq.~\eRes{a}, $|Q|$ is the number of
these polynomials, $S$ labels the subset of projective spaces over which
the integral is performed, $\deg_i$ denotes degrees of homogeneity with
respect to $\CP{n_i}_i$. By definition, this `partial', or `intermediate'
residue will be understood  to vanish if the conditions~\eRes{b,c} are not
satisfied. Henceforth, $|Q|$ will be referred to as the {\it level} of the
residue, and the rational differential forms appearing in these integrals
will be called  the {\it kernel} of the residue.

\bigskip
\centerline{\boxit{5pt}{\advance\hsize by-58pt\vbox{\noindent
   $\Res^S_Q[\f]$ is a `nowhere zero holomorphic $q$-form' as in~\eOm{},
   on the $q$-dimen\-sional \CY\ complete intersection
   ${\cal Q}\define\cap_{j\in Q}\{f^j{=}0\}\subset\prod_{i\in S}\CP{n_i}_i$,
   and is {\it paramet\-rized\/} by its dependence on the complementary
   factor $\prod_{i\nin S}\CP{n_i}_i$. Also, $\Res^S_Q[\f]$ vanishes
   identically on the complement of $\cal Q$ within
   $\prod_{i\in S}\CP{n_i}_i$.}}}\bigskip

 Owing to its definition, the residue~\eRes{} shares all the salient
features of $\W$ in~\eOm{}.
 That is, with regard to the integrated $\CP{n_i}_i$, $i\in S$,
$\Res_Q^S\big[\f\big]$ is non-zero and invariant under holonomy precisely
and exclusively over the subspace ${\cal Q}\subset\prod_{i\in S}\CP{n_i}_i$.
 With regard to the un-integrated\ft{The erudite Reader will have noticed
the similarity between `un-integrated' here, and `un-projected' in~\rLGO.
Indeed, this is not an accident: upon performing a contour integral over
any one of the $n{+}1$ coordinates of a $\CP{n}=U(n{+}1)/[U(n)\8U(1)]$, the
result must, by $U(n{+}1)$-covariance, also be independent (up to an overall
scale) of the remaining $n$ coordinates of that \CP{n}---implying a
projection {\it along} that \CP{n}. This relation will become even more
manifest in \SS~4.}
$\CP{n_i}_i$, $i\nin S$, $\f$ can be chosen so that
$\Res_Q^S\big[\f\big]$ is holomorphic and of degree
$\deg_i(\f)-\sum_{j\in Q}\deg_i(f^j)$. Thereby, it is $\db$-closed and
(obviously) not exact. The well-known formula~\eOm{} is then simply the
special case with $\f=1$, and when $Q$ and $S$ include all polynomials and
all \CP{n_i}'s, respectively. Also, the well-known polynomial deformations
correspond to the special case when $\f$ ranges over the polynomials and
$Q=\emptyset=S$.
 Similar generalizations of~\eABGC, however to non-compact \CY\ spaces,
appear in constructing spacetime variable superstring vacua~\rCYCY.
We now turn to the details of this.
\ping

The residue is evaluated much the same as~\eOm{}, employing a change of
variables such as~\eChng, and direct contour integration. In view of the
requirement~\eRes{b}, the degree of the residue is $\vec0$ for all the
$i\in S$, and it is a $q$-form, where
\eqn\eInterQ{ q ~=~ \sum_{i\in S}n_i-|Q|~. }

Consider first the properties of~\eRes{a} over $\CP{n_i}_i$, $i\in S$.
If $q\geq0$ and in a coordinate neighborhood where $x_i^{\m_0^i}\ne0$, for
fixed $\m_0^i$ and all $i\in S$, the residue will evaluate to
\eqna\eRES
 $$
 \Res^S_Q\big[\f\big]~=~
   \big(\prod_{i\in S}x^{\m_0^i}_i\big)\>
    {\rd x^{\n_1}\cdots\rd x^{\n_q}
     \over J^{~\n_1\cdots\n_q}_{(\m_0^i,~i\in S)}}~\f(x_i)~,
\eqno\eRES{a}
 $$
with $J^{~\n_1\cdots\n_q}_{(\m_0^i,~i\in S)}$ the Jacobian of the
coordinate transformation
\eqn\eCHNG{ (x^{\m^i_1},{\cdots},x^{\m^i_{n_i}}; i\in S)~
           \longrightarrow ~(x^{\n_1},{\cdots},x^{\n_q},f^j; j\in Q)~, }
which has been used to single out $x^{\n_1},\ldots,x^{\n_q}$.

In the event that $q<0$, i.e., $\sum_{i\in S}n_i<|Q|$, there are not enough
differentials to perform all the $|Q|$ integrations, and we can of course
perform only $\sum_{i\in S}n_i$ of them. The resulting rational function
will then be holomorphic only if the function $\f$ can be chosen so
as to cancel the remaining poles\ft{With a little forethought,
we note that non-analytic terms would fail to be $\bar\vd$-closed and so
would not contribute to $H_{\bar\vd}^\star({\cal M},\3)$. The Reader may
instead wish to ignore this motivation and regard this as simply weeding
out non-holomorphic contributions.}.
 Such applications of the L'Hospital theorem will be understood as an
extension of the standard residue calculations.

\bigskip
On the other hand, the degree of the residue~\eRes{a} with respect to the
un-integrated coordinates of $\prod_{i\nin S}\CP{n_i}$ is the same as
that of the $m_i^{th}$ derivative of $\f(x_i)$ with respect to
$x_i\,,i\nin S$, written $\vd^{\mv}\f(x_i)$, where (see~\eRes{a})
\eqn\eVGr{ m_i~\define~\sum_{k\in Q}d_{ik},\quad\,i\nin S~,
           \qquad \vd^{\mv} ~\define~
           \big(\prod_{i\nin S}\vec{\vd}_{(i)}^{~\8m_i}\big)~.}
The result of the contour integration~\eRes{a} is finite over
${\cal Q}\define\bigcap_{k\in Q}\{f^k{=}0\}$ since the intersection of
hypersurfaces is smooth by assumption. (Actually, we only care about the
smoothness of the subspace $\cal M$ where {\it all} $K$ hypersurfaces
meet.) Choosing $\f$ to have a sufficiently positive degree over the
un-integrated $\prod_{i\nin S}\CP{n_i}$, implies that $\Res^S_Q[\f]$ can
be represented as a polynomial over $\prod_{i\nin S}\CP{n_i}$.
Therefore, $\Res^S_Q[\f]$ can be written as a linear combination of the
$\mv^{th}$ multi-derivatives of $\f(x)$:
 $$ \eqalignno{
 \Res^S_Q\big[\f\big]
 &=\oint\!{\ldots}\!\oint_{\P_{k\in Q}\G(f^k)}
    {\prod_{i\in S}(x_i\rd^{n_i} x_i) \over \prod_{k\in Q}f^k}\>\f~,
                                                             &\eRES{b}\cr
 &=~\underbrace{
   \big(\prod_{i\in S}x^{\m_0^i}_i\big)\>
    {\rd x^{\n_1}\cdots\rd x^{\n_q}\over
     \vd^{{}^{\mv}}J^{~\n_1\cdots\n_q}_{(\m_0^i,~i\in S)}}}
     _{{\rm a~nowhere-zero}~q-{\rm form}} ~\vd^{\mv}\f(x_i)~,  &\eRES{c}\cr
 &\define\sum_{\rv}\W_{(q)}^{\rv}\,\f^{(\mv)}_{\rv}(x)~. &\eRES{d}\cr
}$$
 The multi-index $\rv$ contains one index for each of the $m_i$ derivatives
with respect to the un-integrated $x_i$ ($i\nin S$). Notably, the
underbraced term, $\W_{(q)}^{\rv}$, is a `nowhere zero holomorphic
$q$-form' on the $q$-dimensional \CY\ space $\cal Q$ embedded in
$\prod_{i\in S}\CP{n_i}_i$ as the common zero-set of $f^j{=}0$, $j\in Q$,
and {\it parametrized\/} by $x_i\in\CP{n_i}_i$, $i\nin S$, and any other
parameter that the $f^j$ depend upon; ${\cal Q}$ is {\it not} \CY\ over the
un-integrated $\CP{n_i}_i$, $i\nin S$.

Alternatively, the multi-derivatives $\vd^{\mv}f^j$, $j\in Q$ and
indexed by $\rv$, may be regarded the defining equations of a \CY\
$q$-fold ${\cal Q}_{\rv}\subset\prod_{i\in S}\CP{n_i}_i$, for each
of which the
\eqn\eInterOm{ \W_{(q)}^{\rv}~ \define
 ~\big(\prod_{i\in S}x^{\m_0^i}_i\big)\>
    {\rd x^{\n_1}\cdots\rd x^{\n_q}\over
     \vd^{{}^{\mv}}J^{~\n_1\cdots\n_q}_{(\m_0^i,~i\in S)}} }
are the holomorphic volume-forms. They are calculated much the same
as~\eOm{}, for which the Jacobian \`a la Eqs.~\eOm{} and~\eRes{} is indeed
$\vd^{\mv}J^{~\n_1\cdots\n_q}_{(\m_0^i,~i\in S)}$, since the Jacobian
$J^{~\n_1\cdots\n_q}_{(\m_0^i,~i\in S)}$ is multi-linear in the $\{f^j,
j{\in}Q\}$, and the multi-order of the multi-derivative
$\vd^{\mv}$ over the un-integrated $\CP{n_i}_i$, $i\nin S$ exactly equals
the multi-degree of $J^{~\n_1\cdots\n_q}_{(\m_0^i,~i\in S)}$ over the
un-integrated space. Therefore, $\W_{(q)}^{\rv}$ are {\it constant} over
the un-integrated factors $\prod_{i\nin S}\CP{n_i}_i$ of the embedding space
${\cal X}=\prod_i\CP{n_i}_i$, while being nowhere-zero and invariant under
holonomy over $\cal Q$.

Yet another way to think about these is along the discussion of the $q<0$
case; that is, for the residue to be holomorphic, we seek suitable
functions $\f$ in a factorized form, such that one factor precisely cancels
the contribution of negative degree over $\CP{n_i}_i$, $i\nin S$, from
the residue kernel. The remaining factor may then be written as
in~\eRES{d}, and is precisely of the promised general form~\eGenForm. In
fact, when $q<0$, the $\W_{(q)}^{\rv}$ is formally a differential form
of negative order, which is naturally identified with a vector or higher
rank (contravariant) tensor field, {\it i.e.}, a ``reparametrization'' or
``gauge'' degree of freedom.\ping

As a quick (and well known) example, consider the \CY\ 3-fold  intersection
of a degree-(3,0) and a degree-(1,3) hypersurface in $\IP^3{\times}\IP^2$;
${\cal M}\in\cnfg{\IP^3\cr\IP^2\cr}{3&1\cr0&3}$, for short~\rRolf.
Now note that the constraint $g(x,y)$, of degree-(1,3), produces a
1-dimensional \CY\ space  (a torus) in \CP2. Its defining equation depends
non-trivially (linearly) on $x\in\IP^3$, so that these tori are fibered
over the \CP3 linearly. Writing $y$ for coordinates on \CP2 and for a
suitable function $\f$ with $\deg_x(\f)\geq1$ and $\deg_y(\f)=0$, the
partial residue is
\eqna\eRolf
 $$ \eqalignno{\Res_2^y[\,\f\,]~
 &= ~{1\over(2\p i)}\oint_{\G(g)}{(y\rd^2y)\over g(x,y)}\>\f~
  = ~\sum_{a=0}^3\W_{(1)}^a\big(\vd_a\f\big)~,              &\eRolf{a}\cr
\W_{(1)}^a~
 &= ~{1\over(2\p i)}\oint_{\G(g)}{(y\rd^2y)\over\vd_ag(x,y)}~
  = ~y^0{\rd y^{\hat\a}\over\vd_a J^{\hat\a}_{(0)}}
  = ~{y^\a\e_{\a\b\g}\rd y^\g\over\vd_a J^{\g}_{(\a)}}~,    &\eRolf{b}\cr
J^{\hat\a}_{(0)}~
 &= ~\left|{\vd(y^{\hat\a},g)\over\vd(y^1,y^2)}\right|_{y^0\neq0}~,
  \qquad\hat\a,\hat\b=1,2~,\quad\a,\b,\g=0,1,2~;
 &\eRolf{c}\cr}
 $$
there is no summation on any of the repeated indices in~\eRolf{b},
where the last expression is valid in all coordinate patches. Note that
the free index $\b{=}0,1,2$ on the far right of~\eRolf{b} stems from the
coordinate over which the contour integral was performed and it simply
labels different $U(3)$-covariant coordinate choices for writing the same.

It is easy to verify that $\Res_2^y[\f]$ is
 ({\bf 1})~independent of the choice of the coordinate patch (here
$y^0\neq0$), as should be manifest from the last expression in~\eRolf{b},
 ({\bf 2})~holomorphic over \CP3,
 ({\bf3})~identically zero outside $\{g{=}0\}\subset\IP^2$,
 ({\bf 4})~non-zero and finite on the cubic tori
$\{g(x,y){=}0\}\subset\IP^2$.

\subsec{Racks and racks of residues}\subseclab\sRacks\noindent
 \phrasing{a rigadoon}{radoppiando}
Instead of determining the degree-$\vec 0$ forms (${\cal O_M}$-valued
cohomology) as done in the previous subsection, we now turn to
degree-$d_j$ forms, for any $j=1,\ldots,K$. To this end, we shift the
degree labels in~\eKoSeq\ so as to produce ${\cal O}(\vec{d}_j)$ at the far
right. The degree-$d_j$ `ghost-for-ghost\3' sequence is then
\eqn\eKoSdj{\vcenter{\vbox{\offinterlineskip
\halign{&#&~\hfil$#$\hfil&$#$\hfil&$#$\hfil&$#$&\hfil$#$\hfil
&$#$&\hfil$#$\hfil&%
                          $#$&\hfil$#$\hfil~&~$#$\hfil\cr
  &{\cal O}(\sum_{l\ne j}\vec{d}_l)
&\matrix{\nearrow\cr\rightarrow\cr\searrow\cr}
&\matrix{{\cal O}(\sum_{l\ne1,j}\vec{d}_l)\cr\vdots\cr
         {\cal O}(\sum_{l\ne j,K}\vec{d}_l)\cr}
&\Stack{\to}{-3}{\searrow}{2}{\searrow\vrule width0pt depth15pt}
       \mkern-23mu
\Stack{\vrule height17pt width0pt\nearrow}{2}{\nearrow}{-1}{\to}
&\!\!\Stack{\cdots}{3}{\vdots}{3}{\cdots}
&\Stack{\to}{-3}{\searrow}{2}{\searrow\vrule width0pt depth15pt}
       \mkern-23mu
\Stack{\vrule height17pt width0pt\nearrow}{2}{\nearrow}{-1}{\to}
&\matrix{{\cal O}(\vec{d}_j{-}\vec{d}_K)\cr\vdots\cr
         {\cal O}(\vec{d}_j{-}\vec{d}_1)\cr}
&\matrix{\searrow\cr\rightarrow\cr\nearrow\cr}
      &{\cal O}(\vec{d}_j)&\To~{\cal O_M}(\vec{d}_j)~,&\cr
\noalign{\bigskip}
 {\rm or,}\cr
  &(\bigotimes_{l\neq j}{\cal E}^*_{f^l})
   &\matrix{\nearrow\cr\rightarrow\cr\searrow\cr}
  &\matrix{(\bigotimes_{l\ne 1,j}{\cal E}^*_{f^l})\cr
  \vdots\cr(\bigotimes_{l\ne j,K}{\cal E}^*_{f^l})\cr}
&\Stack{\to}{-3}{\searrow}{2}{\searrow\vrule width0pt depth15pt}
       \mkern-23mu
\Stack{\vrule height17pt width0pt\nearrow}{2}{\nearrow}{-1}{\to}
    &\!\!\Stack{\cdots}{3}{\vdots}{3}{\cdots}
&\Stack{\to}{-3}{\searrow}{2}{\searrow\vrule width0pt depth15pt}
       \mkern-23mu
\Stack{\vrule height17pt width0pt\nearrow}{2}{\nearrow}{-1}{\to}
    &\matrix{({\cal E}^*_{f^1}\otimes{\cal E}_{f^j})\cr
             \vdots\cr({\cal E}^*_{f^K}\otimes{\cal E}_{f^j})}
     &\matrix{\searrow\cr\rightarrow\cr\nearrow\cr}
      &{\cal E}_{f^j}&\To~{\cal E}_{f^j}\big|_{{\cal M}}~, &\cr}}}}
where $j$ is being omitted from the above products because
${\cal E}^*_j\8{\cal E}_j=\IC$.
 Remember that these sequences denote the entire process of listing all
${K\over2}2^K$ equivalence relations and assigning the $2^K{-}1$ `ghost',
`ghost-for-ghost' {\it etc}.\ variables, as sketched in \SS\sRhyme. Also,
${\cal E}_{f^j}$ denotes (the sheaf of) functions of degree $\vec{d}_j$
and ${\cal E}^*_{f^j}$ is its dual. The duals stem from division by
$f^j(x_i)$, so that the multiplication indicated by arrows would produce
holomorphic functions on ${\cal X}=\prod_i\CP{n_i}$ of degree $\vec{d}_j$,
that is, ${\cal O}_{\cal X}(\vec{d}_j)$.

Whenever the degrees are non-negative, there clearly is a
straightforward restriction. Moreover, if there is a mapping in the
sequence, there will also be a mapping between the restrictions of
the corresponding functions. For example,
in~\eKoSdj\ there exists a sequence
\eqn\eFxMap{
     \big[\,{\cal O}(\vec{d}_j{-}\vec{d}_j) = {\cal O}(\vec0)\,\big]~
           \tooo{~~{\cdot}f^j(x_i)~~}~{\cal O}(\vec{d}_j)~
           \To~{\cal O_M}(\vec{d}_j)~. }
As discussed in \SS\sRhyme, the degree-$\vec0$ function (constant) here
acts as one complex variable worth of a `ghost' degree of freedom, so
that degree-$\vec{d}_j$ functions on $\cal M$ are obtained as
degree-$\vec{d}_j$ functions on $\cal X$, taken however, modulo a
complex multiple of $f^j(x_i)$:
\eqn\eFx{ \d f^j(x_i)~~ \cong~~ \d f^j(x_i)~ + ~\l\,f^j(x_i)~, }
where $\d f^j(x_i)$ denotes a general variation of $f^j(x_i)$---thus a
general function with the same degrees. In the light of the
definition~\eRes{}, these equivalence classes may be considered as the
`zeroth level' contributions to $H^0({\cal M}, {\cal E}_{f^j})$ and thereby
to $H^1({\cal T})$.

However, there are in general further non-vanishing contributions to the
cohomology classes $H^\star({\cal M}, {\cal E}_{f^j})$, i.e., to
$H^\star({\cal T})$. In order to check our arguments and results below, we
recall some of facts from the Bott-Borel-Weil theorem. In particular,
\eqn\eBBWOk{ \dim H^q(\CP{n},{\cal O}\6(k)) ~=
 ~\cases{ \d_{q,0}{n+k\choose n} & if $0\leq k$~,\cr
          \noalign{\vglue2mm}
          0 & if ${-}(n{+}1)<k<0$~,\cr
          \noalign{\vglue2mm}
          \d_{q,n}{-k-1\choose n} & if $k\leq{-}(n{+}1)$~,\cr}}
where the middle case may be subsumed under the last one, since
\eqn\eXXX{ {\ttt{-k-1\choose n}} ~\define~ 0~,\qquad
           {\rm for}~{-}(n{+}1)<k<0~. }
Consider then any ${\cal O}(\vec{q})$ in the sequence~\eKoSdj. This will
give rise to a non-zero result in the Koszul complex if $q_i$, the
components of $\vec{q}$, satisfy either $q_i\ge0$ or $q_i<{-}n_i$, that is,
if no $q_i$ lies within ${-}1,\ldots,{-}n_i$.
 Let $R$ denote the subrange of the index $i$ for which $q_i\ge0$,
and let $\bar{R}$ denote the complementary subrange of $i$ for which
$q_i<{-}n_i$. Applying Eq.~\eBBWOk\ for each $\CP{n_i}_i$ separately and
then putting it all together, we obtain
\eqn\eBBWCI{ \dim H^q\big({\cal X}, {\cal O}(\vec{q})\big) ~=~
 \prod_{i\in R}{n_i+q_i\choose q_i}
      \prod_{i\in\bar R}{-q_i-1\choose n_i}~, \qquad
  q=\sum_{i\in\bar R}n_i~, }
and this contributes $q$ steps below ${\cal O}(\vec{q})$ in the
chart~\eKoCh. Note that $q_i\geq{-}(n_i{+}1)$ owing to the \CY\
condition~\eCYcond, so that the product over $i\in\bar{R}$ is non-zero and
then equal to one only if $q_i={-}(n_i{+}1)$, for all $i\in\bar{R}$.
In fact, the Bott-Borel-Weil theorem says more: the nonzero
${\cal O}(k)$-valued cohomology groups are generated by degree-$k$
monomials for $k\geq0$; the ${\cal O}(\vec{q})$-valued cohomology is then
simply the product of such factors, or zero if
${-}n_i\leq q_i \leq{-}1$ for some $i$.
 \ping

Our task now is to demonstrate that exactly the same result is obtained
by collecting all non-zero and holomorphic partial residues of the
type~\eRes{}!

Consider therefore the rational polynomials appearing in the
sequence~\eKoSdj, and search for the cohomology on $\cal M$ valued in
$\d f^j(x_i)$---polynomials of the homogeneity of $f^j(x_i)$. Starting
in~\eKoSdj\ from ${\cal O}(\vec{d}_j)$ and going to the left, we divide
one-by-one by the defining polynomials $f^j(x_i)$, to obtain
\eqn\eRatnl{ {\d f^j(x_i)\over \prod_{k\in Q} f^k(x_i)}~, }
where $Q$ is the subset labeling the polynomials with which we have
divided. As $|Q|$ is the number of defining polynomials in the
denominator, this contribution is in the $|Q|^{th}$ column of a
chart like~\eKoCh. The degrees of homogeneity of~\eRatnl\ are
$\vec{d}_j{-}\sum_{k\in Q}\vec{d}_k$.

In order for~\eRatnl\ to produce an analytic contribution to
 $H^\star({\cal M},\ldots)$, the degree with respect to each \CP{n_i} must
be made non-negative. To that end, we multiply~\eRatnl\ by the top
differentials of those $\CP{n_i}_i$ (labeled by the $i\in S$) with
respect to which~\eRatnl\ has negative degrees.
 Now, owing to the homogeneity of the $\CP{n_i}_i$, including only a proper
factor of $(x_i\rd^{n_i}x_i)$ makes no sense: no such factor is
invariant under (projective) coordinate reparametrizations, \ie, with
respect to $\PGL{n_i{+}1,\IC}\approx SU(n_i{+}1;\IC)$.
 This then defines the residue
\eqna\eResdj
 $$ \Res^S_Q\big[\d f^j\big] ~=~
 {1\over(2\p i)^{|Q|}}\oint\!{\ldots}\!\oint_{\prod_{k\in Q} \G(f^k)}
 {\prod_{i\in S} (x_i\rd^{n_i} x_i)
            \over \prod_{k\in Q} f^k(x_i)}\>\d f^j(x_i)~.
\eqno\eResdj{a}
 $$
By construction,
\eqn\eDegRes{ \deg_i\big(\Res^S_Q\big[\d f^j\big]\big)~ =
 ~\cases{d_{ij}+(n_i{+}1)-\sum_{k\in Q} d_{ik}~ \geq ~0~, & $i\in S$,\cr
         \noalign{\vglue2mm}
         d_{ij}-\sum_{k\in Q} d_{ik}~ \geq ~0~,       & $i\nin S$,\cr}}
so that the degrees are non-negative over the un-integrated $\CP{n_i}_i$,
$i\nin S$. As for the {\it integrated} $\CP{n_i}_i$, $i\in S$, only the
residues with the strict equality for all $i\in S$ produce well-defined
($U(n_i{+}1)$-covariant, $i\in S$) forms for
$H^\star({\cal M},{\cal E}_j)$, and we require
 $$ \deg_i\big(\Res^S_Q\big[\d f^j\big]\big)~ =
 ~d_{ij}+(n_i{+}1)-\sum_{k\in Q} d_{ik}~,\quad i\in S~,
\eqno{\eDegRes'}
 $$
to vanish, whereupon the requirements~\eRes{b,c} are met. This recovers the
`vanishing' part of the Bott-Borel-Weil theorem (upon the `filtration'
into $H^\star({\cal M},\3)$) for ${\cal O}(k)$ bundles~\eBBWOk. That is,
the contribution vanishes if some of the degrees ends up in the
${-}1,\ldots,{-}n_i$ region; the subset of indices $S$ in~\eResdj{a} was
labeled $\bar{R}$ for the Bott-Borel-Weil theorem.
 The Bott-Borel-Weil theorem also provides for the degrees to be more
negative than ${-}(n_i{+}1)$, but this cannot occur if we restrict to
\CY\ subspaces of ${\cal X}=\prod_i\CP{n_i}_i$.

Now, if $q=0,1,2, 3$ and in a coordinate neighborhood where
$x_i^{\m_0^i}\ne0$, the residue will evaluate, as in~\eRES{a}, to
 $$
 \Res^S_Q\big[\d f^j\big]~=~
   \big(\prod_{i\in S}x^{\m_0^i}_i\big)\>
    {\rd x^{\n_1}\cdots\rd x^{\n_q}
     \over J^{~\n_1\cdots\n_q}_{(\m_0^i,~i\in S)}}
    ~\d f^j(x_i)~,
\eqno\eResdj{b}
 $$
with $J^{~\n_1\cdots\n_q}_{(\m_0^i,~i\in S)}$ the appropriate Jacobian.

If $q<0$, only $\sum_{i\in S}n_i$ contour integrals can be performed, and
the result is a rational function with poles of total order $|q|$, placed
$|q|$ steps to the left from the the top row of the lower left quadrant of
a chart such as~\eKoCh. This contributes to
$H^\star({\cal M},{\cal E}_{f^j})$ only if the function $\d f^j$ can be
chosen so as to cancel the remaining $|q|$ poles and provide a non-zero and
complex-analytic result via the application of the L'Hospital theorem as
discussed above.
 This application of the L'Hospital theorem does not reduce the number of
differentials (the order of the form) and cancels a pole merely through
judicious choice of $\d f(x_i)$. Therefore, the evaluation via L'Hospital
theorem will not affect the `filtration' and will not move the contribution
within the lower left quadrant of the corresponding chart, such as~\eKoCh.
The contribution stays $|q|$ places to the left of the top right-most
element in the lower left quadrant, and represents $|q|^{th}$ order
`ghost-for-ghost\3' variables. Indeed, as differentials of negative order,
these may be identified with vector or higher rank (contravariant) tensor
variables and therefore represent reparametrizations, see \SS~2.4.

Finally, if $q>3$, the residue~\eResdj{a} would seem to contribute to
non-existent cohomology groups on the 3-fold $\cal M$. However, such
residues either occur in pairs and together with maps of the type~\eFxMap\
such that both the domain and the image `gauge each other away', or such
residues end up being `gauged away' through coordinate reparametrizations
(see below). Moreover, it can be shown as in Ref.~\rPDM, that even the case
$q=3$ can be avoided, except of course for the holomorphic 3-form~\eABGC\
in the degree-$\vec0$ case. That is, complete intersections where there
exist `additional' non-zero residues with $q=3$ turn out to be completely
equivalent to others where no such residue occurs.

The residue is evaluated as in~\SS\sTheRes, and takes the final form
 $$ \Res^S_Q\big[\d f^j\big]~
 =  \W_{(q)}^{\rv} \d f^{j(\mv)}_{\rv}(x)~,
\eqno\eResdj{c}
 $$
where $\W_{(q)}^{\rv}$ is defined in~\eInterOm, and $m_i$ and
$\vd^{\mv}$ are as in~\eVGr.
 Note that $m_i>0$ for at least one $\CP{n_i}_i$ (otherwise, the complete
intersection configuration~\eGCnfg\ would become block-diagonal and so
correspond to either $T^2\times K3$ or $T^6$), whence the variation of the
$\mv^{th}$ gradient, $\d f^{j(\mv)}$, is of too small a degree to
contribute by itself to $H^\star({\cal M},{\cal E}_{f^j})$.
 A `missing piece', i.e., a certain universal quantity of the appropriate
`missing' degree is needed for constructing a proper contribution to
$H^\star({\cal M},{\cal E}_{f^j})$. When comparing with \LGO{} results, this
`missing piece' is identified with the twisted vacuum, the charges of which
indeed complement the charges of the monomials $\d f^{j(\mv)}(x)$ so as
to provide marginal, charge-$({\pm}1,1)$ states. When comparing with the
Koszul calculation, this `missing piece' is identified with (a product of)
Levi-Civita alternating symbols $\e^{\cdots}$, which in turn precisely
corresponds to the twisted vacua~\rCfC.
 In some cases at least, this `missing piece' may be identified with a
radical monomial~\refs{\rCfC,\rPeriods}; more about this below.

\bigskip
To summarize, all the residue kernels are determined completely from the
`ghost-for-ghost-\3' sequence such as the one in the upper left quadrant
of the chart~\eKoCh: each step to the left implies division by one of the
defining polynomials; these reciprocals of polynomials are then multiplied
by the (covariant) top-differentials of the embedding space factors so as
to enable a residue integral to calculate the residue. Integrating
variations of the defining polynomials using these kernels, we form a rack
of polynomial-valued residues [placed into the lower left quadrant
of~\eKoCh], contributing thus to
$H^\star({\cal M},\oplus_j{\cal E}_{f^j})$, and thereby to
$H^\star({\cal M},{\cal T})$; see below.

\subsec{Reparametrizing residues}\subseclab\sReps\noindent
 \phrasing{a rondo}{rallentando}
The erudite reader will undoubtedly have protested by now that the
contributions obtained thus far are overabundant, simply on account
of not having discounted the coordinate reparametrization degrees of
freedom. Of course, the situation with the polynomial deformations~\eFx\
 is quite well understood~\refs{\rPhilip,\rPDM,\rBeast}. The same
Jacobian ring structure (=polynomials modulo the ideal of gradients) shows
up again in the \LGO{} analysis~\refs{\rLGO,\rChiRi,\rGVW,\rSKEW} and then
in a somewhat modified form also for the $E_6$ {\bf1}'s~\rSKEW. Let us
therefore turn to the action of coordinate reparametrizations on the
residues discussed above.

\bigskip
\noindent{$\2{\hbox{Coordinate reparametrizations}}$.~~}\ignorespaces
The action of coordinate reparametrizations of \CP{n} on functions over
\CP{n} is generated by the operators $\ell^\m(x)\vd_\m$, where $\ell^\m(x)$
is an $(n{+}1)$-vector of linear combinations of the homogeneous coordinates
on \CP{n}; write $\ell^\m(x)=x^\n\l_\n^\m$. The trace part of the matrix
$\l_\n^\m$ is easily seen to correspond to (a complex multiple of) the
Euler homogeneity operator; when acting on homogeneous functions, the trace
part of $\l_\n^\m$ then merely duplicates the overall rescalings of the
kind already accounted for by imposing equivalence relations such as~\eFx.
Therefore---when acting on homogeneous functions---the matrix $\l_\n^\m$
will be required to be traceless.
 On the product ${\cal X}=\prod_i\CP{n_i}_i$, the reparametrization group is
the direct product of the individual reparametrization groups, and so is
generated by $\bigoplus_i\ell^{\n_i}(x_i)\vd_{\n_i}$.
 Notably, all these reparametrization operators (tangent vectors) are
homogeneous and of degree zero.

For the polynomial deformations $\d f^j(x_i)$, i.e., the zeroth level
residues, the resulting equivalence may be written (on $\cal M$) as
\eqn\eLinRep{ \d f^j(x) ~ \cong ~\d f^j(x)~ +
     \sum_{\n_i}^N \ell^{\n_i}(x)\vd_{\n_i}~ f^j(x)~. }
These equivalence relations may be used to eliminate (`gauge away')
$\sum_{i=1}^N n_i(n_i+2)$ parameters from~\eFx, the remaining ones then
representing the reparametrization class.

On comparing this equivalence relation~\eLinRep\ with the earlier~\eEqRel\
and~\eOneEq, a mapping notation akin to~\eOneMap\ immediately suggests
itself:
\eqn\eLinMap{ \bigoplus_{i=1}^N\ell^{\n_i}(x_i)\vd_{\n_i}
           ~\tooo{~\rd f^j~}~ \d f^k(x_i)~, }
where the $\ell{\cdot}\vd$ is regarded as a (tangent) vector field\ft{The
contraction $\rd x^\m{\cdot}\vd_\n=\d_\n^\m$ cancels the differential
against the derivative, so that
$\ell^{\n_i}\vd_{\n_i}{\cdot}\rd x^{\m_j}\vd_{\m_j}f^k =
 \ell^{\n_i}\vd_{\n_i}f^k = \d f^k$ produces a linear reparametrization
of $f^k$.}, and is mapped to holomorphic functions of degree
$\deg(f^j)$. The two groups of mappings, \eOneMap\ and~\eLinMap, provide
the very core of the Koszul calculations. Also, the equivalence
relation~\eOneEq\ may be regarded as a special case of~\eLinRep,
corresponding to the trace part of $\l_{\n_i}^{\m_i}$ in
$\ell^{\m_i}(x){=}\l_{\n_i}^{\m_i}x^{\n_i}$. Therefore, both equivalence
relations may be regarded as generated by the gradients of the defining
polynomials, whence the immediate connection to the results of
Ref.~\refs{\rPhilip,\rLGO,\rChiRi,\rGVW}. Indeed, the effective identity
of the ideals was observed in Ref.~\rCQG; here we see the source of this.

\bigskip
The natural analogue of this action of coordinate reparametrizations
to the more general residues such as those in~\eResdj{c} is
straightforward upon the following realization.
 The previous subsection showed that the residue operator entails the
natural restriction of holomorphic functions (polynomials) to the subspace
${\cal M}\subset{\cal X}$. The same then better be true also of vector
fields. Note that $\vd_{\n_i}$ may continue to serve as a basis for tangent
vectors; it is the coefficients $\ell^{\n_i}(x_i)$ which will be restricted
to the submanifold through residue integrals. Without much ado, then,
\eqn\eResRep{ {\eqalign{
 \oint\!{\ldots}\!\oint_{\prod_{k\in Q}\G(f^k)}
{\prod_{i\in S} (x_i\rd^{n_i} x_i)
            \over \prod_{k \in Q} f^k}
                   \sum_{\n_i} \ell^{\n_i}(x)\vd_{\n_i}
 &=~ \sum_{\n_i} \Res^S_Q\big[\ell^{\n_i}\big]\vd_{\n_i}~,   \cr}}}
with the $\Res^S_Q\big[\ell^{\n_i}\big]$ holomorphic or zero,
is the appropriate restriction of the tangent vectors to the subspace
$\cal M$, and so the natural reparametrization operator for the
residue~\eResdj{c}.

Indeed, this operator is obtained with the same residue kernel
as~\eResdj{c}, whereby the coordinate reparametrization action is perfectly
analogous to~\eLinRep, except that all terms are placed {\it within the
residue symbol}. This also means that the operations of linear
reparametrization and evaluation of the residue commute. In fact, requiring
that this always be so prevents the residue reparametrization operators
from acting on (residue) functions unless they are both of the same residue
level and moreover have the same residue kernel. This `selection rule' is
the residue analogue of the $j$-twisting selection rule in \LGO{s}. The
effective identity of these `selection rules' of rather disparate origins
seems to be borne out in practice, although we are aware neither of a
rigorous proof nor of a counter-example. (See however \SS~4.)

To summarize, we have the residue analogue of the coordinate
reparametrization equivalence relation~\eLinRep:
\eqn\eRLinRep{
 \Res^S_Q\big[\d f^j(x)\big] ~\cong~
 ~\Res^S_Q\big[\d f^j(x)\big]~ +
 ~\sum_{k,\n_i}
  \Res^S_Q\big[\ell^{\n_i}\big]\big(\vd_{\n_i}f^j(x)\big)~, }
where all the terms have their integral expressions, given above. In the
`mapping' notation:
\eqn\eRLinMap{
     \bigoplus_{i=1}^N\Res^S_Q\big[\ell^{\n_i}\big]\vd_{\n_i}
     ~\tooo{~\rd f^j~}~
      \Res^S_Q\big[\d f^k(x_i)\big]~, }
Thus, the holomorphic (tangent) vector fields
$\Res^S_Q\big[\ell^{\n_i}\big]\vd_{\n_i}$ serve to eliminate
(`gauge away') some of the parameters in~\eResdj{c}, the residue symbol
again accounting for the proper restriction to the subspace $\cal M$.
The mapping~\eLinMap\ and the equivalence relations~\eLinRep\ are now
clearly the special `zeroth level' case ($|Q|=0=|S|$) of~\eRLinMap\
and~\eRLinRep, respectively; the maps themselves, i.e., generators of the
equivalence relations remain the same: $f^j$ and $\rd f^j$, i.e.,
(linear combinations of) the gradients of $f^j$.

This, however, is not the whole reparametrization story. It is
possible to list all the residue kernels, racking them in a chart such
as~\eKoCh, just as was done above for the polynomial-valued residues.
Recall that the degree of homogeneity of the linear reparametrization
operators is $\vec0$, and note that the original paradigm~\eABGC\ may
be regarded as the residue of the constant 1, with the residue kernel being
$\prod_{i=1}^N(x_i\rd^{n_i}x_i)/\prod_{j=1}^K f^j$. We
thus immediately turn to
\eqn\eKahlers{ \oint_{\G_{f^1}}\!\!\ldots\oint_{\G_{f^K}}
        {\prod_{i=1}^N(x_i\rd^{n_i}x_i)\over \prod_{j=1}^Kf^j(x)}
          \sum_{\m_j,\n_j}x_j^{\m_j}\l_{\m_j}^{\n_j}\vd_{\n_j}~, }
and note that the canonical contractions
\eqn\eCC{ \vd_{\n_j}\rd x_i^{\m_i}=\d^i_j\,\d^{\m_i}_{\n_j}~, }
lower the overall degree of the differential by one and produce (upon
taking the residues) $N$ independent 2-forms of homogeneity $\vec0$; one
corresponding to each of \CP{n_i}.
 Since $\sum_{\m_j,\n_j}x_j^{\m_j}\l_{\m_j}^{\n_j}\vd_{\n_j}$ are local
tangent vectors, these 2-forms are tangent bundle valued. The (Serre-)
duals of these are then cotangent bundle valued 1-forms, that is,
(1,1)-forms.

The trace-part of each matrix $\l_{\m_i}^{\n_i}$ contributes precisely
nothing, in virtue of the skew-symmetry of $(x_i\rd^{n_i}x_i)$, with which
$x_j^{\m_j}\l_{\m_j}^{\n_j}\vd_{\n_j}$ become contracted via~\eCC:
\eqn\eZap{ {\eqalign{ (x\rd^n x){\cdot}x^\m\l_\m^\n\vd_\n
 &={1\over (n{+}1)!} \e_{\r_0\cdots \r_n}
                         x^{\r_0}\,\rd x^{\r_1}\cdots\rd x^{\r_n}\cdot
       x^\m(\inv{n}\d_\m^\n{\rm Tr}[\BM{\l}]+\0\l_\m^\n)\vd_\n~,\cr
 &={{\rm Tr}[\BM{\l}] \over n\,(n{+}1)!}~
                        \underbrace{\e_{\r_0\cdots \r_j\cdots \r_n}
                         x^{\r_0}x^{\r_j}}_{=0}~\rd x^{\r_1}\cdots
                         \widehat{\rd x^{\r_j}}
                         \cdots\rd x^{\r_n}{\cdot}~,\cr
 \noalign{\vglue-3mm}
 &~~~+{1\over (n{+}1)!}~\underbrace{\e_{\r_0\cdots \r_j\cdots \r_n}
                         x^{\r_0}\0\l_\m^{\r_j}x^\m}_{\ne0}~
                         \rd x^{\r_1}\cdots\widehat{\rd x^{\r_j}}
                         \cdots\rd x^{\r_n}{\cdot}~,\cr}}}
where we have suppressed the subscript $i$ for clarity and the `hat'
labels omitted factors; the traceless part of matrix $\0\l_\m^\n$
does contribute.
 However, unlike the other ${\cal T_X}$-valued residues which act as
reparametrizations of some polynomial-valued residue or another, these
residues {\it typically} `stand by themselves' as there is {\it typically}
nothing they could act on (see however below). All of them are ${\cal
T_X}$-valued 2-forms on the 3-fold $\cal M$ --- one set, parametrized by
$\0\l_{\m_i}^{\n_i}$, for each factor $\CP{n_i}_i$.
 It may seem as each of these would depend on each of the off-diagonal
matrix elements, but this is not so: each one such element is merely a
coordinate-reparametrized copy of another. That is, modulo coordinate
reparametrizations themselves, the ${n_i+1\choose2}$ various possible
choices are all equivalent, whence such residues produce precisely one such
2-form for each factor $\CP{n_i}_i\subset{\cal X}$.
 Tangent-bundle valued 2-forms being dual on a \CY\ 3-fold to (1,1)-forms,
and there being precisely one per each factor $\CP{n_i}_i$ --- each of the
residues~\eKahlers\ corresponds to (the dual of the pullback of) the
K\"ahler form on each of the $\CP{n_i}_i$'s!

We employ here the standard results~\refs{\rGrHa,\rBeast} which enable us
to identify the kernel and the cokernel of the map~\eRLinMap:
\eqn\eRLM{ H^q({\cal Q},{\cal T_Q}) ~\tooo{~\i~}~
     \bigoplus_{i=1}^N\Res^S_Q\big[\ell^{\n_i}\big]\vd_{\n_i}
     ~\tooo{~\rd f^j~}~
      \Res^S_Q\big[\d f^k(x_i)\big] ~\tooo{~\D~}~
      H^{q+1}({\cal Q},{\cal T_Q})~, }
where $q=\sum_{i\in S}n_i-|Q|$, and ${\cal Q}$ is the simultaneous zero-set
$f^j{=}0$, $j\in Q$; ultimately, once $Q$ includes all constraints,
$\cal Q$ becomes the desired complete intersection $\cal M$.
In this extension of~\eRLinMap, $\i$ identifies with elements of
$H^q({\cal Q},{\cal T_Q})$ those elements of
$\Res^S_Q\big[\ell^{\n_i}\big]\vd_{\n_i}$ which are annihilated by
$\rd f^j$ (a.k.a.\ the {\it kernel\/} of $\rd f^j$).
Also, the equivalence class of $\Res^S_Q\big[\d f^k(x_i)\big]$ modulo the
$\rd f^j$-multiples of $\Res^S_Q\big[\ell^{\n_i}\big]\vd_{\n_i}$  (a.k.a.\
the {\it cokernel\/} of $\rd f^j$) contributes, via the `degree-changing
map' $\D$, to $H^{q+1}({\cal Q},{\cal T_Q})$.
\bigskip

Equivalently, we may  consider, dually, the cotangent bundle
valued residues, replacing $\sum_{\n_j}x_j^{\m_j}\l_{\m_j}^{\n_j}\vd_{\n_j}$ in
the above calculations
with
\eqn\eDuals{ \sum_{\n_i}(\g_i)_{\n_i}(x)\rd x_i^{\n_i}~,\qquad
             i=1,\ldots,N~. }
For these to be of homogeneity $\vec0$, the function
$(\g_i)_{\n_i}(x)$ and must have degree ${-}1$ over $\CP{n_i}_i$, but be
constant over the remaining factors in the embedding space. It is easy to
see that only the `diagonal' rational differentials have a residue along
corresponding linear hypersurfaces in \CP{n_i}. On each factor
$\CP{n_i}_i$ and upon coordinate reparametrizations, such `diagonal'
rational (in fact, logarithmic) differential forms may be written as, say,
${\rd x_i^0\over x_i^0}=\rd\log(x_i^0)$ and the corresponding linear
hypersurface ($\approx\CP{n_i-1}$) where the residue is supported is
defined by $x_i^0=0$.
 The ${\rd x_i^0\over x_i^0}$ are then the de~Rham duals of hyperplanes in
$\CP{n_i}_i$, i.e., representatives of the K\"ahler classes on \CP{n_i},
and may also serve as their pull-backs on $\cal M$.
 We will not pursue this alternative calculation here any further.

Back to the $\cal T_X$-valued residues, it is a tedious but straightforward
exercise to show in a case by case analysis, that no
additional non-zero residue contribution to the cohomology on $\cal M$ can
occur. In the dual calculation with~\eDuals, the explicit appearance of
some particular $\rd x_i^{\n_i}$ precludes multiplication by
$(x_i\rd^{n_i}x_i)$, whereas multiplication by some $(x_j\rd^{n_j}x_j)$,
$j\ne i$, does not help produce a contour integral to pick out the
residue --- except for the cases which are duals to the $\cal T_X$-valued
residues listed above.

A detailed comparison then with the Koszul computation is again
straightforward, but rather laborious and will not be presented here. It
however reinforces that the residue map (symbol) naturally provides a
concrete realization of the Koszul calculations. Simple examples such as
this one certainly {\it can} be analyzed by simply listing all possible
residue kernels, all relevant rational/radical polynomials, and then
determining the final list of effective deformations. In general, however,
this approach quickly gets out of hand and the application of the standard
Koszul machinery seems difficult to avoid. The residue map can then be used
to provide a concrete realization of any particular cohomology element.

Finally, although they may be combined, note that the gauge-equivalence
in~\eLinRep\ is different from that one in~\eEqRel. The former is a
consequence of the global symmetry
$\PGL{n{+}1,\IC}\approx{\rm GL}(n{+}1,\IC)/C^*$ of projective spaces. That
is, in the underlying 2-dimensional field theory as studied by Witten~\rPhases,
there is a ${\rm GL}(n{+}1,\IC)$ global field reparametrization symmetry,
of which the diagonal (projective) $C^*$ is gauged.
 The equivalence relation~\eEqRel, however, stems from the imposition of
constraints $f^j(x_i)=0$, from which the whole construction of the
Koszul sequence~\eKoSeq\ and the subsequent calculations~\eKoCh\ are
developed. Where appropriate, we have indicated the corresponding
BRST-type treatment of such constraints, but are not specifying here the
details of this approach any further.
\ping

\bigskip
\noindent{$\2{\hbox{Too many K\"ahler forms}}$.~~}\ignorespaces
Finally, it remains to discuss a type of `reparametrizations' which can
occur only when two or more hypersurface is being intersected to define
$\cal M$, the submanifold under study, and then only for
$H^2({\cal M},{\cal T})$. Such models do not have a straightforward \LGO{}
analogue \`a la Refs.~\refs{\rLGO,\rGVW,\rChiRi}.

In certain configurations~\eGCnfg\ft{These are configurations which have a
``decomposing 2-leg'', see~\rPDM\ or \SS~2.1.2 of~\rBeast.}, not all
(pullbacks of) K\"ahler forms of the
factor $\CP{n_i}_i$'s are independent elements of $H^2({\cal M},{\cal T})$.
This may be easiest to follow by considering an example:
\eqn\eOLD{ {\cal M}~\in~\Cnfg{\CP3\cr\CP1\cr\CP1}{4&0\cr0&2\cr1&1\cr}~,
 \qquad\cases{f(x,z)=0~, & of degree (4,0,1),\cr
              \noalign{\vglue2mm}
              g(y,z)=0~, & of degree (0,2,1),\cr}}
where the generic manifold has $b_{1,1}=2$, $b_{2,1}=86$ and $\EU=-168$.

The ghost-for-ghost- sequence is
\eqn\eKSOLD{\vcenter{\vbox{\offinterlineskip
\halign{&#&~\hfil$#$\hfil&$#$&\hfil$#$\hfil&$#$&\hfil$#$\hfil&%
                          $#$&\hfil$#$\hfil~&~$#$\hfil\cr
  &{\cal O}(p{-}4, q{-}2, r{-}2)
   &\stack{\nearrow}{0}{\searrow}
    &\stack{{\cal O}(p, q{-}2, r{-}1)}{5}
             {{\cal O}(p{-}4, q, r{-}1)}
       &\stack{\searrow}{0}{\nearrow}
        &{\cal O}(p,q,r)&\To~{\cal O_M}(p,q,r)&\cr}}}~, }
and the full rack of $\d f$- and $\d g$-valued residues and similarly the
${\cal T}_x$,- ${\cal T}_y$- and ${\cal T}_z$- valued residues are easy to
calculate along the lines described above. The novel feature occurs when
accounting for all the equivalence relations of the type~\eRLinRep, that
is, of all the maps of the type~\eRLinMap. In particular, there are four
contributions to $H^2({\cal M},{\cal T_X})$ for the above example:
\eqn\eKahls{ \Res_{f,g}^{x,y,z}\big[\vq_x\big]~,\qquad
             \Res_{f,g}^{x,y,z}\big[\vq_y\big]~,\qquad
             \Res_{f,g}^{x,y,z}\big[\vq_z\big]~, }
and
\eqn\eTwoVs{ \Res_f^x\big[\vq_z\big] ~=~
             \Res_f^x\big[z^r\l_r^s\big]\,\vd_s ~=~
             {\mit\L}^s\vd_s~, }
where
\eqn\eXXX{ \vq_x~\define~ x^a\l_a^b\vd_b~,\quad
           \vq_y~\define~ y^\a\l_\a^\b\vd_\b~,\quad
           \vq_z~\define~ z^r\l_r^s\vd_s~. }
This last contribution~\eTwoVs\ provides a straightforward (2-parameter)
reparametrization of $\Res_f^x[\d g]$, just as in~\eRLinRep. Simply
on account of the degrees, we can fix this to be
\eqn\egRep{
 \Res^x_f\big[\d g(y,z)\big] ~\cong~
 ~\Res^x_f\big[\d g(y,z)\big]~ +
 ~\Res^x_f\big[z^r\l_r^s\big]\big(\vd_s g(y,z)\big)~, }
and note that $\Res_f^x[\d g]$ is a quadric over $\CP1_y$, which we
may write as $\d g'_z$. Modulo the two gradients $\l^s(\vd_s g)$, this
produces a 1-parameter equivalence class, which we may write as
$\{\d g'_z/g'_z\}$. More precisely, as in~\eResdj{d}, we can write this in
more detail as
\eqna\eXXX
 $$\twoeqsalignno{
 \Res^x_f\big[\d g(y,z)\big]
 &=\bigg[{\inv{3!}x^a\e_{abcd}\rd x^c\rd x^d \over
                      \vd_s\vd_bf(x,z)}\bigg]
    \d\big(\vd_s g(y,z)\big)
&&\define{\mit\D}^s\d\big(\vd_s g(y,z)\big)~,
 &\eXXX{a}\cr
 \noalign{\vglue-1.5mm}
 \Res^x_f\big[z^r\l_r^s\big]
 &=\bigg[{\inv{3!}x^a\e_{abcd}\rd x^c\rd x^d \over
                      \vd_r\vd_bf(x,z)}\l_r^s\bigg]
&&\define{\mit\L}^s~,
 &\eXXX{b}\cr
 &\quad\hbox{no summation over $b$}\cr
}$$
where the quantities in the square brackets, ${\mit\D}^s$ and ${\mit\L}^s$,
are nowhere-zero 2-forms on the \CY\ quartics in $\CP3_x$,
$\{\vd_s f(x,z){=}0\}$, one for each $s=1,2$.
$\d(g'_z)=\d\big({\mit\D}^r\vd_r g(y,z)\big)$ being a quadric over
$\CP2_y$, it depends on ${3\choose2}{=}3$ parameters and
$g'_z=\big({\mit\L}^r\vd_r g(y,z)\big)$ providing two gauge degrees of
freedom, ${\mit\L}^r$, the quotient $\{\d(g'_z)/g'_z\}$ is 1-dimensional.

Having resolved this equivalence relation, we would seem to remain with
three elements in $H^2({\cal M},{\cal T_X})$---the duals of (the pullbacks
of) the K\"ahler forms~\eKahls. Also, the 1-dimensional cohomology group of
polynomial-valued 2-forms $\{\d(g'_z)/g'_z\}$, should then produce a
1-dimensional contribution to
$H^3({\cal M},{\cal T})=H^{2,3}({\cal M})$ much as polynomial-valued
0-forms contribute to $H^1({\cal M},{\cal T})$; see~\eRLM.
However, on general grounds, we know that $H^{2,3}({\cal M})=0$, whence
precisely one linear combination of the three residues~\eKahls\ {\it must}
provide one last equivalence relation so as to gauge away
$\{\d g'_z/g'_z\}$ completely. In the process, we will remain with only two
independent elements for $H^2({\cal M},{\cal T_X})$, and will therefore
have also $\dim\,H^2({\cal M},{\cal T_M})=2$, i.e., $b_{2,2}=b_{1,1}=2$.

To see from where such a map is induced, note that the $\vq_x$- and
$\vq_z$-valued residues may be written in a `cascade' fashion:
\eqn\eXXX{ \Res_{f,g}^{x,y,z}\big[\vq_x{\oplus}\vq_z\big] ~=~
 \Res_f^x\Big[\Res_g^y\big[(z\rd z)(\vq_x{\oplus}\vq_z)\big]\Big]~. }
That is, the $\G(g)$-contour integral may be calculated first, over
$\CP1_y$, which produces a quantity of degree $(0,0,1)$ and so is a
well-defined `partial residue' of the type~\eRes{}. This would not be
possible for $\Res_{f,g}^{x,y,z}[\vq_y]$, since the $\vd_y$ contracts with
$\rd y$ and we cannot integrate over $\CP1_y$ to define the `inside'
residue. Therefore, the equivalence relation~\egRep\ must be enlarged,
so as to correspond to the combined map
\eqn\eMMap{
 \matrix{ \Res_f^x\Big[ \Res_g^y\big[\> (z\rd z)
           \cdot(\vq_x{\oplus}\vq_z)\>\big]\Big]\cr
            \hfill\Res_f^x\big[\> z^r\l_r^s\vd_s\>\big]\cr}\quad
 \matrix{ \ddd\tooo{~~\vd_zg~~}\cr
           \noalign{\vglue1mm}
           \ddd\tooo{~~\>\rd g\>~~}\cr
           \noalign{\vglue1mm}}\quad
           \Res_f^x\big[\>\d g(y,z)\>\big]~, }
where the maps have been determined solely from degrees of homogeneity.
Note that all these contributions happen {\it within the $\Res_f^x[~~]$
operator}, and we can therefore determine the equivalence relation
corresponding to the upper map from considering the `inner' $\Res_g^y[~~]$.
Using that
\eqn\eXXX{ (z\rd z) \cdot \vq_z ~=~
           \inv2\e_{pq}z^p\rd z^q \cdot z^r\l_r^s\vd_s ~=~
           \inv2\e_{ps}z^p z^r\l_r^s~, }
we note that, using~\eResdj{},
\eqn\eZz{ {\eqalign{
     \Res_g^y\big[\>(z\rd z)\cdot\vq_z\>\big]
 &=~ \bigg[{\e_{\a\b}y^\b\over\vd_r\vd_\a g(y,z)}\bigg]_{g=0}
       \big(\l_{(r}^s\e^{\phantom{s}}_{q)s}z^q\big) ~=~ Z(z)~,\cr
 y^\b &=\hbox{{\it const.}, no summation over $\a$} ~,\cr }}}
is constant over $\CP3_x\,$, nowhere-zero over
$\{\vd_r\vd_\a g{=}0\}\subset\CP1_y$, and linear over
$\CP1_z$, so that upon multiplication with $\vd_zg$, it can be added to $\d
g$ within the $\Res_f^x[~~]$ symbol in the target of the maps in~\eMMap. The
equivalence relation~\egRep\ therefore becomes enlarged to
\eqn\egdgRep{
 \Res^x_f\big[\d g(y,z)\big] ~\cong~
 ~\Res^x_f\big[\d g(y,z)\big]~ +
 ~\Res^x_f\big[z^r\l_r^s+Z^s(z)\big]\big(\vd_s g(y,z)\big)~, }
where $Z^s=\e^{rs}\l_{(r}^p\e^{\phantom{s}}_{q)p}z^q
          =z^r(\l^s_r{-}\inv2\d^s_r{\rm Tr}[\BM{\l}])$, $s=1,2$, are two
linearly independent terms in the linear function obtained as the
residue~\eZz. This then provides for completely `gauging away' the
contribution to polynomial-valued 2-forms, whence we recover $H^3({\cal
M},{\cal T})=H^{2,3}({\cal M})=0$ and also
$b_{2,2}=b_{1,1}=2$, since only $\Res^{x,y,z}_{f,g}[\vq_x]$ and
$\Res^{x,y,z}_{f,g}[\vq_y]$ remain from~\eKahls.\ping

This residue calculation is in complete agreement with the Koszul
calculation, even to the extent that there is a formal 1--1 correspondence
in the {\it form\/} of the representatives. For example, the fact that all
the representatives in~\eMMap\ occur within $\Res_f^x[~~]$ is paralleled
by the fact that all Koszul representatives corresponding to those
in~\eMMap, occur with an overall factor of $\e^{abcd}$, the Levi-Civita
alternating symbol on $\CP3_x$. Indeed, $\Res_f^x[~~]$ involves integration
over the 3-form $(x\rd^3x)$, which in turn includes $\e_{abcd}$ in the
definition~\ePVol.

Moreover, a similarly detailed 1--1 correspondence can be established
between all of the residue and Koszul representatives, whereby we believe
to have demonstrated the effective identity between these two methods. Our
goal will now be to see if the residue calculations might be extended
beyond what is known about the Koszul calculations and also to explore
whether the similarly detailed agreement with the \LGO{} methods persists
beyond the overlap with the Koszul calculations, where the identity follows
owing to earlier results~\refs{\rCQG,\rCfC}. Before that, however, a few
examples are perhaps in order, to illustrate the various types of `higher
order' contributions.

\newsec{A Representative Ragout}\seclab\SRagout\noindent
\subsec{A reconnoissance residue raffle}\subseclab\sWarp\noindent
 \phrasing{a rhumba}{rasguaedo}
Consider the `warped' model of Ref.~\rCQG:
\eqn\eWarp{ {\cal M} \in \Cnfg{\CP3\cr\CP2\cr\CP1\cr}
                              {3&1&0\cr0&2&1\cr0&0&2\cr}~:\qquad
                         \cases{f(x)=0~,   &of degree (3,0,0),\cr
                                \noalign{\vglue1mm}
                                g(x,y)=0~, &of degree (1,2,0),\cr
                                \noalign{\vglue1mm}
                                h(x,y)=0~, &of degree (0,1,2),\cr}
}
where the generic manifold $\cal M$ has $b_{1,1}=9$, $b_{2,1}=33$ and
$\EU=-48$. The ghost-for-ghost-\3\ sequence is
\eqn\eKoSeqII{\vcenter{\vbox{\offinterlineskip
\halign{&#&~\hfil$#$\hfil&$#$&\hfil$#$\hfil&$#$&\hfil$#$\hfil&%
                          $#$&\hfil$#$\hfil~&~$#$\hfil\cr
  &{\cal O}\mtx(p{-}4,q{-}3,r{-}2)
   &\Stack{\nearrow}{0}{\to}{0}{\searrow}
    &\Stack{{\cal O}\mtx(p{-}1,q{-}3,r{-}2)}{1}
           {{\cal O}\mtx(p{-}3,q{-}1,r{-}2)}{1}
           {{\cal O}\mtx(p{-}4,q{-}2,r)}
     &\Stack{\to}{3}{\searrow}{14}{\searrow\vrule width0pt depth18pt}
       \mkern-23mu
      \Stack{\vrule height22pt width0pt\nearrow}{15}{\nearrow}{1}{\to}
      &\Stack{{\cal O}\mtx(p,q{-}1,r{-}2)}{1}
             {{\cal O}\mtx(p{-}1,q{-}2,r)}{1}
             {{\cal O}\mtx(p{-}3,q,r)}
       &\Stack{\searrow}{0}{\to}{0}{\nearrow}
        &{\cal O}\mtx(p,q,r)&\To~{\cal O_M}\mtx(p,q,r)&\cr}}}~, }
where we listed the degrees vertically, akin to~\eWarp.

Among the polynomial-valued residues, at zero level, there are
\eqna\eZLR
 $$
 \d f(x)~,\qquad \d g(x,y)~,\qquad \d h(y,z)~,
\eqno\eZLR{a}
 $$
which are the usual polynomial deformations, and
 $$
 {\d f(x)\over f(x)}\to\l_f~,\qquad
 {\d g(x,y)\over g(x,y)}\to\l_g~,\qquad
 {\d h(y,z)\over h(y,z)}\to\l_h~,
\eqno\eZLR{b}
 $$
each of which is holomorphic only if the variation is chosen to be
proportional to the polynomial in the denominator, whence the rations
are constants: $\l_f,\l_g,\l_h$. From the $\cal T_X$-valued residues, at
zero level, there are
 $$
 \vq_x\define x^a\0\l_a^b\vd_b~,\qquad
 \vq_y\define y^\a\0\l_\a^\b\vd_\b~,\qquad
 \vq_z\define z^r\0\l_r^s\vd_s~,
\eqno\eZLR{c}
 $$
which generate the usual coordinate reparametrizations. The  matrices
$\0\l$ being traceless, they combine with \eZLR{b} to produce the familiar
equivalence class~\rPhilip
\eqn\eXXX{
\left[\matrix{\d f(x)\cr\d g(x,y)\cr\d h(y,z)\cr}\right] ~\cong~
 \left[\matrix{\d f(x)\cr\d g(x,y)\cr\d h(y,z)\cr}\right] ~+~
 \big(x^a\l_a^b\vd_b+y^\a\l_\a^\b\vd_\b+z^r\l_r^s\vd_s\big)
 \left[\matrix{f(x)\cr g(x,y)\cr h(y,z)\cr}\right]~, }
where now the $\l$ matrices are no longer traceless. These produce the
well known
\eqn\eXXX{ [{\ttt{6\choose3}{-}16}]+
           [{\ttt{4\choose1}{4\choose2}{-}9}]+
           [{\ttt{3\choose1}{3\choose2}{-}4]}=24 }
polynomial deformation contributions to $H^1({\cal M},{\cal T_M})$.

Among higher-level polynomial-valued residues, we find nonzero only:
\eqna\eHLRP
 $$ \twoeqsalignno{
 {\d g(x,y)\over h(y,z)}
 &\to~\Res_h^z\big[\d g\big] ~=~ {\mit\D}^\a\d(\vd_\a g_y)~\sim~\d g'_y~,
&&\quad\deg{=}(1,1,0)~, &\eHLRP{a}\cr
 {\d f(x)\over g(x,y) h(y,z)}
 &\to~\Res_{g,h}^{y,z}\big[\d f\big] ~=~ {\mit\D}^a\d(\vd_af)~,
&&\quad\deg{=}(2,0,0)~, &\eHLRP{b}\cr }
 $$
where ${\mit\D}^\a$ and ${\mit\D}^a$ are the {\it nowhere-zero} 0- and
1-forms calculated from the respective residues in the coordinate patch
where, e.g., $y^0,z^0=const$.
\eqna\eOneF
 $$ \twoeqsalignno{
 {\mit\D}^\a  &\define
   \bigg[{z^0\over\vd_\a J_{(0)}}\bigg]_{h=0}~, &\quad
 J_{(0)}      &\define \left[{\vd h\over
                              \vd z^1}\right]_{z^0\neq0}~,  &\eOneF{a}\cr
 {\mit\D}^a   &\define
   \bigg[{y^0z^0\rd\h\over\vd_aJ^\h_{(0,0)}}\bigg]_{\SSS g=0\atop
                                                     \SSS h=0}~, &\quad
 J^\h_{(0,0)} &\define
            \left[{\vd(\h,g,h)\over
                   \vd(y^1,y^2,z^1)}\right]_{y^0,z^0\neq0}~.&\eOneF{b}\cr
}$$
Among higher-level ${\cal T_X}$-valued residues, we find nonzero only:
\eqna\eHLRT
 $$ \twoeqsalignno{
 {y^\a\l_\a^\b\vd_\b\over h(y,z)}
 &\to~\Res_h^z\big[\vq_y\big] ~=~ {\mit\L}^\b\vd_\b~,
&&\quad\deg({\mit\L}^\b){=}(0,0,0)~, &\eHLRT{a}\cr
 {x^a\l_a^b\vd_b\over g(x,y)h(y,z)}
 &\to~\Res_{g,h}^{y,z}\big[\vq_x\big] ~=~ {\mit\L}^b\vd_b~,
&&\quad\deg({\mit\L}^b){=}(0,0,0)~,  &\eHLRT{b}\cr }
 $$
and the three duals of (the pullbacks of) the K\"ahler forms:
 $$ \Res_{f,g,h}^{x,y,z}\big[\vq_x\big]~,\qquad
    \Res_{f,g,h}^{x,y,z}\big[\vq_y\big]~,\qquad
    \Res_{f,g,h}^{x,y,z}\big[\vq_z\big]~,
\eqno\eHLRT{c}
 $$
where
\eqn\eXXX{ \vq_x \define x^a\0\l_a^b\vd_b~,\qquad
           \vq_y \define y^\a\0\l_\a^\b\vd_\b~,\qquad
           \vq_z \define z^r\0\l_r^s\vd_s~, }
and the matrices $\0\l$ are traceless. The ${\mit\L}^\b$ and ${\mit\L}^b$
are the {\it nowhere-zero} 0- and 1-forms calculated from the respective
residues in the coordinate patch where, e.g., $y^0,z^0=const$.
\eqn\eXXX{
 {\mit\L}^\b~\define~
     \bigg[{z^0\over\vd_\a J_{(0)}}\l_\a^\b\bigg]_{h=0}~, \qquad
 {\mit\L}^b~\define~
     \bigg[{y^0z^0\rd\h\over\vd_aJ^\h_{(0,0)}}\l_a^b\bigg]_{\SSS g=0\atop
                                                             \SSS h=0}~, }
and where $\vd_\a J_{(0)}$ and $\vd_aJ^\h_{(0,0)}$ are the same as
in~\eOneF. For example, the 1-forms ${\mit\D}^a$ and ${\mit\L}^a$ are both
nowhere zero holomorphic 1-forms on the torus embedded as the simultaneous
zero-set $\vd_ag(x,y)=0=h(y,z)$ in $\CP2_y{\times}\CP1_z$, i.e., on a member
of $\cnfg{\IP^2\cr\IP^1\cr}{2&1\cr0&2\cr}$.

To summarize, the degree-(3,0,0) polynomial-valued cohomology is obtained
from:
\eqn\eWrpChF{\vcenter{\vbox{\offinterlineskip\ninepoint
\halign{&#&~\hfil$#$\hfil&$#$&\hfil$#$\hfil&$#$&\hfil$#$\hfil&%
                          $#$&\hfil$#$\hfil~&\vrule#&~$#$\hfil\cr
  &{\cal O}\mtx({-}1,{-}3,{-}2)
   &\Stack{\nearrow}{0}{\to}{0}{\searrow}
    &\Stack{{\cal O}\mtx(~~2,{-}3,{-}2)}{1}
           {{\cal O}\mtx(~~0,{-}1,{-}2)}{1}
           {{\cal O}\mtx({-}1,{-}2,~~0)}
     &\Stack{\to}{3}{\searrow}{14}{\searrow\vrule width0pt depth18pt}
       \mkern-23mu
      \Stack{\vrule height22pt width0pt\nearrow}{15}{\nearrow}{1}{\to}
      &\Stack{{\cal O}\mtx(~~3,{-}1,{-}2)}{1}
             {{\cal O}\mtx(~~2,{-}2,~~0)}{1}
             {{\cal O}\mtx(0,0,0)}
       &\Stack{\searrow}{0}{\to}{0}{\nearrow}
        &{\cal O_X}\mtx(3,0,0)&&\To{\cal O_M}\mtx(3,0,0)&\cr
  &\omit&\omit&\omit&\omit&\omit&\omit&\omit&height4pt&\omit&\cr
 \noalign{\hrule}
  &\omit&\omit&\omit&\omit&\omit&\omit&\omit&height2pt&\omit&\cr
  &0&&0&&\l_f&\tooo{~f~}&\d f&& \To\{\d f/f{\cdot}\l_f\}\in H^0&\cr
  &\omit&\omit&\omit&\omit&\omit&\omit&\omit&height3pt&\omit&\cr
  &0&&0&&0&&0&&\To{\mit\D}^a\d(\vd_af)\in H^1&\cr
  &\omit&\omit&\omit&\omit&\omit&\omit&\omit&height3pt&\omit&\cr
  &0&&0&&0&&0&&\To H^2=0&\cr
  &\omit&\omit&\omit&\omit&\omit&\omit&\omit&height3pt&\omit&\cr
  &0&&{\mit\D}^a\d(\vd_af)&&0&&0&&\To H^3=0&\cr
  &\omit&\omit&\omit&\omit&\omit&\omit&\omit&height3pt&\omit&\cr
  &0&&0&&0&&0&&\To H^4({\cal M})\id0&\cr
  &\omit&\omit&\omit&\omit&\omit&\omit&\omit&height3pt&\omit&\cr
  &0&&0&&0&&0&&\To H^5({\cal M})\id0&\cr
  &\omit&\omit&\omit&\omit&\omit&\omit&\omit&height3pt&\omit&\cr
  &0&&0&&0&&0&&\To H^6({\cal M})\id0&\cr
  &\omit&\omit&\omit&\omit&\omit&\omit&\omit&height3pt&\omit&\cr
  & \hbox to0pt{\hglue35pt\ninepoint
    The residue ${\mit\D}^a\d(\vd_af)$ is defined in Eq.~\eHLRP{b}.\hss}
        &\omit&\omit&\omit&\omit&\omit&\omit& width0pt height10pt&\omit&\cr
}}}} 
the degree-(1,2,0) polynomial-valued cohomology from:
\eqn\eWrpChG{\vcenter{\vbox{\offinterlineskip
\halign{&#&~\hfil$#$\hfil&$#$&\hfil$#$\hfil&$#$&\hfil$#$\hfil&%
                          $#$&\hfil$#$\hfil~&\vrule#&~$#$\hfil\cr
  &{\cal O}\mtx({-}3,{-}1,{-}2)
   &\Stack{\nearrow}{0}{\to}{0}{\searrow}
    &\Stack{{\cal O}\mtx(~~0,{-}1,{-}2)}{1}
           {{\cal O}\mtx({-}2,~~1,{-}2)}{1}
           {{\cal O}\mtx({-}3,~~0,~~0)}
     &\Stack{\to}{3}{\searrow}{14}{\searrow\vrule width0pt depth18pt}
       \mkern-23mu
      \Stack{\vrule height22pt width0pt\nearrow}{15}{\nearrow}{1}{\to}
      &\Stack{{\cal O}\mtx(~~1,~~1,{-}2)}{1}
             {{\cal O}\mtx(0,0,0)}{1}
             {{\cal O}\mtx({-}2,~~2,~~0)}
       &\Stack{\searrow}{0}{\to}{0}{\nearrow}
        &{\cal O_X}\mtx(1,2,0)&&\To{\cal O_M}\mtx(1,2,0)&\cr
  &\omit&\omit&\omit&\omit&\omit&\omit&\omit&height4pt&\omit&\cr
 \noalign{\hrule}
  &\omit&\omit&\omit&\omit&\omit&\omit&\omit&height2pt&\omit&\cr
  &0&&0&&\l_g&\tooo{~g~}&\d g&& \To\{\d g/\6(g\l_g)\}+\d g'_y\in H^0&\cr
  &\omit&\omit&\omit&\omit&\omit&\omit&\omit&height6pt&\omit&\cr
  &0&&0&&\d g'_y&&0&&\To H^1=0&\cr
  &\omit&\omit&\omit&\omit&\omit&\omit&\omit&height3pt&\omit&\cr
  &0&&0&&0&&0&&\To H^2=0&\cr
  &\omit&\omit&\omit&\omit&\omit&\omit&\omit&height0pt&\omit&\cr
  &\vdots&&\vdots&&\vdots&&\vdots&&\qquad\vdots&\cr
  &\omit&\omit&\omit&\omit&\omit&\omit&\omit&height3pt&\omit&\cr
  & \hbox to0pt{\hglue35pt\ninepoint
    The contribution $\d g'_y$ is defined in Eq.~\eHLRP{a}.\hss}
        &\omit&\omit&\omit&\omit&\omit&\omit& width0pt height10pt&\omit&\cr
}}}} %
and the degree-(0,1,2) polynomial-valued cohomology from:
\eqn\eWrpChH{\vcenter{\vbox{\offinterlineskip
\halign{&#&~\hfil$#$\hfil&$#$&\hfil$#$\hfil&$#$&\hfil$#$\hfil&%
                          $#$&\hfil$#$\hfil~&\vrule#&~$#$\hfil\cr
  &{\cal O}\mtx({-}4,{-}2,~~0)
   &\Stack{\nearrow}{0}{\to}{0}{\searrow}
    &\Stack{{\cal O}\mtx({-}1,{-}2,~~0)}{1}
           {{\cal O}\mtx({-}3,~~0,~~0)}{1}
           {{\cal O}\mtx({-}4,{-}1,~~2)}
     &\Stack{\to}{3}{\searrow}{14}{\searrow\vrule width0pt depth18pt}
       \mkern-23mu
      \Stack{\vrule height22pt width0pt\nearrow}{15}{\nearrow}{1}{\to}
      &\Stack{{\cal O}\mtx(0,0,0)}{1}
             {{\cal O}\mtx({-}1,{-}1,~~2)}{1}
             {{\cal O}\mtx({-}3,~~1,~~2)}
       &\Stack{\searrow}{0}{\to}{0}{\nearrow}
        &{\cal O_X}\mtx(0,1,2)&&\To{\cal O_M}\mtx(0,1,2)&\cr
  &\omit&\omit&\omit&\omit&\omit&\omit&\omit&height3pt&\omit&\cr
 \noalign{\hrule}
  &\omit&\omit&\omit&\omit&\omit&\omit&\omit&height2pt&\omit&\cr
  &0&&0&&\l_h&\tooo{~h~}&\d h&& \To\{\d h/\6(h\l_h)\}\in H^0&\cr
  &\omit&\omit&\omit&\omit&\omit&\omit&\omit&height3pt&\omit&\cr
  &0&&0&&0&&0&&\To H^1=0&\cr
  &\omit&\omit&\omit&\omit&\omit&\omit&\omit&height0pt&\omit&\cr
  &\vdots&&\vdots&&\vdots&&\vdots&&\qquad\vdots&\cr
  &\omit&\omit&\omit&\omit&\omit&\omit&\omit&height3pt&\omit&\cr
  & \hbox to0pt{\hglue25pt\ninepoint
    This produces only `polynomial deformations'.\hss}
        &\omit&\omit&\omit&\omit&\omit&\omit& width0pt height10pt&\omit&\cr
}}}} %

The Reader should now have no difficulty reproducing the analogous charts
for the $\vq_x$,-  $\vq_y$- and  $\vq_x$-valued residues. Note that the
various contributions listed in~\eZLR{c}, \eHLRT{a} and~\eHLRT{b} occur in
the same {\it place} as the polynomials and residues on which they act;
this illustrates the selection rule encoded in~\eRLinRep, wherein all
residues must be constructed with the same residue kernel, and so the same
index-sets $S,Q$. So, the~\eHLRT{a} act on~\eHLRP{a}, and the~\eHLRT{b} act
on~\eHLRP{b} as reparametrizations:
\eqna\eRRepW
 $$\eqalignno{
 \Res_h^z\big[\d g\big] &\cong ~\Res_h^z\big[\d g\big]
 + \Res_h^z\big[\vq_y\big]{\cdot}\rd_yg~, &\eRRepW{a}\cr
 \Res_{g,h}^{y,z}\big[\d f\big] &\cong ~\Res_{g,h}^{y,z}\big[\d f\big]
 + \Res_{g,h}^{y,z}\big[\vq_x\big]{\cdot}\rd_xf~, &\eRRepW{b}\cr
}$$
which evaluate to
 $$\eqalignno{
 \d g'_y &\cong ~\d g'_y + \l^\a(\vd_\a g)~, &\eRRepW{a'}\cr
 {\mit\D}^a\d(\vd_af) &\cong ~{\mit\D}^a\d(\vd_af)
 + {\mit\L}^a(\vd_a f)~. &\eRRepW{b'}\cr
}$$
The former of these two equivalence classes are bilinear in $x,y$, and
taken modulo $\vd_\a g$, leaving $4{\cdot}3-3=9$ elements for
$H^1({\cal M},{\cal T_M})$. The latter of these are quadratic in $x$,
taken modulo $\vd_af$, leaving ${5\choose2}-4=6$ elements for
$H^2({\cal M},{\cal T_M})$; see~\eRLM.

Note also that the $\vq_x$,-  $\vq_y$- and  $\vq_x$-valued residues
in~\eHLRT{c} occur in positions which are void in the polynomial-valued
charts~\eWrpChF--\eWrpChH. By the selection rule of~\eRLinRep, the
residues~\eHLRT{c} cannot act on anything and `stand on their own', as
three elements of $H^2({\cal M},{\cal T_X})$. Under the obvious embedding
map ${\cal T_M}\to{\cal T_X}$, these are isomorphic to three corresponding
elements in $H^2({\cal M},{\cal T_M})=H^{2,2}({\cal M})$, which are (the
duals of the pullbacks of) the K\"ahler forms on \CP3, \CP2 and \CP1.

\bigskip
Putting all these together, we have obtained:
\item{\bf-} all the $\cal T_M$-valued 1-forms: 24 zeroth-level and 9
         first-level;
\item{\bf-} all the $\cal T_M$-valued 2-forms: 6 second-level and 3
         third-level;
\item{\bf-} that all contributions are in precise 1--1 correspondence with
         the Koszul computation and also with the \LGO{} analysis, except
         that the higher-level residues~\eRRepW{} represent only the
         `monomial' part of the massless modes~\refs{\rLGO,\rSKEW},
         excluding the `twisted vacuum' part.
\bigskip

To remedy this last observation and increase the degree of the
representatives without changing the number of parameters, as in
Ref.~\rCfC, we seek a ``universal'' scalar multiplier, that is, a scalar
which may only depend on the defining equations~\eWarp, and determinants of
their derivative matrices. The representatives~\eRRepW{a'} clearly have
degree (1,1,0) and the scalar multiplier must have degree (0,1,0).
Without much ado,
\eqn\eSqRtDDH{ \sqrt{\det\big[\vd_r\vd_s h(y,z)\big]} }
precisely fits the bill. In a concrete example for a Landau-Ginzburg
potential \`a la Refs.~\refs{\rLGO,\rGVW} $W = f(X) + g(X,Y) + h(Y,Z)$,
we may choose:
\eqn\eLGW{
   f(X) = \sum_{r=0}^3 X_r^{~3}~,\quad
 g(X,Y) = \sum_{\a=0}^2 X_\a Y_\a^{~2}~,\quad
 h(Y,Z) = \sum_{r=0}^1 Y_r Z_r^{~2}~. }
Then, $\sqrt{\det\big[\vd_r\vd_s h(Y,Z)\big]}=\sqrt{Y_0Y_1}$. It is
straightforward to verify that not only is the scaling weight of this
object equal to the scaling weight of $\ket{\10}_{NS}^{(6)}$, the
$6^{th}$ twisted Neveu-Schwarz-vacuum in the Landau-Ginzburg picture, but
the `warp' symmetry charges~\rCQG\ agree as well.

The representatives~\eRRepW{b'} clearly have degree (2,0,0) and the scalar
multiplier must have degree (1,0,0). This object is in fact a little more
difficult to spot, as it relies on the fact that the $X,Y,Z$ fields are
coupled. That is, the superpotential $W=f(X)+g(X,Y)+h(Y,Z)$ may be decoupled
into four irreducible models:
\eqn\eSPotW{ W = \sum_{r=0}^1\Big[X_r^{~3}+X_r Y_r^{~2}+Y_r Z_r^{~2}\Big]~
        + \Big[X_2^{~3}+X_2Y_2^{~2}\Big] + \Big[X_3^{~3}\Big]~. }
Then, it makes perfect sense to consider
${1\over6}\sqrt{\det\big[\vd_r\vd_s f(X)\big]}=\sqrt{X_0X_1}~$
which again perfectly fits the bill; both the scaling weight (degree) and
the warp charge equal those of the twisted vacuum.

The translation of these radicals into the Koszul language is again
fairly straightforward. We note that the Koszul representatives
corresponding to~\eHLRP{a} and~\eHLRP{b} carry an additional factor:
$\e^{rs}$ and $\e^{\a\b\g}\e^{rs}$, respectively. These being
skew-symmetric, no contraction with the defining polynomial coefficient
tensors is possible directly (as those are symmetric). However, their
(direct) square can; the product
\eqn\eXXX{ \e^{pq}\e^{rs}~h_{\a pr}h_{\b qs} y^\a y^\b }
is in fact a scalar. Since $h_{\a pr}y^\a=\vd_p\vd_rh(y,z)$, this product
is simply the determinant of the hessian (with respect to $z$) of $h(y,z)$.
This shows that $\e^{pq}\e^{rs}$ is dual to $h_{\a pr}h_{\b qs} y^\a y^\b$,
whence $\e^{rs}$ is, rather formally, dual to
$\sqrt{h_{\a pr}h_{\b qs} y^\a y^\b}$.
The second ``universal'' scalar multiplier follows in the same vein,
noting that
\eqn\eXXX{ \e^{pq}\e^{rs}\,h_{\a\,pr}h_{\d\,qs}~
            \e^{\a\b\g} \e^{\d\l\k}\,g_{a\,\b\l}g_{b\,\g\k}\>x^ax^b }
is a scalar, whence
$\sqrt{\e^{pq}\e^{rs}\e^{\a\b\g}\e^{\d\l\k}}$ is dual to
$\sqrt{h_{\a\,pr}h_{\d\,qs}g_{a\,\b\l}g_{b\,\g\k}\>x^ax^b}$. Amusingly,
with the above specific choice of polynomials~\eSPotW, this contraction
yields zero, unless for example $h(Y,Z)$ is shifted by a cross-coupling
term $Y_2 Z_0 Z_1$ (for which the above square-root does produce
$\sqrt{X_0X_1}$, as above). For a generic choice of superpotential, the
radical is of course nonzero.

\subsec{Reaming, refining and reducing residues}\subseclab\sReam\noindent
 \phrasing{a raga}{rubato}
Of course, the wealth of complete intersections also features models
where somewhat unusual representatives or relations amongst those
occur. The following examples are meant to provide further practical
guidance.

Consider for example the family of complete intersections
\eqn\eExII{
 {\cal M} \in \Cnfg{\CP5\cr\CP1\cr}{4&1&1\cr0&1&1\cr}~,\qquad
 \cases{ f(x)   &$\define f_{abcd}\,x^a x^b x^c x^d = 0$,\cr
         g(x,y) &$\define g_{a\b}\,x^a y^\b         = 0$,\cr
         h(x,y) &$\define h_{a\b}\,x^a y^\b         = 0$,\cr} }
where $b_{1,1}=2$ and $b_{2,1}=86$. The Koszul (ghost-for-ghost-\3)
sequence for degree-$(p,q)$ functions here becomes
\eqn\eKoSeqII{\vcenter{\vbox{\offinterlineskip
\halign{&#&~\hfil$#$\hfil&$#$&\hfil$#$\hfil&$#$&\hfil$#$\hfil&%
                          $#$&\hfil$#$\hfil~&~$#$\hfil\cr
  &{\cal O}(p{-}6,q{-}2)
   &\Stack{\nearrow}{0}{\to}{0}{\searrow}
    &\Stack{{\cal O}(p{-}2,q{-}2)}{5}
           {{\cal O}(p{-}5,q{-}1)}{5}
           {{\cal O}(p{-}5,q{-}1)}
     &\Stack{\to}{-3}{\searrow}{2}{\searrow\vrule width0pt depth15pt}
       \mkern-23mu
      \Stack{\vrule height17pt width0pt\nearrow}{2}{\nearrow}{-1}{\to}
      &\Stack{{\cal O}(p{-}1,q{-}1)}{5}
             {{\cal O}(p{-}1,q{-}1)}{5}
             {{\cal O}(p{-}4,q)}
       &\Stack{\searrow}{0}{\to}{0}{\nearrow}
        &{\cal O}(p,q)&\To~{\cal O_M}(p,q)&\cr}}}~. }

Owing to the commensurate degrees of $g(x,y)$ and $h(x,y)$, there
occur several atypical cohomology representatives which may be
regarded as residues in the above generalized sense. For example,
variations of $h(x,y)$ are listed from~\eKoSeqII\ with $(p,q)=(1,1)$:
\eqn\eKoSeqIIG{\vcenter{\vbox{\offinterlineskip
\halign{&#&~\hfil$#$\hfil&$#$&\hfil$#$\hfil&$#$&\hfil$#$\hfil&%
                          $#$&\hfil$#$\hfil~&~$#$\hfil\cr
  &{\cal O}({-}5,{-}1)
   &\Stack{\nearrow}{0}{\to}{0}{\searrow}
    &\Stack{{\cal O}({-}1,{-}1)}{5}
           {{\cal O}({-}1,0)}{5}
           {{\cal O}({-}1,0)}
     &\Stack{\to}{-3}{\searrow}{2}{\searrow\vrule width0pt depth15pt}
       \mkern-23mu
      \Stack{\vrule height17pt width0pt\nearrow}{2}{\nearrow}{-1}{\to}
    &\Stack{{\cal O}(0,0)}{5}
           {{\cal O}(0,0)}{5}
           {{\cal O}({-}3,1)}
     &\Stack{\searrow}{0}{\to}{0}{\nearrow}
      &{\cal O}(1,1)&\To~{\cal O_M}(1,1)&\cr}}}~. }
The two ${\cal O}(0,0)$ sheaves of degree-$(0,0)$ functions correspond
to the two residue kernels: ${\d h(x,y)\over h(x,y)}{=}\l_h$ and
 ${\d h(x,y)\over g(x,y)}$.
The former is obviously of zeroth level and provides for the usual
one-parameter equivalence as discussed above
\eqn\eHxyIIh{ \d h(x,y)~~ \cong~~ \d h(x,y)~ + ~\l_h\,h(x,y)~. }
The other kernel ${\d h(x,y)\over g(x,y)}$, however, is also of zeroth
level and makes perfect sense---on $\cal M$---as a holomorphic object, if
this $\d h(x,y)$ is again chosen to be proportional to $h(x,y)$. While the
rational function $h/g$ is {\it not} holomorphic on $\CP5{\times}\CP1$,
it is on $\cal M$. To see this, note that the intersection of hypersurfaces
$\{g{=}0\}\cap\{h{=}0\}$ contains $\cal M$ as the locus of $f{=}0$ therein.
 However, by definition, both $g(x,y)$ and $h(x,y)$ vanish in
$\{g{=}0\}\cap\{h{=}0\}$, so that their (limiting) ratio may be evaluated
using the L'Hospital theorem. Such ratios are also considered as residues
and we define
\eqn\eHbyG{
  \Res_{{\cal M}}\Big[{h(x,y)\over g(x,y)}\Big] ~\define~
  \lim\nolimits_{{\cal M}}\Big[{h(x,y)\over g(x,y)}\Big]~
  = ~\hbox{\it const.}~, }
noting that this is a holomorphic degree-$(0,0)$ function on the compact
manifold $\cal M$. This of course provides for another one-parameter
equivalence, and combining with~\eHxyIIh,
\eqn\eHxyIIgh{ \d h(x,y)~~ \cong~~ \d h(x,y)~ + ~\l_h\,h(x,y)~
                                       + ~\l_g\,g(x,y)~. }
The analogous is then true of variations of $g(x,y)$:
\eqn\eGxyIIgh{ \d g(x,y)~~ \cong~~ \d g(x,y)~ + ~\k_g\,g(x,y)~
                                       + ~\k_h\,h(x,y)~. }
Note that this is almost obvious from the fact that Eq.~\eFx\ is not
written in a covariant fashion. Instead, on writing
\eqn\eFX{ \d f^j(x_i)~~ \cong~~ \d f^j(x_i)~ + ~\l^j_k\,f^k(x_i)~, }
where $\l^j_k$ is understood to be zero if it cannot be holomorphic owing
to the relative degrees, such cross-reparametrizations become obvious.
However, merely `covariantizing' Eq.~\eFx\ will not suffice in general and
examples of further generalizations through higher level residues will
occur below.

Next, there is also an unusual first level residue from the
degree-(4,0) sequence
\eqn\eKoSeqIIF{\vcenter{\vbox{\offinterlineskip
\halign{&#&~\hfil$#$\hfil&$#$&\hfil$#$\hfil&$#$&\hfil$#$\hfil&%
                          $#$&\hfil$#$\hfil~&~$#$\hfil\cr
  &{\cal O}({-}2,{-}2)
   &\Stack{\nearrow}{0}{\to}{0}{\searrow}
    &\Stack{{\cal O}(2,{-}2)}{5}
           {{\cal O}({-}1,{-}1)}{5}
           {{\cal O}({-}1,{-}1)}
     &\Stack{\to}{-3}{\searrow}{2}{\searrow\vrule width0pt depth15pt}
       \mkern-23mu
      \Stack{\vrule height17pt width0pt\nearrow}{2}{\nearrow}{-1}{\to}
    &\Stack{{\cal O}(3,{-}1)}{5}
           {{\cal O}(3,{-}1)}{5}
           {{\cal O}(0,0)}
     &\Stack{\searrow}{0}{\to}{0}{\nearrow}
      &{\cal O}(4,0)&\To~{\cal O_M}(4,0)&\cr}}}~. }
{}From ${\cal O}(2,{-}2)\sim{\d f\over gh}$, the residue integrand
 $(y\rd y)\,\d f(x)\over g(x,y)h(x,y)$
is formed which has degree-(2,0). This residue may be evaluated by
contour-integration about {\it one} of the hypersurfaces, either
$\{g{=}0\}$ or $\{h{=}0\}$:
\eqn\eFxIInew{{\eqalign{
 \int_{\G_g\,{\rm or}\,\G_h}{(y\rd y)\,\d f(x)\over
                              g(x,y)h(x,y)}\bigg|_{{\rm on}{\cal M}}
 &=~ {\d f(x)\over\det[\vd_ag\,\vd_\b h]}\bigg|_{{\rm on}{\cal M}}\cr
 &=~\W_{({-}1)}^{ab}\>\d(\vd_a\vd_b f)
 ~=~\d f''(x)~.\cr}}}
The second equality follows on noting that the determinant
$\det[\vd_ag\,\vd_\b h]$ does not vanish on $\cal M$, owing to
smoothness. Alternatively, this is a straightforward consequence of the
general formulae in \SS\sTheRes. The subscript ${-}1$ reminds us that this
contribution ends up in the next to the right-most place in the top line
of the lower left quadrant of the chart \`a la~\eKoCh, and so is a `ghost'
variable. This in turn produces a further equivalence relation to
which the variations of $f(x)$ must be subject:
\eqn\eFxII{ \d f(x)~~ \cong~~ \d f(x)~ + ~\m_f\,f(x)~
                          + ~\det[\vd_\a g\,\vd_\b h]\,\d f''(x)~, }
where again the $\det[\vd_y g\,\vd_y h]$ factor has been introduced
for correct degree of homogeneity. This exemplifies additional
reparametrizations from ${\rm Aut}\big({\oplus}_j{\cal O}(\vec{d}_j)\big)$
which go beyond a mere `covariantization' of~\eFx\ into~\eFX. Note also
that the copy of $\det[\vd_\a g\,\vd_\b h]$ in~\eFxIInew\ is {\it not}
canceled by the one in~\eFxII; the former produces the (double)
derivative $\d f''(x)$.

The alert Reader will have noticed how ``following the filtration'' and the
resulting index $q$ (here ${-}1$) unambiguously determined the fate of this
contribution as generating a 21-parameter equivalence class of variations
of $f(x)$ rather than, say, an independent source of polynomial-valued
0-forms. Recall that $q$, as defined in~\eInterQ, counts the order of the
differential form minus the order of the pole, subtracting the latter on
account of contour integrations with which to evaluate the residue. The present
case then illustrates the $q<0$ case of our general result in \SS\sTheRes.

Finally, as a double-check, we use the Bott-Borel-Weil
theorem to calculate the cohomology corresponding to the
sequence~\eKoSeqIIF:
\eqn\eChExIIF{\vcenter{\vbox{\offinterlineskip
\halign{&#&~\hfil$#$\hfil&$#$&\hfil$#$\hfil&$#$&\hfil$#$\hfil&%
                          $#$&\hfil$#$\hfil~&\vrule#&~$#$\hfil\cr
  &{\cal O}({-}2,{-}2)
   &\Stack{\nearrow}{0}{\to}{0}{\searrow}
    &\Stack{{\cal O}(2,{-}2)}{5}
           {{\cal O}({-}1,{-}1)}{5}
           {{\cal O}({-}1,{-}1)}
     &\Stack{\to}{-3}{\searrow}{2}{\searrow\vrule width0pt depth15pt}
       \mkern-23mu
      \Stack{\vrule height17pt width0pt\nearrow}{2}{\nearrow}{-1}{\to}
    &\Stack{{\cal O}(3,{-}1)}{5}
           {{\cal O}(3,{-}1)}{5}
           {{\cal O}(0,0)}
     &\Stack{\searrow}{0}{\to}{0}{\nearrow}
      &{\cal O}(4,0)&&{\cal O_M}(4,0)&\cr
  &\omit&\omit&\omit&\omit&\omit&\omit&\omit&height6pt&\omit&\cr
 \noalign{\hrule}
  &\omit&\omit&\omit&\omit&\omit&\omit&\omit&height6pt&\omit&\cr
  &0&&0&&H^0{\approx}\IC&\to&H^0{\approx}{\cal O}(4,0)&&\To H^0&\cr
  &\omit&\omit&\omit&\omit&\omit&\omit&\omit&height6pt&\omit&\cr
  &0&&H^1{\approx}{\cal O}(2,0)&\mkern9mu
                                 \smash{\raise10pt\hbox{+}}\mkern-9mu
                                &0&&0&&\To H^1&\cr
  &\omit&\omit&\omit&\omit&\omit&\omit&\omit&height6pt&\omit&\cr
  &0&&0&&0&&0&&\To H^2&\cr
  &\omit&\omit&\omit&\omit&\omit&\omit&\omit&height6pt&\omit&\cr
  &0&&0&&0&&0&&\To H^3&\cr
  &\omit&\omit&\omit&\omit&\omit&\omit&\omit&height6pt&\omit&\cr
  &0&&0&&0&&0&&\To H^4\id0&\cr
  &\omit&\omit&\omit&\omit&\omit&\omit&\omit&height6pt&\omit&\cr
  &0&&0&&0&&0&&\To H^5\id0&\cr
  &\omit&\omit&\omit&\omit&\omit&\omit&\omit&height6pt&\omit&\cr
  &0&&0&&0&&0&&\To H^6\id0&\cr
  &\omit&\omit&\omit&\omit&\omit&\omit&\omit&height6pt&\omit&\cr}}}}
Indeed, the non-zero cohomology as obtained here from the
Bott-Borel-Weil theorem and also the general features of spectral
sequences (filtering and induced `differential' maps) are precisely
reflected in the residue calculations above.\ping

Rather similar to this is the model
\eqn\eExIII{
 {\cal M} \in \Cnfg{\CP5\cr\CP1\cr}{3&2&1\cr0&1&1\cr}~,\qquad
 \cases{ f(x)   &$\define f_{abcd}\,x^a x^b x^c  = 0$,\cr
         g(x,y) &$\define g_{ab\b}\,x^a x^b y^\b = 0$,\cr
         h(x,y) &$\define h_{a\b}\,x^a y^\b      = 0$,\cr} }
where $b_{1,1}=2$ and $b_{2,1}=62$. From the sequence for variations
of $g(x,y)$, there occurs a zeroth level residue with the kernel
 $g(x,y)\over h(x,y)$, which now has degree-(1,0) and is easily seen
to be a linear function over \CP5, upon applying the L'Hospital
theorem.
 Also, from the sequence for variations of $f(x)$, there occurs a
first level residue with the kernel
 $(y\rd y)\d f(x)\over g(x,y) h(x,y)$, which is of degree-(0,0). This is
evaluated similarly to~\eFxIInew, except that this now produces $(\d f''')$
rather than $(\d f'')$. $(\d f''')$ is constant and induces the additional
{\it one}-parameter equivalence class in:
\eqn\eFxIII{ \d f(x)~~ \cong~~ \d f(x)~ + ~\m_f\,f(x)~
                          + ~\det[\vd_\a g\,\vd_\b h]\,(\d f''')~, }
in place of the 21-parameter equivalence class in~\eFxII.

\bigskip
A remark about (possibly) related \LGO{s} is in order. The model~\eExII\
naively allows the construction of a \LGO{} \`a
la~Refs.~\refs{\rLGO,\rGVW,\rChiRi} for which
$f(x)+g(x,y)+h(x,y)$ serves as the superpotential. However, the chiral
superfields $X^a$ and $Y^\a$ having scaling charges $1\over4$ and
$3\over4$, respectively, this \LGO{} would seem to have central charge 6,
which does not correspond correctly to the fact that~\eExII\ describes a
complex 3-dimensional \CY\ manifold. Worse yet, the model~\eExIII\ does
not even allow a consistent (non-zero) scaling charge assignment for a
\LGO{} as described in~\rGVW. Both of these fall in the category of `split
models', for which \LGO{s} are ill-defined~\rSpliTwo. Hopefully, the
framework of Ref.~\rPhases\ may provide a resolution in such no-go
situations.

\newsec{Reflected and Ramified Residues}\seclab\SResW\noindent
 \phrasing{a radiant ritornello}{rinforzando}
The above results are applicable to all complete intersections in products
of flag-spaces --- the simplest of which, $\CP{n}$'s, were explicitly
studied above. However, most of the
\LGO{s}~\refs{\rLGO,\rChiRi,\rGVW,\rWBRG,\rSKEW}, or their gauged
generalizations~\rPhases, naturally apply to
weighted projective hypersurfaces. In many situations it is also of interest to
study models realized  in a quotient of a (weighted) CICY. It is therefore
interesting to see if the above residue calculations admit a `weighted'
generalization.

\subsec{Ratifying the residue recipe}\subseclab\sTheResW\noindent
 \phrasing{a ramulose recoupe}{riprendando}
For simplicity let us consider a hypersurface ${\cal M}$, defined by
$P(x){=}0$ in a quotient of a single weighted projective space
$\CP4_{(k_0,\ldots,k_4)}$ by the group $H$. The general case of complete
intersections in products of weighted projective spaces follows
straightforwardly. Since $\CP4_{(k_0,\ldots,k_4)} = \CP4/j$ where
$j\simeq(\ZZ_d:k_0,\ldots,k_4)$, it is natural to consider a quotient by
$G=H{\times}j$; the notation implies the action\ft{We continue indexing
coordinates with superscripts, as appropriate for contravariance, hoping
that the Reader will have little if any difficulty in distinguishing
superscripts from exponents.}
\eqn\eActQ{ j(x^0,x^1,x^2,x^3,x^4) ~=~
  (\l^{k_0}x^0,\l^{k_1}x^1,\l^{k_2}x^2,\l^{k_3}x^3,\l^{k_4}x^4)~,\quad \l^d=1}
The essential novelty with weighted projective hypersurfaces, and then
necessarily also the complete intersections in products of weighted
flag-spaces, owes to the inherent singularity of these spaces. This stems
from the unequal scaling weights of the quasihomogeneous coordinates (and
more generally, the twist-charges with respect to $G=H{\times}j$); each
singular subspace is fixed by at least one element of the quotient group
$G$.

We then find  the fixed-point sets of $G$ in the usual manner, \ie\ for
each element $g\in G$, the fixed-point set is
$\S_g=\{x^\m\,|\,P{=}0\,,\>(g{-}1)x^\m{=}0\}$.
Let $N_g$ label the coordinates $x^\m$, $\m\in N_g$ on which $g$ acts
non-trivially\ft{The fixed-point sets must be counted with appropriate
multiplicity, one for each element of the symmetry group separately:
if $g^n=1$ for $n$ a {\it prime}, then there will be a total of $n{-}1$
coinciding fixed point sets. If $n$ is not a prime, one must take special
care of the fixed-point sets of $g^m$ when $m$ divides $n$.}.
 Thus, now there {\it does} exist a natural way to split the top
differential form,
\eqn\eTotV{ (x\rd^4x) ~\define~ \sum_{\m,\n,\r,\s,\t}{k_\m\over5!}
 \e_{\m\n\r\s\t} x^\m\rd x^\n\rd x^\r\rd x^\s\rd x^\t~. }
However, due to the unequal scaling weights, the coordinate differentials
in the various permutations  in~\eTotV\ are, in general, of uneven order
over the $\S$. Therefore, we choose to work instead on the affine space
$\IC^5_{(k_0,\ldots,k_4)}$, where the top differential may be written as
\eqn\eAffV{ \rd^5x ~\define~
 \rd x^0{\wedge}\rd x^1{\wedge}\rd x^2{\wedge}\rd x^3{\wedge}\rd x^4~, }
and remember later to reduce the order of the obtained differentials by one
and also to return the explicit $\e_{\cdots}$'s. Then,
\eqn\eSigV{ \rd_\|^{5-|N_g|}x ~\define~ \bigwedge_{\m\nin N_g} \rd x^\m~,
            \qquad x^\m\in\S_g }
is the affine version of the top differential on $\S$, and
\eqn\eNorm{ \rd_\perp^{|N_g|}x ~\define~ \bigwedge_{\m\in N_g} \rd x^\m~,
            \qquad x^\m\nin\S_g }
is the affine version of the top differential on the normal bundle to
$\S\subset\CP4_{(k_0,\ldots,\k_4)}$. Note that both top differentials are
invariant under $g$. In this way the $x^\m$ are not all at the same footing
and hence we can try to write down residue expressions  where it is not
necessary to include all $x^\m$ in the differential as was the case for the
homogeneous projective spaces.

In fact, we are facing a situation not at all unlike the one with the
complete intersections in products of homogeneous projective spaces, in
\SS~\sTheRes. There, we were restricting the residue integrals to a proper
factor $\prod_{i\in S}\CP{n_i}_i\subset\prod_{{\rm all}\,i}\CP{n_i}_i$.
Now, we restrict the residue integrals to a coordinate subset {\it within}
a given weighted projective space. That is, in the affine version we again
restrict to a proper factor $\IC^5_{(k_0,\cdots,k_4)}$. Upon
\hbox{re-}projectivization, however, this factorization is no longer
global: the subset $\S_g$ and its normal bundle no longer form a {\it
global} holomorphic tensor product, although locally this factorization
prevails. The natural generalization of~\eRes{} therefore becomes
\eqn\eResW{ \Res_P^{N_g}\big[\d P\big] ~\define~
          {1\over(2\p i)}\oint_{\G(P)}{\rd_\perp^{|N_g|}x\over P}\>\d P~, }
The residue is evaluated as before, using a change of variables as outlined
in \SS~\sTheRes, and we obtain a holomorphic $q+1=|N_g|{-}1$ form, (before
\hbox{re-}projectivization!)
\eqna\eRESw
 $$
\Res_P^{N_g}\big[\d P\big] ~=~
    {\rd x^{\n_1}\cdots\rd x^{\n_{q+1}}\over
    J^{\n_1\cdots \n_{q+1}}}\>\d P~,
\eqno\eRESw{a}
 $$
with $J^{\n_1\cdots \n_{q+1}}$ the Jacobian for the coordinate
transformation
\eqn\echngW{ (x^{\m_1},{\cdots},x^{\m_{N_g}}; \m_i\in N_g)~
           \longrightarrow ~(x^{\n_1},{\cdots},x^{\n_{q+1}},P)~. }
However, we also note that the degree of the residue~\eResW\ with respect
to the coordinates $x^\m$, $\m\nin N_g$ is the same as that of a
multi-derivative of $P$ with respect to $x^\m$, which we
again denote by $\vd^{\mv} P$ where now $\mv=(m_\m,\m=0,\ldots,4)$ satisfy
$(\mv{\cdot}\kv)=\sum_{\m\nin N_g} k_\m$ with $\kv=(k_\m,\m=0,\ldots,4)$.
Thus, we can again write $\Res_P^{N_g}\big[\d P\big]$, formally, as a
general variation of the $\mv^{th}$ gradient of $P$:
 $$ \eqalignno{
 \Res_P^{N_g}\big[\d P\big]
 &={1\over(2\p i)}\oint_{\G(P)}{\rd^{|N|}x\over P}\>\d P     &\eRESw{b}\cr
 &=~\underbrace{{\rd x^{\n_1}\cdots\rd x^{\n_{q+1}}\over
                 \vd^{{}^{\mv}}J^{~\n_1\cdots\n_{q+1}}}}_{{\rm nowhere-zero}}
                               \d\big(\vd^{\mv}P(x_i)\big)~, &\eRESw{c}\cr
 &\define\sum_{\rv}\W_{(q+1)}^{\rv}\,\d P^{(\mv)}_{\rv}(x)~. &\eRESw{d}\cr
}$$
The multi-index $\rv$ again contains one index for each of the $\mv$
derivatives. Each $\W_{(q+1)}^{\rv}$ will become, upon
\hbox{re-}projectivization, the `nowhere zero holomorphic $q$-form' on
the $q$-dimensional \CY\ space ${\cal Q}_{\rv}$ (given below), and is
again {\it parametrized\/} by the $x_\m,\m\nin N$, and any other parameter
that $P$ depends upon. This is indeed very similar to the situation with
Eqs.~\eRES{}--\eInterOm. Here however the actual number of partial
derivatives in the multi-derivative $\vd^{\mv}$ depends on the weights,
$k_\m$, of the coordinates; the weight of $\vd_\m$ clearly being $-k_\m$.

Alternatively, the multi-derivatives $\vd^{\mv}P$, indexed by $\rv$, again
may be regarded the defining equations of a \CY\ $q$-fold ${\cal Q}_{\rv}$,
embedded in $\CP{|N_g|-1}_{(k_{\n_\star},\n_\star\in N_g)}$, the
\hbox{re-}projectivized normal bundle of $\S_g$.
For each ${\cal Q}_{\rv}$,
\eqn\eInterOm{ \W_{(q+1)}^{\rv}~ \define
    {\rd x^{\n_1}\cdots\rd x^{\n_{q+1}}\over
     \vd^{{}^{\mv}}J^{~\n_1\cdots\n_q}} }
becomes the holomorphic volume-form upon projectivization. Indeed, each
${\cal Q}_{\rv}$ has vanishing first Chern class, since
\eqn\eDegDP{ \deg(\vd^{\mv}P) ~=~ \deg(P) - (\mv{\cdot}\kv)
 ~=~ d-\sum_{\n\nin N_g}k_\n ~=~ \sum_{\m\in N_g}k_\m~. }

To find the dimension of this contribution one would first have to
find the number of monomials in $x_\m,\m\nin N_g$ of degree~\eDegDP,
and then project onto those residue representatives which are invariant
under the action of all elements of $G$, not just $g$. It is then important
to remember to take into account the non-trivial transformation of
$\e_{\cdots}$ which appears in the differential in~\eNorm, which is
equivalent to that of $\prod_{\m\nin N_g}\rd x^\m$.
Finally, depending on whether $|N_g|$ is two or three this will contribute
either to the complex structure or to the K\"ahler deformations in perfect
analogy with the discussion in \SS~\SReap.

Next we consider the reparametrization operator-valued residues. Apart from
the contribution from the original projective space in the form of
\eqn\eProjK{ \vq_{\CP4}\define
              x^\m\l_\m^\n\vd_\n~, \qquad \m,\n=0,\ldots,4~. }
there will be contributions from each of the fixed point sets $N_g$,
associated to the action of the element $g\in G$ which takes the
form
\eqn\eNorB{ \vq_{\S_g} \define x^\m\l_\m^\n\vd_\n~, \qquad\m,\n\nin N_g~. }
Equivalently, and writing $|g|$ for the order of $g$
\eqn\eNORb{ \vq_{\S_g} ~=~ {\cal P}_g(x^\m)\,\l_\m^\n\vd_\n~,\qquad
            {\cal P}_g\define\inv{|g|}(g{+}g^2{+}\ldots{+}g^{|g|})~,
            \quad g^{|g|}=1~, }
that is, ${\cal P}_g$ projects on the $g$-invariant set.
The $\vq_{\S_g}$ are in fact the K\"ahler forms inherited from the
fixed-point sets themselves, much the same as the K\"ahler forms of each
$\CP{n_i}_i$ factor in the discussion of the homogeneous complete
intersections; see \SS~\sReps. Therefore, in place of a single K\"ahler
form with homogeneous projective spaces, we now obtain rather naturally
the multi-component residue class\ft{In case $|g|$ is not prime one would
have study the fixed point set more carefully in which case not all of
the $\vq_{\S_g}$ may be independent.}
\eqn\eXXX{ \oint_{\G(P)} {\rd^5x\over P}
            \big(\,\bigoplus_{g\ne1}\vq_{\S_g}\oplus\vq_{\CP4}\,\big)
  ~=~\oint_{\G(P)} {\rd^5x\over P}
            \big(\,\bigoplus_{{\rm all}\,g}\vq_{\S_g}\,\big). }
since $\CP4_{(k_0,\ldots,k_4)}$ is the fixed-point set of the identity.
The differential order is decreased by one each owing to:
({\bf1})~contraction between a $\rd x^\m$ and a derivative in
 $\bigoplus_{g\ne1} \vq_{\S_g}\oplus\vq_{\CP4}$,
({\bf2})~evaluation of the contour-integral,
({\bf3})~\hbox{re-}projectivization.

\subsec{A reassembling rally}\subseclab\sResExW\noindent
 \phrasing{a rapid romanesca}{recueilli}
As an illustration, we consider the simple and well-studied model, the
family of quasihomogeneous octics in $\CP4_{(1,1,2,2,2)}$, denoted as
$\CP4_{(1,1,2,2,2)}[8]$. The projectivization symmetry is
\eqn\eProjQ{ j ~=~ (\ZZ_8 : 1,1,2,2,2)~, }
and we consider no additional quotient.
 The embedding projective space, $\CP4_{(1,1,2,2,2)}$, is singular at the
subspace $\S\approx\CP2$, found at $x^0=0=x^1$ and parametrized by the
weight-2 coordinates $x^2,x^3,x^4$, is fixed under the action of
$j^4=(\ZZ_2 : 1,1,0,0,0)$, and each point is a local $\ZZ_2$-quotient
singularity.
 An octic quasihomogeneous hypersurface, ${\cal M}=\{P{=}0\}$, in
$\CP4_{(1,1,2,2,2)}$ cannot, in general, avoid meeting this singular plane
and will intersect it in a curve $C$; thus, $\cal M$ is said to have
inherited a curve of local $\ZZ_2$-quotient singularities.

 There exists a blow-up of $\cal M$ along the curve, sometimes denoted
$\Tw{\cal M}$, in which the singular curve $C$ is replaced by a ``ruled
surface'' $E$, obtained by fibering a $\CP1$ over the curve $C$. This
complex 2-fold, $E$, is a {\it divisor} in $\Tw{\cal M}$ and contributes a
new and non-trivial class to $H_4(\Tw{\cal M})$. It is also {\it isomorphic}
to a class in $H^{1,1}(\Tw{\cal M})$, both having a common dual in
$H_2(\Tw{\cal M})$. Together with the (pull-back of the) K\"ahler class of
$\CP4_{(1,1,2,2,2)}$, this provides for $\dim H^{1,1}(\Tw{\cal M})=2$.
Next, note that the octic when restricted to $\S$ becomes a quartic in
$x^2,x^3,x^4$, and so has genus 3; the three handles provide a dual pair of
$S^1$'s each, together with 1-forms supported on each of these, and so
$\dim H^{1,0}(C)=3$. In $E$, therefore, there are three dual pairs of
3-cycles of the form $S^1{\times}\CP1$, and produce three new and
non-trivial elements for $H^{2,1}(\Tw{\cal M})$ [the duals being in
$H^{1,2}(\Tw{\cal M})$]. The remaining 83 elements in
$H^{2,1}(\Tw{\cal M})\approx H^1(\Tw{\cal M},{\cal T}_{\Tw{\cal M}})$ are
easy to find as linear reparametrization classes of octic quasihomogeneous
polynomials.
 Therefore, the blow-up $\Tw{\cal M}$ is a smooth \CY\ space with
$b_{1,1}=2$, and $b_{2,1}=86$, and so $\EU={-}168$. This fully agrees with
the \LGO{} calculation \`a la Refs.~\refs{\rLGO,\rChiRi,\rGVW} and
also~\rSKEW~: there are two twisted $(a,c)$ vacua---matching
$b_{1,1}=2$, and 83 untwisted and 3 twisted $(c,c)$ states---matching
$b_{2,1}=86$ and also the `twistedness' of three of these.\ping

We now turn to the residues. In the case at hand, the factors of~\eAffV
\eqn\eSigV{ \rd_\|^3x ~\define~ \rd x^2{\wedge}\rd x^3{\wedge}\rd x^4~, }
\eqn\eNorm{ \rd_\perp^2x ~\define~ \rd x^0{\wedge}\rd x^1~, }
are the affine top differential on $\S$, and the affine top differential on
the normal bundle to $\S\subset\CP4_{(1,1,2,2,2)}$, respectively.

Thus, in addition to the residues that the above analysis covers, we now
also have to consider
\eqn\eExc{ \oint_{\G(P)} {\rd_\perp^2x\over P} \d P~, \qquad{\rm and}\qquad
           \oint_{\G(P)} {\rd_\|^3x\over P} \d P~, }
providing a degree-2 and a degree-6 contribution. As before, these may be
written as
\eqn\eThree{ \bigg[\sum_{i,j,{k_\star}=0}^1
    {\e_{ij}\rd x^j\over \vd_i\vd^{\mv}P}\bigg]_\perp\>
               \d\big(\vd^{\mv}P\big)_\|~,\quad \mv\cdot\kv=6\,, }
and
\eqn\eXXX{ \bigg[\sum_{i,j,k,l=2}^4
    {\e_{ijk}\rd x^j\rd x^k\over \vd_i\vd^{\mv}P}\bigg]_\|\>
           \d\big(\vd^{\mv}P\big)_\perp~,\quad \mv\cdot\kv=2\,, }
respectively. Here, the subscript ``$\|$'' denotes restriction to local
coordinates on $\S$ while ``$\perp\,\S$'' labels a restriction to local
fibre coordinates of the normal bundle of $\S\subset\CP4_{(1,1,2,2,2)}$.
The manifest factorization is very similar to that in \SS~\SReap, as
discussed in~\sTheResW. In both cases, the general variation of the
multi-derivative of $P$ simply becomes a general polynomial of degree 2
and 6, respectively.

However, only the former of these contributes. To see this, note that
$\d\big(\vd^{\mv}P\big)_{\|\,\S}$ is a degree-2 polynomial
in coordinates of the exceptional set $\S$, and which are also fixed by
$j^4$. Clearly, any linear combination of $x^2,x^3,x^4$ will do, whence a
3-parameter family of such contributions.
 By contrast, $\d\big(\vd^{\mv}P\big)_{\perp\,\S}$ is a degree-6 polynomial in
coordinates which are normal to the exceptional set, $x^0,x^1$, both of
which are however `projected out' by $j^4$.
 Finally, note that the former contribution is a 1-form before returning
from the affine $\IC^5_{(1,1,2,2,2)}$ to the projective
$\CP4_{(1,1,2,2,2)}$, whence we conclude that~\eThree\ supplies, upon
\hbox{re-}projectivization, three polynomial-valued 0-forms. As usual,
these then contribute to the
$H^1(\Tw{\cal M},{\cal T}_{\Tw{\cal M}})\approx H^{2,1}(\Tw{\cal M})$.
Note that there is no action of coordinate reparametrizations on this;
there are no appropriate residues of $x^i\l_i^j\vd_j$. These three
contributions {\it exactly} correspond to the massless twisted $(c,c)$
states of the form $X_i\ket{{3\over4},{3\over4}}^4_{NS}$, $i=2,3,4$, and
also correspond to the three (2,1)-forms we have described above.

Next we reconsider the reparametrization operator-valued residues, and see
that, owing to the unequal scaling weights, there are two separate classes
of reparametrization (degree-0 and $j$-invariant) operators:
\eqn\eNorB{ \vq_{\CP4}  \define x^i\l_i^j\vd_j~, \qquad i,j=0,\ldots,4~, }
and
\eqn\eTanB{ \vq_\S \define x^i\l_i^j\vd_j~, \qquad i,j=2,3,4~. }
The mixed operators
\eqn\eMixB{ \vq_{\rm mixed} \define
             x^i x^j\l_{ij}^k\vd_k~, \qquad i,j=0,1\quad k=2,3,4~, }
merely contribute to the reparametrization of~\eTanB, and are anyway
`projected out' by $j^4$. Therefore, in place of a single K\"ahler form
with homogeneous projective spaces, we now obtain a 2-component residue
class
\eqn\eKahW{ \oint_{\G(P)} {\rd^5x\over P}
            \big(\,\vq_{\CP4} \oplus \vq_\S\,\big) }
and so two 2-forms: the differential order is decreased by three, as in
the general case considered above. These two 2-forms again match the
massless twisted $(a,c)$ states $\ket{{-}1,1}^4_{NS}$ and
$\ket{{-}1,1}^7_{NS}$, and also the two (1,1)-forms described above.

Together with the usual residues obtained as described in the preceding
sections, this then completes the complete residue representation of the
massless {\bf27} and {\bf27*} states for $\CP4_{(1,1,2,2,2)}[8]$. It is in
complete and precise 1--1 correspondence with both the geometrical
description given above, the \LGO{} description (when restricted to the
complex deformation moduli space) \`a la
Refs.~\refs{\rLGO,\rChiRi,\rGVW} and so also with the results \`a la
Ref.~\rSKEW. The non-exceptional part of the analysis is in a similarly
detailed 1--1 correspondence with the Koszul computation, which is then
generalized through the inclusion of the exceptional residues~\eExc\
and the second contribution in~\eKahW. Note however that this still does
not provide a weighted Koszul calculation by itself: the exceptional
contributions are found via the exceptional residues. A `purely' weighted
Koszul calculation (and so also a weighted Bott-Borel-Weil Theorem) can
hopefully be developed in the context of equivariant cohomology, but this
is well beyond the scope of this article.

\newsec{Residue Rings}\seclab\SRings\noindent
 \phrasing{a rip-roaring revelry}{revivando}
The foregoing has established a 1--1 correspondence between the
above residue calculations and the Koszul calculations of
Refs.~\refs{\rEndT,\rBeast}. It should be clear that this correspondence
provides a residue representation not only for the $E_6$ {\bf27}'s and
{\bf27\con}'s, but also to the $E_6$ {\bf1}'s. Calculations of the
{\bf1}-spectrum is typically rather more involved~\refs{\rEndT,\rSKEW} and
will not be detailed here.

On the other hand, the ring structure of the moduli fields for the complex
structure  as determined from the Koszul calculations and from the \LGO{}
analysis turns out to be remarkably similar~\rCQG, and in fact has to be
the same; changing the K\"ahler structure in going from the
Landau-Ginzburg phase to the
large radius limit cannot affect the ring structure of the complex
structure moduli. In particular it means that we can use selection rules
based on ``quantum symmetries'' obtained at the Landau-Ginzburg point, in the
\CY\ phase. That is, as long as this symmetry is not broken by deforming in a
direction given by a moduli from one of the fixed point sets. In other
words we can blow up the fixed point set in a way that keeps the shape of
the blow up and only affects the size of it, by varying the toric divisor
which came from the reparametrization of the fixed point set.  Of course,
since the model is no longer at the specially symmetric point in the
K\"ahler moduli space, we can no longer use the quantum symmetry there and
hence there will be no straightforward selection rules for the Yukawa
couplings among the $(1,1)$ forms.

The present residue representation is clearly simply extending the
polynomial deformation analysis of Ref.~\rPhilip, and so inherits
the same ring structure. That is, the Yukawa couplings for the
non-polynomial deformations, such as those specified in~\eRLinRep,
are easily calculated---by exactly the same method as in
Ref.~\rPhilip. The only difference being that the radical factors
will have to occur in such products that the Yukawa coupling product
would become (modulo the reparametrization ideals~\eLinRep\
and~\eRLinRep) proportional to the `top-degree' polynomial. For example,
in the model~\eWarp, this `top-degree' polynomial is proportional to
\eqn\eQExI{ \det\Big[\vd^2\big(f(x){\oplus}g(x,y){\oplus}h(y,z)\big)\Big]
 ~\cong~ \bigg|\ppd{f(x)}{x^a}{x^b}\bigg|{\cdot}
          \bigg|\ppd{g(x,y)}{y^\a}{y^\b}\bigg|{\cdot}
           \bigg|\ppd{h(y,z)}{z^r}{z^s}\bigg|~, }
modulo the Jacobian ideal (generated by gradients of $f\6(x)$, $g\6(x,y)$
and $h\6(y,z)$).

Notice that the additional 9 representatives~\eRRepW{a'} were multiplied by
the radical~\eSqRtDDH. It is then straightforward that only couplings with
an even (including zero) number of such `radical deformations' may be
non-zero. This produces a straightforward selection rule which, matches the
effect of the selection rule based on the quantum symmetry in the
corresponding \LGO{}. However, this is not at all surprising but should be
{\it expected}, since on general grounds a variation in the K\"ahler
moduli space does not affect the $(c,c)$-ring. In fact, following the
analysis in Ref.~\rCfC, it can be shown that this extension of the by now
standard Yukawa coupling calculation perfectly agrees with the
corresponding \LGO{} calculations. It also matches the general Yukawa
coupling formula obtained for the Koszul calculation~\refs{\rEndT,\rBeast}.
Moreover, as in Ref.~\rCfC, the calculations can be performed both for the
model~\eWarp\ ``as is'', and also for its `ineffectively split' variant in
which the additional 9 representatives~\eRRepW{a'} become ordinary
polynomial deformations whereupon the standard calculations apply
straightforwardly.

The {\bf27\con} Yukawa couplings, on the other hand, are easiest to
determine using the `dual' description~\eDuals, where they become
2-forms and the usual (`topological') Yukawa coupling may be calculated
straightforwardly. Of course, the instanton-corrected Yukawa couplings
are best calculated using mirror symmetry, for which techniques are
being vigorously developed~\refs{\rCdGP,\rPeriods,\rCdFKM,\rHKTYI,\rCFKM,
\rSpokes,\rHKTYII}.

\newsec{Resolvents and Radicals}\seclab\SRad\noindent
 \phrasing{a retrograde ricercar}{ritartando}
We wish to present an alternative and perhaps even more heuristic
derivation of the above results. Let us therefore focus, for the moment, on
the deformations of the complex  structure which may be realized as
deformations of the defining polynomial  system. This does exhaust all
deformations of complex structure for all homogeneous hypersurfaces, but
not so for their quasihomogeneous (weighted) cousins. Consider then, for
the moment, a simple $n$-dimensional hypersurface
${\cal M}\define\{P{=}0\}\subset{\cal X}$, where $\cal X$ is some
homogeneous space (or product thereof).

As is well known, the choice of $P$ determines the complex structure of its
zero-set, $\cal M$. $P$ also determines the holomorphic volume-form $\W$ on
$\cal M$ via Eq.~\eOm{} and, indeed, the choice of $\W$ among all elements
of $\oplus_{p+q=n}H^{p,q}({\cal M})$ is equivalent to the choice of the
complex structure. The variations of $\W$ also correspond to variations of
the complex structure (see Eq.~\eSpecGeom) and we presently examine this
by direct calculation.

To that end, deform $P \to P-t^\a \d P_\a$, and calculate
 $$ \pd{\W}{t^\a} ~=~
 \oint_{\G(P)}{(x\rd^{n+1}x)\over P}\Big({\d P_\a\over P}\Big)~.
\eqno\ePDM
 $$
Iterating this $n$ times,
\eqn\eThreeDer{ {\vd^n\W\over\vd t^{\a_1}\cdots\vd t^{\a_n}} ~=~
             {1\over n!}\oint_{\G(P)}{(x\rd^{n+1}x)\over P}
             {\d P_{\a_1}\over P}\cdots{\d P_{\a_n}\over P}~, }
which becomes the ``Yukawa'' $n$-point coupling upon multiplying $\ba\W$
and integrating over the manifold~\rPhilip.
Thus, the homogeneity degree-$\vec0$ rational polynomials $\d P_\a/P$,
taken however modulo terms which merely rescale $\W$, may be identified
with elements of $H^1({\cal T})$, i.e., with tangent vectors $\nabla_\a\W$
to the moduli space. Such polynomial deformations of the complex structure
have been studied in great detail so far both from the geometrical point of
view where $P-t{\cdot}\d P$ is the (deformed) defining equation of a
manifold,  and also from the \LGO{} point of view, where $P-t{\cdot}\d P$
is the (deformed) superpotential. An iteration of~\ePDM\ and integration by
parts leads to the Picard-Fuchs equations~\refs{\rPicFux,\rCF}, which
provide additional information for the Special Geometry
calculations~\refs{\rCdFKM,\rCFKM}.

We will not pursue these considerations here, but instead turn to a little
more involved example: ${\cal M}\in\cnfg{\IP^4\cr\IP^1\cr}{4&1\cr0&2\cr}$,
defined as the common zero-set of $f(x)$ and $g(x,y)$, which have degree
$(4,0)$ and $(1,2)$, respectively, over the embedding space
$\IP^4{\times}\IP^1$~\rPDM. Write $f_0,g_0$ for the reference choice of
polynomials and consider a deformation of the holomorphic volume-form $\W$:
\eqn\eXXX{ \W ~\define \oiint_{\G(f){\times}\G(g)}
                        {(x\rd^4x)(y\rd y)\over(f_0-\d f)(g_0-\d g)}~, }
so that $\W_0$ refers back to the reference choice of the complex
structure, where $\d f=0=\d g$.

Now expand to first order in $\d f,\d g$:
\eqn\eExp{ \W~ = ~\W_0~ +
 ~\oiint_{\G(f){\times}\G(g)}\!\!\!{(x\rd^4x)(y\rd y)\over f_0~g_0}
            \Big({\d f\over f_0}\Big)~ +
 ~\oiint_{\G(f){\times}\G(g)}\!\!\!{(x\rd^4x)(y\rd y)\over f_0~g_0}
            \Big({\d g\over g_0}\Big)~. }
On the face of it, these two (double) residues are simply the polynomial
deformations of complex structure, as discussed above.

However, note that the first double residue also produces another
contribution when calculated in the following stepwise fashion:
\eqna\eStp
 $$\eqalignno{
 \oiint_{\G(f){\times}\G(g)}\!\!\!{(x\rd^4x)(y\rd y)\over f_0~g_0}
            \Big({\d f\over f_0}\Big)
 &=\oint_{\G(f)}{(x\rd^4x)\over f_0^{~2}}~
    \bigg[\oint_{\G(g)}{(y\rd y)\over g_0}\d f\>\bigg]         &\eStp{a}\cr
 &=\oint_{\G(f)}{(x\rd^4x)\over f_0}\Big({\d f'\over f_0}\Big) &\eStp{b}\cr
}$$
where
\eqn\eXXX{ \d f'~ \define ~\oint_{\G(g)}{(y\rd y)\over g_0}\d f~
           = ~\W_{(0)}^a \vd_af(x) }
and
\eqn\eXXX{ \W_{(0)}^a~ \define ~\oint_{\G(g)}{(y\rd y)\over\vd_a g_0}~, }
are {\it exactly} as defined in \SS\sTheRes, and used throughout the
foregoing analysis! We emphasize, however, that this occurrence of the
partial residue~\eRes{} owes to our insisting that the intermediate
residue integral within the square brackets in~\eStp{a} should have an
independent meaning. On comparison with the twisted states in \LGO{}, we
note that, formally at least,
$\oint_{\G(f)}\!\!{(x\rd^4x)\over f_0}\,\W_{(0)}^a$ plays the r\^ole
of the twisted vacuum, while $\vd_af(x)$ is exactly the monomial part.

In fact, it should be clear that all the polynomial-valued residues can be
recovered by simply expanding the ``main'' residue in such a fashion; the
Reader should encounter no difficulty in recovering all of the above
results in this alternate manner.

{\it A posteriori} at least, this focus on the independent r\^ole of the
intermediate residues may be argued by direct comparison with the results
from other methods. However, considerable further work seems to be
required to recover all the reparametrization relations and degrees of
freedom and in this respect, this approach is presently lacking.
Nevertheless, this approach has the virtue of being a straightforward and
direct study of the deformations of the complex structure, by studying
{\it directly} the deformations of the holomorphic volume-form. The Koszul
computations, with all the charts and maps and `filtration'\3, may then be
viewed as a bookkeeping device for racking all the myriad of possible
contributions to the deformations of $\W$ and their various relations.\ping

The form of Eq.~\eStp{b}, perhaps more forcibly than any argument before,
suggests reinterpreting $\d f'$ as a deformation of the original defining
polynomial $f_0(x)$. Of course, it is not possible simply to deform
$f_0(x) \to f_0(x) + \d f'(x)$.
 From the \LGO{} side, the difference in the degrees of (quasi)homogeneity
of $f_0(x)$ and of $\d f'(x)$ implies that their linear combination as the
superpotential explicitly breaks the quantum symmetry; the IR fixed point
would be determined by $\d f'$, as that one has a lower degree. This is
precisely {\it not} what we are after; it says nothing about marginal
deformations of the model with $f_0(x)$ in the superpotential.
 From the geometrical side, the deformation $f_0(x) \to f_0(x) + \d f'(x)$
simply makes no sense! It is nowhere well-defined on the {\it projective}
embedding space, in view of the different degrees of homogeneity in $f'$
and $f$. Furthermore, it should be clear that the $\d f'$ cannot possibly
provide {\it bona fide} deformations of the defining polynomial, since
$\d f'$ corresponds to deformations of the complex structure which
precisely are {\it not} deformations of the embedding!

Thus --- if the expansion~\eExp, eventually containing $\d f'(x)$ ---
is to be collapsed back somehow differently, so that $\d f'$ would appear
explicitly in a variation of $f_0(x)$, $\d f'$ must come multiplied by a
factor $\mit\D$ of compensating degrees of homogeneity:
\eqn\eRadDef{ f_0(x)~ \longrightarrow ~f_0(x) + {\mit\D}\,\d f'(x) }
In addition, this $\mit\D$ must be `universal', that is, it must be
independent of the deformation parameters. We now observe that for all
complete intersections in products of {\it homogeneous} (rather than
quasihomogeneous) flag spaces, the Koszul calculation obtains certain
Levi-Civita alternating symbols in place of $\mit\D$. These can, in turn,
{\it always} be identified with square-roots of certain precisely
corresponding determinants~\rCfC.

For the quasihomogeneous models, this identification of $\mit\D$ with
radicals is neither straightforward nor is it clear that such radicals
would always exist. In fact, the three exceptional contributions to
$H^1({\cal M},{\cal T_M})$ in~\eThree, for the case
${\cal M}\in\CP4_{(1,1,2,2,2)}[8]$ have degree 2, and need a $\mit\D$ of
degree 6. Straightforwardly, and following the experience from the
homogeneous cases, one tries
\eqn\eSqRt{ \sqrt{\det\big[\vd_i\vd_j P(x)\big]_{\|\,\S}}~. }
That is, $i,j=2,3,4$ are restricted to the coordinates `along' $\S$. This
indeed has the correct degree and has the virtue of being in the same 1--1
correspondence with the $\S$-restriction of the Levi-Civita symbol which
persists for all homogeneous cases~\rCfC. However, strange things may
happen: if $P(x)$ is chosen to be the most popular of all, the Fermat
polynomial,
\eqn\eWRad{ P_F(x)~ = ~(x^0)^8 + (x^1)^8 + (x^2)^4 + (x^3)^4 + (x^4)^4~, }
then
\eqn\eXXX{ {\eqalign{\sqrt{\det\big[\vd_i\vd_j P(x)\big]_{\|\,\S}}~
 &= \sqrt{\det\big[\d_{ij}\,4{\cdot}3(x^i)^2\big]}~,\qquad i,j=2,3,4~,\cr
 &= 24\sqrt{3}\,x^2 x^3 x^4~,\cr }} }
is not radical at all! Moreover, the product of this and the
$\d(\vd_\perp^6P)$ from~\eThree\ becomes a linear combination
\eqn\eXXX{ (x^2 \oplus x^3 \oplus x^4)\,x^2 x^3 x^4~, }
which were already accounted for among the `plain' polynomial
deformations. Of course, for a generic reference defining polynomial
$P(x)$, no such degeneration occurs and the radical~\eWRad\ is indeed
radical. This is not unlike the situation encountered in \SS~\sWarp, where
the analogous radical vanished for a simple choice of defining equations.
The conditions for the required radical to either vanish or degenerate
into a non-radical polynomial seem to be independent of well-definedness;
that means that such degenerations are to be expected in every family of
well behaved \CY\ and/or \LGO{} models. Until it is understood precisely
why such degenerations occur at innocuous albeit special reference defining
polynomials, the identification of the factor $\mit\D$ with such radicals
certainly cannot be regarded {\it universal}.

We emphasize again that the deformation~\eRadDef\ is {\it formal} and does
not correspond to actual deformations of the embedding. By the same token,
the introduction of this universal factor $\mit\D$ and then its
identification with the radicals of the general type~\eSqRt\ is at best
only equally formal and may serve for Yukawa coupling calculations~\rCfC,
and so also for the calculation of periods of $\W$ by direct
integration~\refs{\rPeriods,\rSpokes}.

Eventually, the residues representing these `higher cohomology' and
`twisted' deformations of the complex structure should be obtainable in
the framework of Ref.~\rPhases. Suffice it here to note the following. In
the 2-dimensional field theory, the representatives we have been studying
should correspond to marginal operators. In a correlation function
(the Yukawa coupling), these would appear something like
\eqn\eXXX{ {\eqalign{\V{0|\cdots({\mit\D}\,\d f')\cdots|0}~
 &= ~Z_0^{{-}1}\int D[\f]\cdots({\mit\D}\,\d f')\cdots e^{{-}S_0}~,\cr
 &= ~Z_0^{{-}1}\int D[\f,\j]\cdots(\d f')\cdots e^{{-}(S_0+S_2)}~,\cr}}}
where $\f$ denote the scalar field zero-modes (the contribution from
path-integration over nonzero-modes canceling between fermions and bosons)
and $\d f'$ is the polynomial part of such a radical deformation.
The second equality follows on noting that the additional factor $\mit\D$
(at least in all above examples and certainly all homogeneous complete
intersections in projective spaces~\rCfC) turns out to be a square-root of
a determinant such as~\eSqRt, and so can be `\hbox{re-}exponentiated' by
Gaussian integration over anticommuting
$\j$'s. The so
\hbox{re-}exponentiated term is then
\eqn\eXXX{ S_2~ = ~\int\rd^2\s~ \big(\j^i\,\vd_i\vd_j P(\f)\,\j^j\big)~, }
where ${\mit\D}=\sqrt{\det[\vd_i\vd_j P(\f)]}$ and $P$ is typically one of
the defining polynomials or some part thereof, as in~\eSqRt.
 Clearly, such a term is routinely present in the supersymmetric completion
of the action, and will contribute in the correlation function if the
$\j$'s are not paired with commuting modes, \ie, if the $\j$'s are
Fermionic zero-modes. Even in purely bosonic $\s$-models, such terms can
arise in a BRST-type treatment of constraints, where the $\j$'s would then
be (odd order) ghost variables. In any case, this would seem to provide a
straightforward field-theoretic explanation of the radical deformations.

 %
\bigskip\vfill\noindent
{\it Acknowledgments\/}:
We thank Paul Green and Sheldon Katz for helpful and encouraging
discussions.
 P.B. would like to thank the Institute for Theoretical Physics, Santa
Barbara, the Theory Division, CERN, Geneva and the Aspen Center for Physics,
Aspen where part of this research was carried out.
 P.B.\ was supported  by the American Scandinavian Foundation,
the Fulbright Program, the Robert~A.~Welch foundation,
the NSF grants PHY 8904035, 9009850, and the DOE grant DE-FG02-90ER40542.
 T.H.\ was supported in part by the Howard University FRSG~'93 Program,
and the DOE grants DE-FG02-88ER-25065 and DE-FG02-94ER-40854.

\bigskip\vfill

\listrefs

 %
\bye